\documentclass[fleqn,usenatbib]{mnras}

\usepackage{newtxtext,newtxmath}

\usepackage[T1]{fontenc}
\usepackage{subcaption}
\DeclareRobustCommand{\VAN}[3]{#2}
\let\VANthebibliography\thebibliography
\def\thebibliography{\DeclareRobustCommand{\VAN}[3]{##3}\VANthebibliography}

\usepackage{graphicx}	
\usepackage{amsmath}	

\title[Dynamo transition: non-Helical to Helical flows]{Revisiting \textcolor{black}{Kinematic Fast} Dynamo in 3-dimensional magnetohydrodynamic plasmas:  Dynamo transition from non-Helical to Helical flows}

\author[Biswas \& Ganesh 2022.]{
Shishir Biswas,$^{1, 2}$\thanks{E-mail: shishirbeafriend@gmail.com, shishir.biswas@ipr.res.in}
Rajaraman Ganesh,$^{1, 2}$\thanks{E-mail: ganesh@ipr.res.in}
\\
$^{1}$Institute for Plasma Research, Bhat, Gandhinagar, Gujarat  382428, India\\
$^{2}$Homi Bhabha National Institute, Training School Complex, Anushaktinagar, Mumbai 400094, India\\
}

\date{Accepted XXX. Received YYY; in original form ZZZ}

\pubyear{2022}

\graphicspath{{./Files/}}
\begin{document}
\label{firstpage}
\pagerange{\pageref{firstpage}--\pageref{lastpage}}
\maketitle

\begin{abstract}
Dynamos wherein magnetic field is produced from velocity fluctuations are fundamental to our understanding of several astrophysical and/or laboratory phenomena. \textcolor{black}{Though fluid helicity is known to play a key role in the onset of dynamo action, its effect is yet to be fully understood. In this work, a fluid} flow proposed recently  [Yoshida et al. Phys. Rev. Lett. 119, 244501 (2017)] is invoked such that one may inject zero or finite fluid helicity using a control parameter, at the beginning of the simulation. Using a simple \textcolor{black}{kinematic fast} dynamo model, we demonstrate unambiguously the strong dependency of short scale dynamo on fluid helicity. \textcolor{black}{In contrast to conventional understanding, it is shown that fluid helicity does strongly influence the physics of short scale dynamo. To corroborate our findings, late time magnetic field spectra for various values of injected fluid helicity is presented along with rigorous ``geometric'' signatures of the 3D magnetic field surfaces, which shows a transition from ``untwisted'' to ``twisted'' sheet to ``cigar'' like configurations.} It is also shown that one of the most studied ABC dynamo model is not the ``fastest'' dynamo model for problems with lower magnetic Reynolds number. This work brings out, for the first time, the role of fluid helicity in moving from ``non-dynamo" to ``dynamo" regime systematically.
\end{abstract}

\begin{keywords}
\textcolor{black}{fluid simulation} -- plasmas -- \textcolor{black}{magnetohydrodynamics} -- dynamo -- magnetic fields
\end{keywords}



\section{Introduction}

The theories of HydroDynamics (HD) \citep{Kraichnan:1965, Brachet:1988} and MagnetoHydroDynamics (MHD) \citep{biskamp:2003, beresnyak_mhd:2019, Schekochihin:2022} are often used to analyze the HD turbulence  and magnetized plasma turbulence respectively, which are fundamental to our understanding of the behaviour of astrophysical plasmas present in the Sun or other young stars. The presence of \textcolor{black}{small scale,} mean or large scale magnetic fields are very much common in such astrophysical objects, planets, stars, interstellar medium, galaxy, accretion disks and also in the Sun \citep{amitava:2014a, amitava:2014b, amitava_APJ:2015}. Understanding the origin of these \textcolor{black}{multi} scale magnetic fields, which are in turn responsible for various complex phenomenon in the universe, is of paramount importance. Naturally a relevant question arises \textcolor{black}{as to}, what are the sources of these \textcolor{black}{multi} scale magnetic energy? As first pointed out by Parker, \textcolor{black}{such} magnetic fields are generated by the motions of conducting fluids through a transfer of kinetic to magnetic energy, that is, via Dynamo action \citep{parker:1979}.


\textcolor{black}{Depending on the length scales involved, dynamos are broadly classified into two categories :  Small Scale Dynamo (SSD) and Large Scale Dynamo (LSD) or mean field dynamo. For LSD, it is believed that a lack of reflectional symmetry (or in other words, nonzero fluid helicity) to be a crucial element, whereas for SSD, no correlation has been shown to exist between onset and sustenance of SSD on fluid helicity \citep{rincon_dynamo:2019}.  Depending on the time scale, dynamos may further divided into two categories: Fast dynamos (growth rate can remain finite in the limit $R_m \to \infty$) and Slow dynamos (magnetic diffusion plays a significant role) \citep{rincon_dynamo:2019}. Depending on the feedback strength of the magnetic field on to the fluid motion, dynamos are categorized as linear or non-linear. A linear dynamo is one in which the magnetic field dynamics does not ``back react'' with the velocity field and the velocity field is either given or it obeys the NS equation \citep{rincon_dynamo:2019}, whereas a self-consistent dynamo or a nonlinear dynamo tends to change the flow - as the magnetic field becomes large enough - to further impede magnetic field growth. That is, the flow and the B-field ``back react'' on each other leading to a nonlinear saturation \citep{rincon_dynamo:2019}.}

\textcolor{black}{In the present study, we focus on a kinematic fast dynamo model, wherein velocity and magnetic fields are un-correlated. As indicated earlier, in a fast dynamo model, below a certain value of resistivity, the growth rate becomes insensitive to the magnitude of resistivity, i.e, becomes independent of resistivity.  \citep{childress_STF:1995}.}
A popular example of fast dynamo is the Solar dynamo \citep{choudhuri:1990}.

It is now well acknowledged that, to achieve fast dynamo, an  \textcolor{black}{careful} selection of plasma  flow profile is necessary, as the key mechanism behind the fast dynamo action is that \textcolor{black}{of} stretching, twisting and folding (STF) of magnetic field lines, ``advected'' in the fluid flow. Hence the flow is expected to have ``enough dynamcs'' to amplify the magnetic field exponentially.

\textcolor{black}{As discussed before, depending on
	the length scale involved, dynamos are largely classified into two broad categories
	: Small Scale Dynamo (SSD) and Large Scale Dynamo (LSD). For a LSD or mean
	field dynamo, a lack of reflectional symmetry is required, where as conventional understanding is that, no such condition is expected to be satisfied for SSD \citep{rincon_dynamo:2019}.} Among various astrophysical flows addressed, the Arnold-Beltrami-Childress [ABC] flow is a well known prototype for fast dynamo action for its stretching ability and chaotic transport properties. A detailed dynamo study using ABC flow has been reported by various authors, eg. \citep{arnold:1983, Frish_Dynamo:1984, dombre:1986, galanti:1992, Dorch:2000, Archontis:2003, Bouya:2013}. Along with the exponential growth of magnetic energy, the magnetic energy iso-surface \textcolor{black}{are known to exhibit} special signature called ``cigar'' like \citep{Frish_Dynamo:1986} or ``ribbon'' like \citep{Frish_Dynamo:1986,Archontis:2003} structure. For ABC flow profile, \textcolor{black}{the vorticity field is parallel to the velocity direction,} i.e. $\vec{\nabla} \times \vec{u}$ has a finite parallel component to $\vec{u}$ at all times. This naturally indicates that, finite fluid helicity [$\int_{V} \vec{u} . (\vec{\nabla} \times \vec{u}) dV$] is present in the flow. The fluid helicity effect on dynamo action is often expressed as $\alpha$ effect, and studied extensively using ABC flow family \citep{alpha_effect:2006}.

\textcolor{black}{Study of dynamo instability using a 3D flows becomes numerically demanding and is perhaps relatively more complex. To reduce the complexity in numerical methods often, time dependent two dimensional flows [$\vec{u}(x, y, t)$] have been taken into consideration for the investigation of fast dynamo action. Naturally, the presence of time dependency introduces chaoticity in such flows. Galloway-Proctor flow, sometimes known as GP flow, is one such well-known flow \citep{galloway_nature:1992}. In light of the fact that the flow profile is entirely independent of one spatial coordinate, in this case the z-coordinate, it is conceivable to seek monochromatic solutions for the magnetic field of the form $\vec{B}(x, y, z, t) =\vec{B}(x, y, t) e^{ik_zz}$, so that the problem becomes two-dimensional for a given value of $k_z$. In past several authors \citep{Cattaneo_PRL:1995, Cattaneo_PRL:1996, Hughes_PLA:1996, Brandenburg_PRE:2003} have investigated the dynamo instability using Galloway-Proctor (GP) flow. Recently GP flow \citep{galloway_nature:1992} has been taken into consideration for studying large scale magnetic fields in the presence of velocity shear \citep{tobias_shear:2013}.}

Some of the interesting and relevant questions are: is the \textcolor{black}{small-scale kinematic} fast dynamo action possible only for chaotic ABC flow \textcolor{black}{and time dependent GP flow?} what is the exact role of fluid helicity in the context of \textcolor{black}{small-scale} fast dynamo action? Is there flow field using which one can  systematically inject fluid helicity in the system and clearly demonstrate a non-dynamo to dynamo transition when the fluid flow transits from non-helical to helical flow? \textcolor{black}{Does the fluid helicity affect short-scale dynamo action?}

\textcolor{black}{In past several authors examined the influence of the flow helicity in the context of kinematic fast dynamo action. For example, \citet{Hughes_PLA:1996} show that the small scale, fast dynamo will always exist regardless of the flow helicity distribution using three different types of helicity distribution (non zero local and global helicity, non zero local helicity but zero global helicity, and zero local and global helicity). The flow considered by \citet{Hughes_PLA:1996} is a time dependent 2-dimensional flow [$\vec{u}(x,y,t)$] \citep{galloway_nature:1992}. As previously mentioned, the induction equation for magnetic field supports a monochromatic solution of the form $\vec{B}(x,y,t) \times e^{ik_zz}$, for a given $k_z$ value, the resulting problem for the magnetic field is two-dimensional. The most significant finding was that the helicity distribution of the driving flow did not significantly affect the small-scale fast dynamo action \citep{Hughes_PLA:1996}. Additionally, it was pointed out that as fluid helicity increases, the dynamo growth rate noticeably increases. However, these Authors quickly clarify that this is only coincidence \citep{Hughes_PLA:1996}.}

\textcolor{black}{By keeping these earlier ideas in mind,} we study the \textcolor{black}{fast} dynamo action using a newly proposed Yoshida-Morrison flow or YM flow \citep{EPI2D:2017} \textcolor{black}{to investigate the role of helicity on SSD.} \textcolor{black}{It is important to indicate that the whole class of YM flow depends upon all three spatial coordinates i.e,  $\vec{u}(x,y,z)$ similar to ABC flow. We have considered small homogeneous ambient initial magnetic field in all 3 directions as considered earlier by  \citet{Frish_Dynamo:1984, Dorch:2000, Archontis:2003}.} \textcolor{black}{Another} interesting and useful aspect of YM flow is that, it is possible to inject finite fluid helicity [$\int_{V} \vec{u} . (\vec{\nabla} \times \vec{u}) dV$] in the system, by systematically varying certain physical parameter.  More over this flow has a special property that, it establishes a topological bridge between \textcolor{black}{quasi} 2-dimensional class and  3-dimensional class of flows. \textcolor{black}{Therefore, our present study focuses on various flows with varying kinetic energy, chaotic characteristics, and helicity on the onset and sustenance of SSD.}

In the present work, we propose a new possible route that connects non-dynamo regime to dynamo regime via fluid helicity injection using Yoshida-Morrison (YM) flow as a prototype. \textcolor{black}{Our numerical observation demonstrates that small scale dynamo is feasible for
	this type of initial condition. Our work shows unambiguously that fluid helicity does affect the dynamics of small-scale dynamos. This finding is in complete contrast to the
	earlier results \citep{Hughes_PLA:1996}.} Our findings systematically connect most of the previous works and brings in several new insights. \textcolor{black}{To corroborate our findings, time dependent magnetic energy spectra, for various magnitudes of injected fluid helicity is calculated.} We also show that, how a ``cigar'' like iso-surface, which is a signature of fast dynamo action, emerges naturally starting from a non-dynamo iso-surface. Our numerical investigation also suggests that, for lower magnetic Reynolds number ($R_m$) experiments, the regular ABC dynamo model is not the best dynamo model.

The organization of the paper is as follows. In Sec. 2 we present about the dynamic equations. About our numerical solver and simulation details are described in Sec. 3. The initial conditions, parameter details are shown in Sec. 4. Section 5 is dedicated to the simulation results on induction dynamo action that we obtained from our code and finally the summary and conclusions are listed in Sec. 6.

\section{Governing Equations}\label{Equations}
The governing equations to study \textcolor{black}{Kinematic Fast} dynamo action for the single fluid MHD plasma are as follows,
\begin{eqnarray}
	&& \label{Bfield} \frac{\partial \vec{B}}{\partial t} + \vec{\nabla} \cdot \left( \vec{u} \otimes \vec{B} - \vec{B} \otimes \vec{u}\right) = \textcolor{black}{\frac{1}{R_m}} \nabla^2 \vec{B}\\
	&& \label{div B} \vec{\nabla} \cdot \vec{B} = 0
\end{eqnarray}
\textcolor{black}{where, $\vec{u}$, $B$ and $R_m$ represent the velocity, magnetic fields and magnetic Reynolds number respectively. We define Alfven speed as, $V_A=\frac{u_0}{M_A}$, here $M_A$ is the Alfven Mach number of the plasma and $u_0$ is the maximum fluid speed. \textcolor{black}{The initial magnetic field ($B_0$) is determined as from Alfven speed as, $V_A \propto B_0$.} Time is normalized as $t = t_0\times t'$, $t_0 = \frac{L}{V_A}$. The symbol ``$\otimes$'' represents the dyadic between two vector quantities.}
\label{equations}

For solving the above set of equations at high enough grid resolution, one need a suitable scalable numerical solver.
\section{Simulation Details: \textit{GMHD3D} Solver}
In this Section, we discuss the details of the numerical solver  along with the  benchmarking of the solver carried out by us.
In order to study the plasma dynamics governed by MHD equations described above, we have very recently upgraded an already existing well bench-marked single GPU MHD solver \citep{rupak_thesis:2019}, developed in house at Institute For Plasma Research to multi-node, multi-card (multi-GPU) architecture for better performance \citep{GTC}. The newly upgraded GPU based magnetohydrodynamic solver [\textit{GMHD3D}] is now capable of handling very large grid sizes. \textit{GMHD3D} is a multi-node, multi-card, three dimensional (3D), weakly compressible, pseudo-spectral, visco-resistive magnetohydrodynamic solver \citep{GTC}. It uses pseudo-spectral technique to simulate the dynamics of 3D magnetohydrodynamic plasma in a cartesian box with periodic boundary condition. By this technique one  calculates the spatial derivative to evaluate non-linear term in governing equations with a standard $\frac{2}{3}$ de-aliasing rule \citep{dealiasing:1971}. OpenACC FFT library [AccFFT library \citep{Accfftw:2016}] is used to perform Fourier transform and Adams-bashforth time solver, for time integration. For 3D iso-surface visualization, an open source Python based data converter to VTK [Visualization Tool kit] by ``PyEVTK'' \citep{VTK} is developed, which converts ASCII data to VTK binary format. After dumping the state data files to VTK, an open source visualization softwares, VisIt 3.1.2 \citep{visit} and Paraview \citep{paraview} is used to visualize the data.
As mentioned earlier, we have upgraded a well benchmarked single GPU solver to  multi-node, multi-card (multi-GPU) architecture, for this present work, the new solver's accuracy with the single GPU solver has been cross-checked and it is verified that the results match upto machine precision. Further few more benchmarking studies have been performed such as, the 3D induction dynamo effect by  \citep{Frish_Dynamo:1986, Dorch:2000, Archontis:2003}, have been reproduced with ABC flow at grid resolution $64^3$ (Details of these are not presented here.).

To study the induction dynamo action, an accurate selection of ``drive'' velocity field is imperative, which we discuss in the Section to follow.   

\section{Initial Condition}
The study of dynamo action or more precisely \textcolor{black}{kinematic fast} dynamo action in the presence of chaotic 3D Arnold-Beltrami-Childress [ABC] flow has been explored by various authors \citep{Frish_Dynamo:1984, dombre:1986, galanti:1992, Dorch:2000, Archontis:2003, Bouya:2013}. The velocity field for ABC flow is as follows:
\begin{equation}\label{ABC Flow}
	\begin{aligned}
		u_x &= u_0 [ A \sin(k_0z) + C \cos(k_0y) ]\\
		u_y &= u_0 [ B \sin(k_0x) + A \cos(k_0z) ]\\
		u_z &= u_0 [ C \sin(k_0y) + B \cos(k_0x) ]
	\end{aligned}
\end{equation}
Recently Yoshida et al. \citep{EPI2D:2017} proposed a new intermediate class of flow, that establishes a topological bridge between quasi-2D  and 3D flow classes.  The flow is formulated as follows:
\begin{equation} 
	\vec{u} = \alpha \times u_0 \times \vec{u}_+ + \beta \times u_0 \times \vec{u}_-
\end{equation}
with
\begin{align} 
	\vec{u}_+ &= \begin{bmatrix} 
		B sin(k_0y) - C cos(k_0z) \\
		0\\
		A sin(k_0x) \\
	\end{bmatrix}
\end{align}
and 
\begin{align} 
	\vec{u}_- &= \begin{bmatrix} 
		0\\
		C sin(k_0z) - A cos(k_0x)\\
		-B cos(k_0y) \\
	\end{bmatrix}
\end{align}
so that,
\begin{equation}\label{Yoshida_flow}
	\begin{aligned}
		u_x &= \alpha u_0 [ B \sin(k_0y) - C \cos(k_0z) ]\\
		u_y &= \beta u_0 [ C \sin(k_0z) - A \cos(k_0x) ]\\
		u_z &= u_0 [ \alpha A \sin(k_0x) - \beta B \cos(k_0y) ]
	\end{aligned}
\end{equation}
where \textcolor{black}{$k_0$}, $\alpha, \beta$, A, B and C are  arbitrary real constants. As indicated earlier, we dub this flow (Eq. \ref{Yoshida_flow}) as Yoshida-Morrison flow or YM flow. \textcolor{black}{We consider the value of $k_0, \alpha$, A, B and C to be unity for this present study.} The variation of $\beta$ value in YM flow leads to new classes of flows.

\begin{figure*}
	\centering
	\begin{subfigure}{0.32\textwidth}
		\centering
		\includegraphics[scale=0.083]{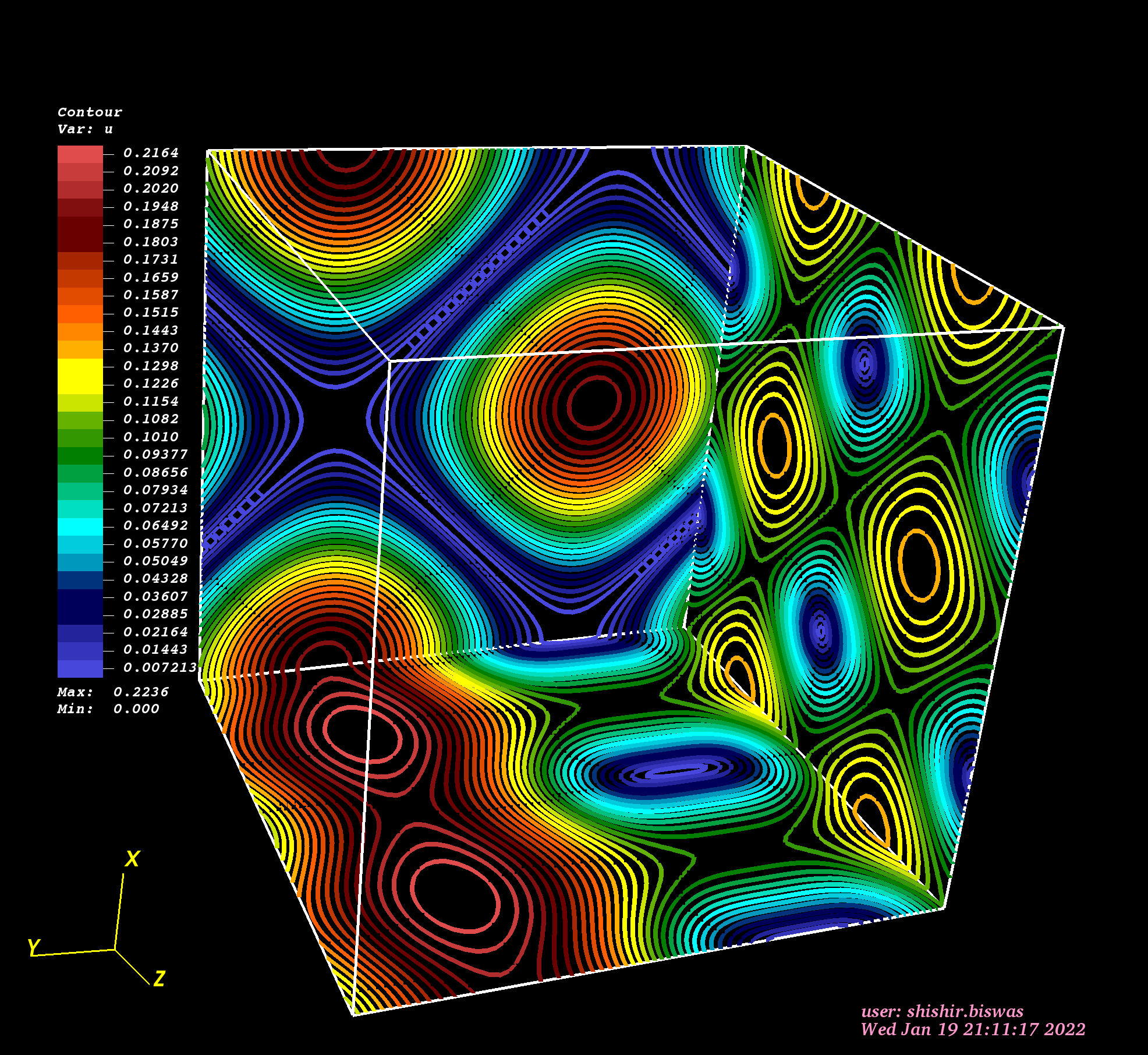}
		\caption{}
		\label{initial flow beta 0}
	\end{subfigure}
	\begin{subfigure}{0.32\textwidth}
		\centering
		\includegraphics[scale=0.083]{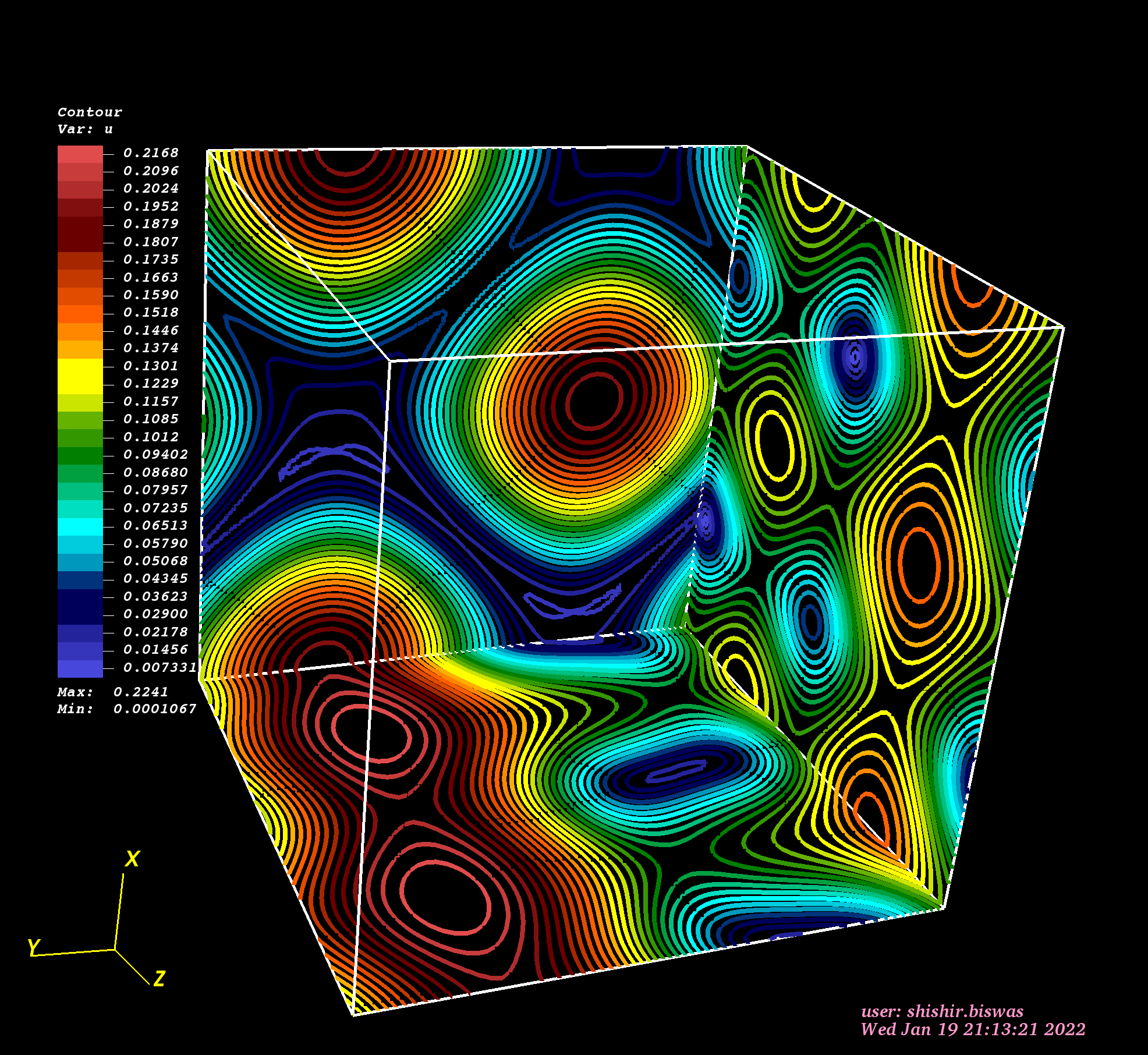}
		\caption{}
		\label{initial flow beta 0p2}
	\end{subfigure}
	\begin{subfigure}{0.32\textwidth}
		\centering
		\includegraphics[scale=0.083]{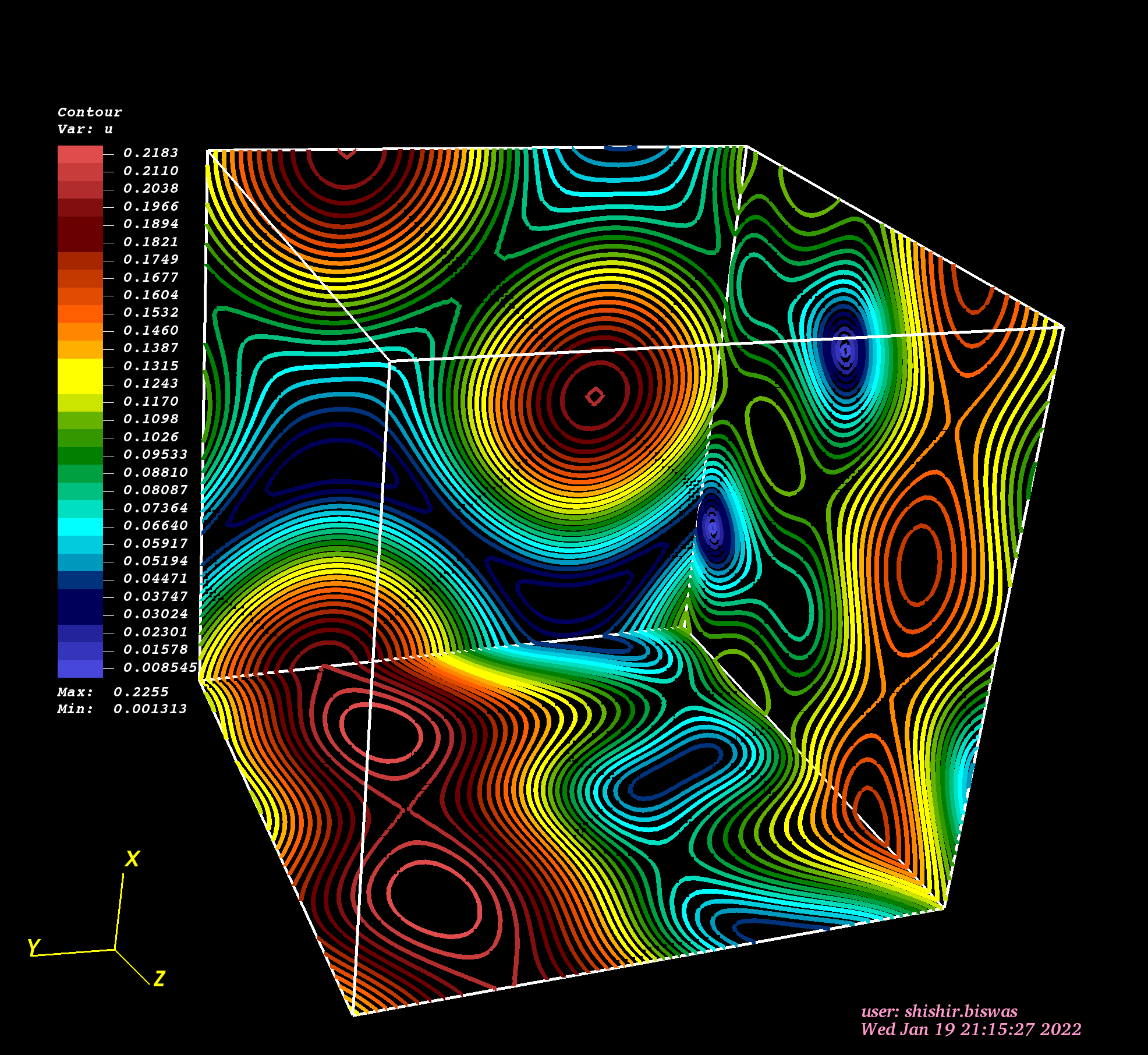}
		\caption{}
		\label{initial flow beta 0p4}
	\end{subfigure}
	\begin{subfigure}{0.32\textwidth}
		\centering
		\includegraphics[scale=0.083]{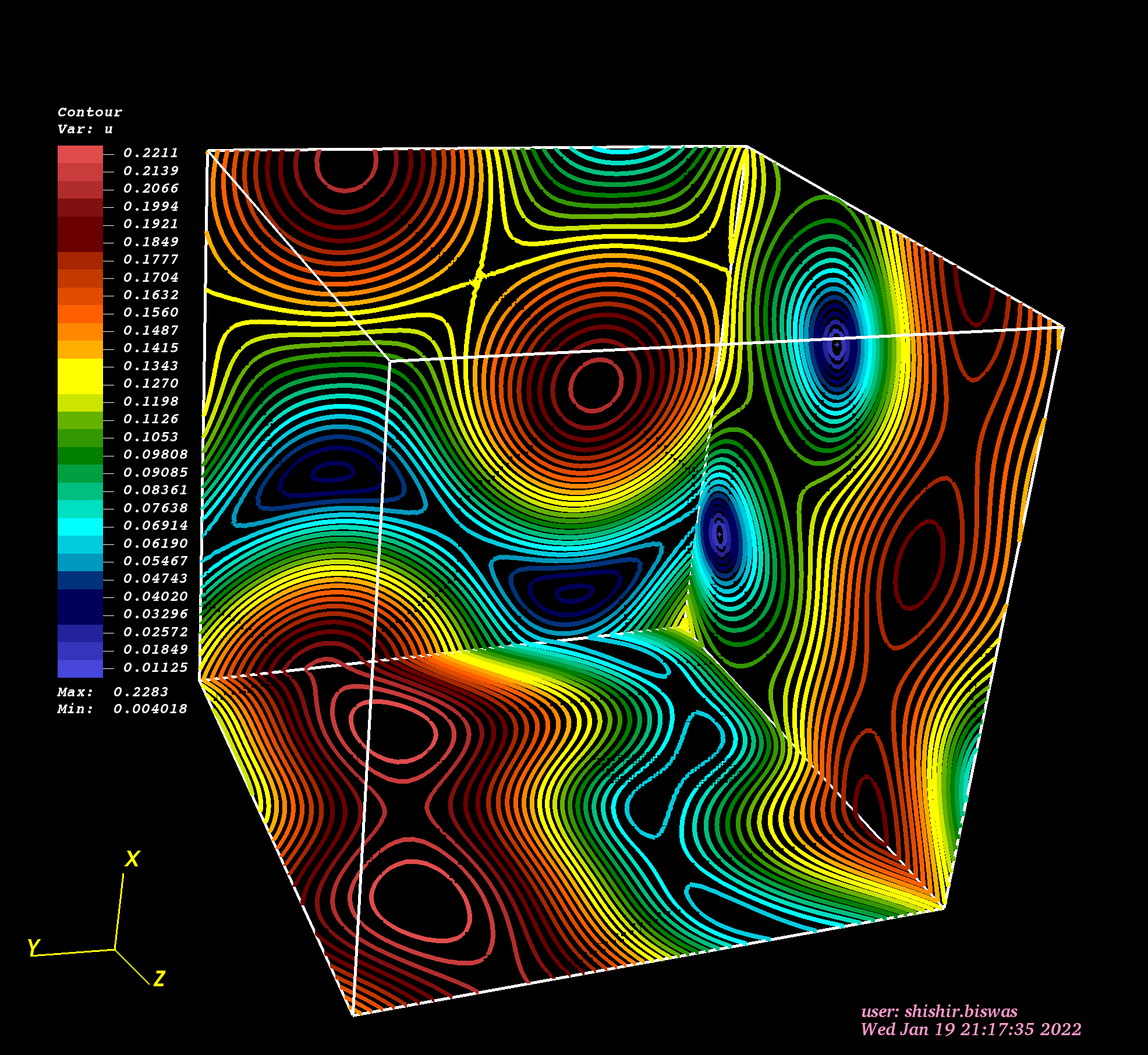}
		\caption{}
		\label{initial flow beta 0p6}
	\end{subfigure}
	\begin{subfigure}{0.32\textwidth}
		\centering
		\includegraphics[scale=0.083]{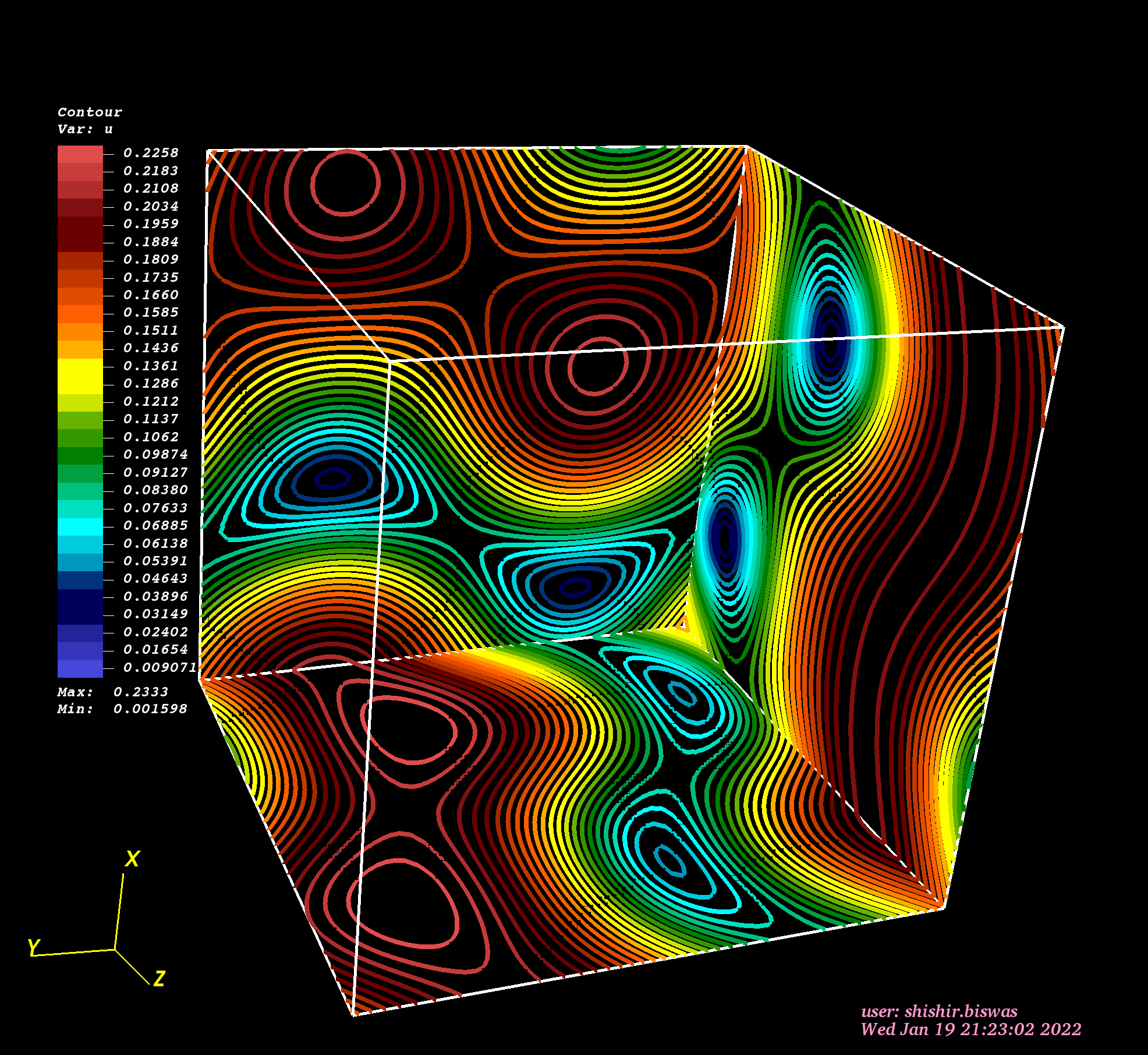}
		\caption{}
		\label{initial flow beta 0p8}
	\end{subfigure}
	\begin{subfigure}{0.32\textwidth}
		\centering
		\includegraphics[scale=0.083]{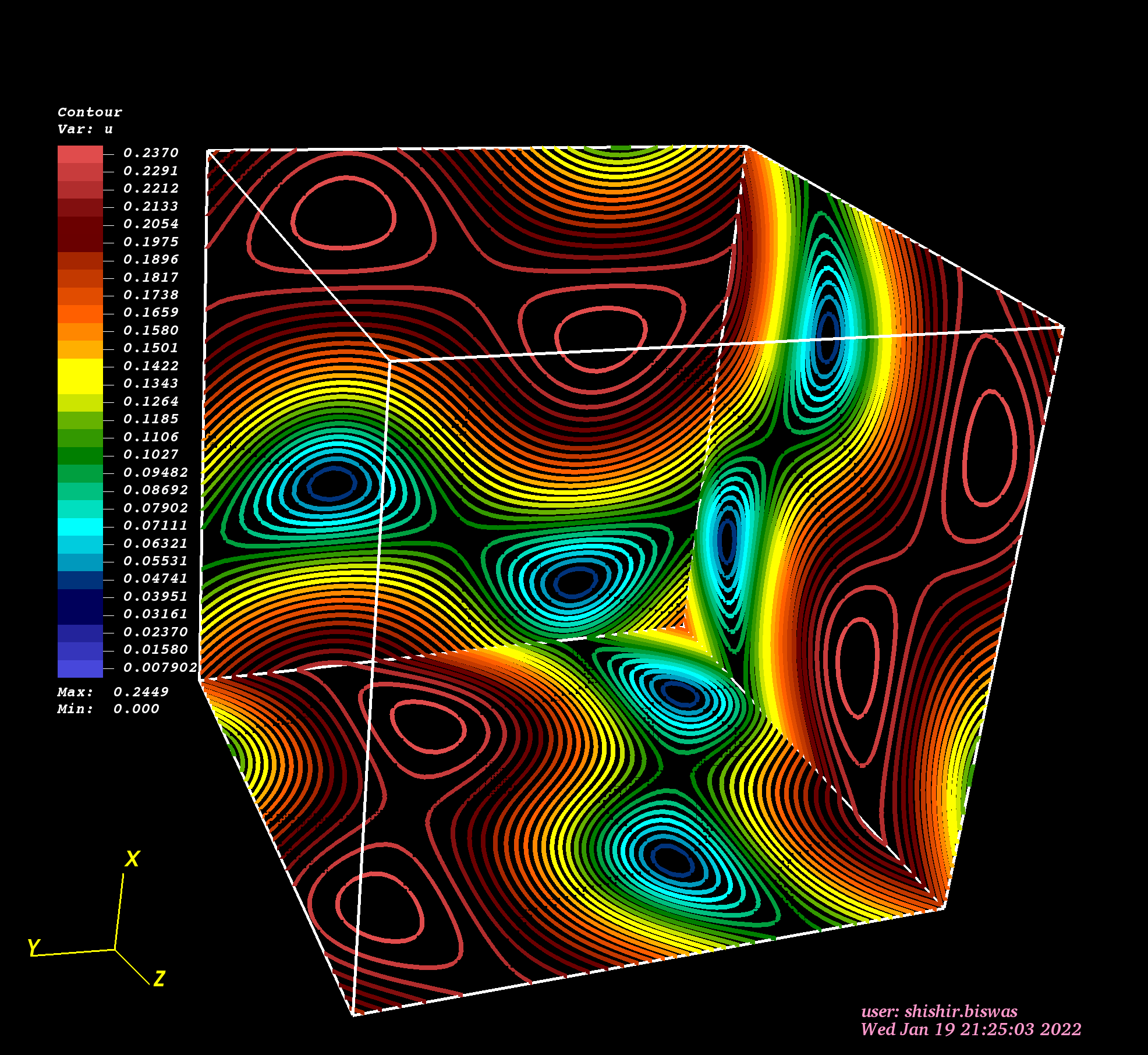}
		\caption{}
		\label{initial flow beta 1p0}
	\end{subfigure}
	\caption{Driving velocity contours of Yoshida-Morrison (YM) flow. The velocity contour visualization is shown for (a) $\beta = 0.0$, (b) $\beta = 0.2$, (c) $\beta = 0.4$, (d) $\beta = 0.6$, (e) $\beta = 0.8$, (f) $\beta = 1.0$, where $\beta = 0.0$ is least chaotic and $\beta = 1.0$ is most chaotic. It is important to identify that for the increasing $\beta$ values the velocity field lines are becoming chaotic and finally the visible separatrix has been formed for most helical (i.e. $\beta = 1.0$) case.}
	\label{initial flow}
\end{figure*}

For example, for \textcolor{black}{$\beta = 0$}, Yoshida et al. \citep{EPI2D:2017} classifies it as EPI-2D flow [See Fig. \ref{initial flow beta 0}] which is given by :
\begin{equation}\label{EPI2D}
\textcolor{black}{
	\begin{aligned}
		u_x &= u_0 [ \sin(y) - \cos(z) ]\\
		u_y &= 0\\
		u_z &= u_0 [ \sin(x)]
	\end{aligned}
}
\end{equation}
\textcolor{black}{This flow (i.e, Eq. \ref{EPI2D}) is dependent on all the 3 spatial coordinates (i.e, $x, y, z$), whereas only two flow components are nonzero. Thus EPI-2D flow is quasi-2D in nature.}  

As can be expected, for $\beta = 1$ the above Eq. \ref{Yoshida_flow} becomes well known ABC like flow [See Fig.\ref{initial flow beta 1p0}], 
\begin{equation}\label{ABC_like}
\textcolor{black}{
	\begin{aligned}
		u_x &= u_0 [ \sin(y) -  \cos(z) ]\\
		u_y &= u_0 [  \sin(z) -  \cos(x) ]\\
		u_z &= u_0 [  \sin(x) -  \cos(y) ]
	\end{aligned}
}
\end{equation}
(The only mathematical difference between the flow described in Eq. \ref{ABC_like} and the ABC flow [Eq. \ref{ABC Flow}] are that the two terms are being subtracted in the latter while added in the former. Nevertheless, as will be shown below, the flow in Eq. \ref{ABC_like} is identical in its properties to ABC flow [Eq. \ref{ABC Flow}].) As $\beta$ is varied from $0$ to $1.0$, a whole set of intermediate class of flows emerge, such that a normalized fluid helicity is exactly $0.0$ for $\beta = 0$ and is maximum for $\beta =1.0$ (i.e, ABC-like flows).  For a comprehensive study, we explore various values of $\beta$ between $0$ to $1$, keeping all other parameters identical [See Fig. \ref{initial flow beta 0p2}, \ref{initial flow beta 0p4}, \ref{initial flow beta 0p6}, \ref{initial flow beta 0p8}].

\begin{figure*}
	\centering
	\includegraphics[scale=0.55]{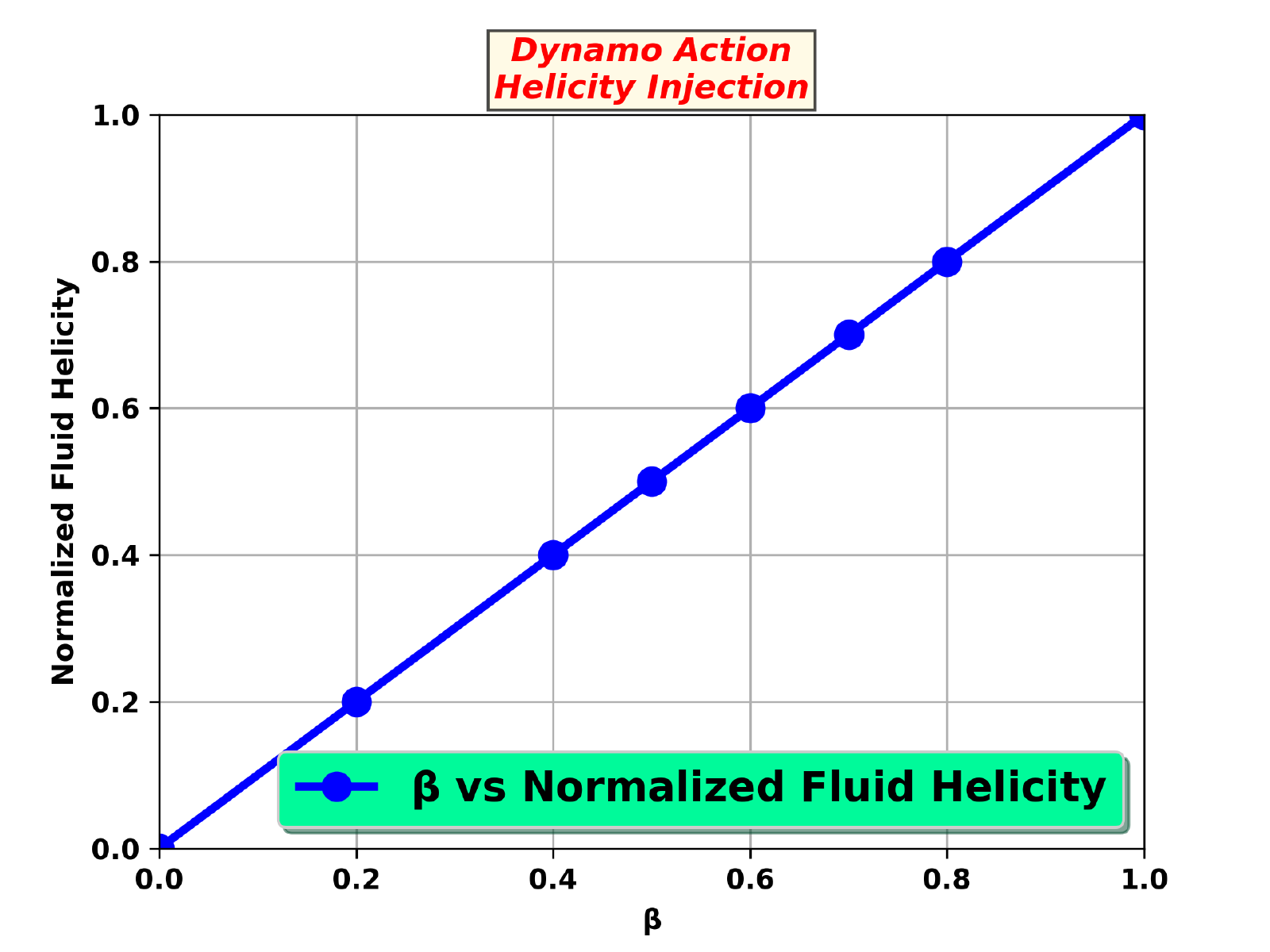}
	\caption{Normalized fluid helicity  as a function of $\beta$ for Yoshida-Morrison (YM) flow. \textcolor{black}{Here $\beta$ is real constant}. $\beta = 0.0$ corresponds to non-helical (zero normalized fluid helicity) flow  and $\beta = 1.0$ corresponds to maximum helical (maximum normalized fluid helicity) flow. For detail calculation and normalization [See. Appendix \ref{Appen A}].}
	\label{fluid helicity}
\end{figure*}

By increasing $\beta$ values we inject normalized fluid helicity in the system [See. Appendix \ref{Appen A} for details]. In Fig. \ref{fluid helicity}, normalized fluid helicity [$\int_{V} \vec{u} . (\vec{\nabla} \times \vec{u}) dV$] is calculated, for YM flow. It is observed that for increasing $\beta$ values normalized fluid helicity also increases. It is also interesting to note from YM flow velocity contour visualization [See Fig. \ref{initial flow beta 0}, \ref{initial flow beta 0p2}, \ref{initial flow beta 0p4}, \ref{initial flow beta 0p6}, \ref{initial flow beta 0p8}, \ref{initial flow beta 1p0}] that for increasing $\beta$ values, the velocity contours become increasingly chaotic and visible separatrix forms emerge, which is a crucial ingredient of chaotic flows. \textcolor{black}{The maximum fluid velocity is $u_0 = 1.0$ and Alfven Mach number is $M_A = 10.0$. Following \citep{Frish_Dynamo:1984}, the initial magnetic field is considered as very small but uniform in nature.} With these given velocity fields, we perform our simulation, the details of which is given next.\\

\subsection{Parameter Details}
We evolve the above set of equations discussed in Section \ref{Equations}, for class of YM flow profile, in a triply periodic box of length $L_x = L_y = L_z = 2\pi$ with time steeping $(dt) = 10^{-4}$ and grid resolution $256^3$. We have performed grid size scaling study [See. Appendix \ref{Appen B}] using Arnold-Beltrami-Childress (ABC) flow \citep{Frish_Dynamo:1984} \textcolor{black}{for} different magnetic Reynolds numbers. It is obvious that $256^3$ grid resolution is more than enough for this problem [See. Appendix \ref{Appen B}]. With these initial conditions and parameter spaces we present our
numerical simulation results.

\section{Simulation Results}
We consider a recently proposed general class of flow named YM flow (Eq. \ref{Yoshida_flow}) \citep{EPI2D:2017} for this present study, as indicated earlier and evolve the homogeneous initial B-field in time, using Eq.\ref{Bfield} for a given set of values of $\beta$ and $u_0$ using our GPU solver \textit{GMHD3D}. An interesting and useful aspect of YM flow is that, it is possible to inject finite fluid helicity [$\int_{V}\vec{u}.(\vec{\nabla} \times \vec{u}) dV$] in the system [See Fig. \ref{fluid helicity}], by systematically varying certain physically meaningful parameter, i.e. $\beta$.

\begin{figure*}
	\begin{subfigure}{0.49\textwidth}
		\centering
		\includegraphics[scale=0.55]{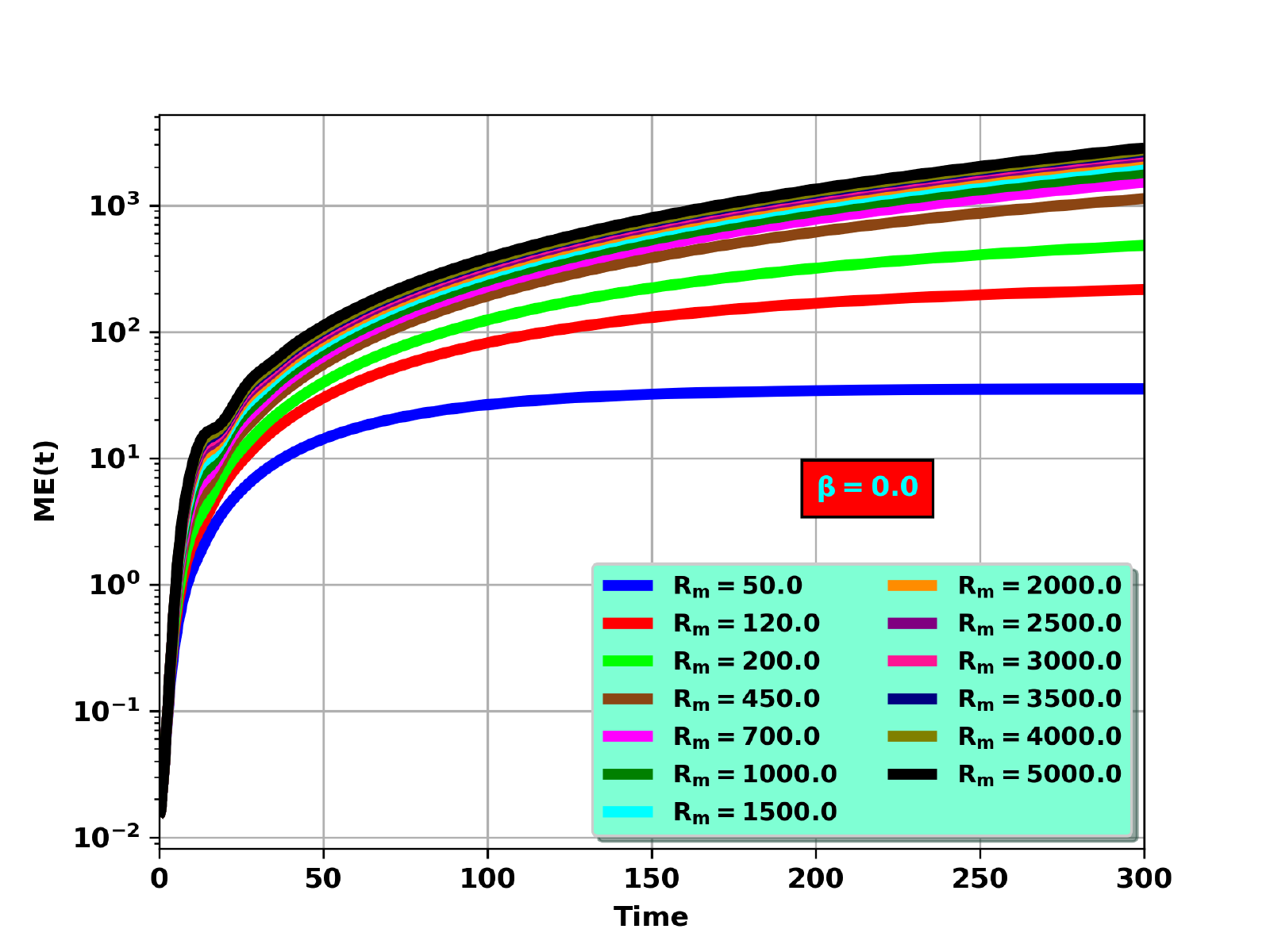}
		\caption{}
		\label{beta_0p0_energy}
	\end{subfigure}
	\begin{subfigure}{0.49\textwidth}
		\centering
		\includegraphics[scale=0.055]{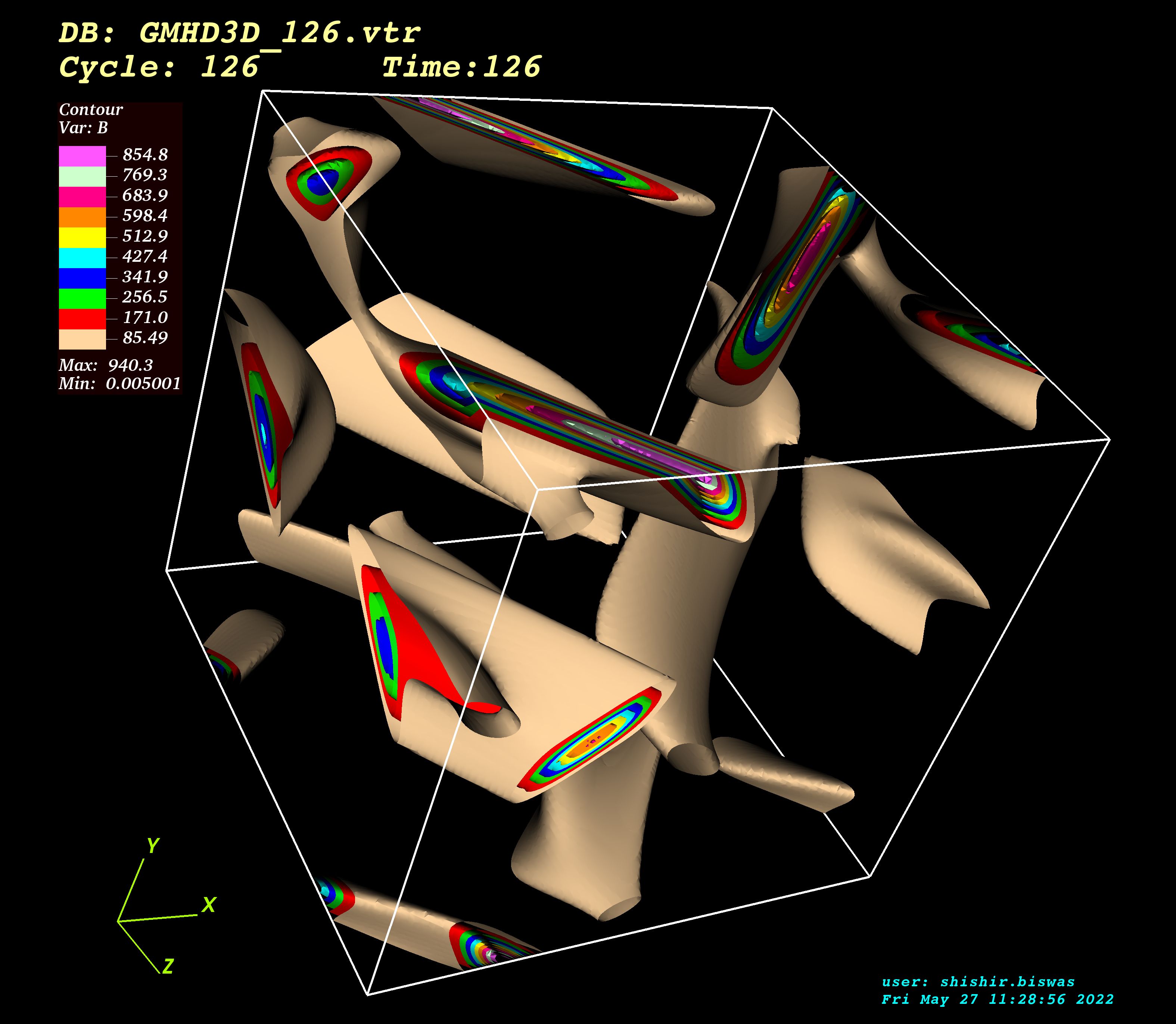}
		\caption{}
		\label{beta_0p0_iso}
	\end{subfigure}
	\begin{subfigure}{0.49\textwidth}
		\centering
		\includegraphics[scale=0.060]{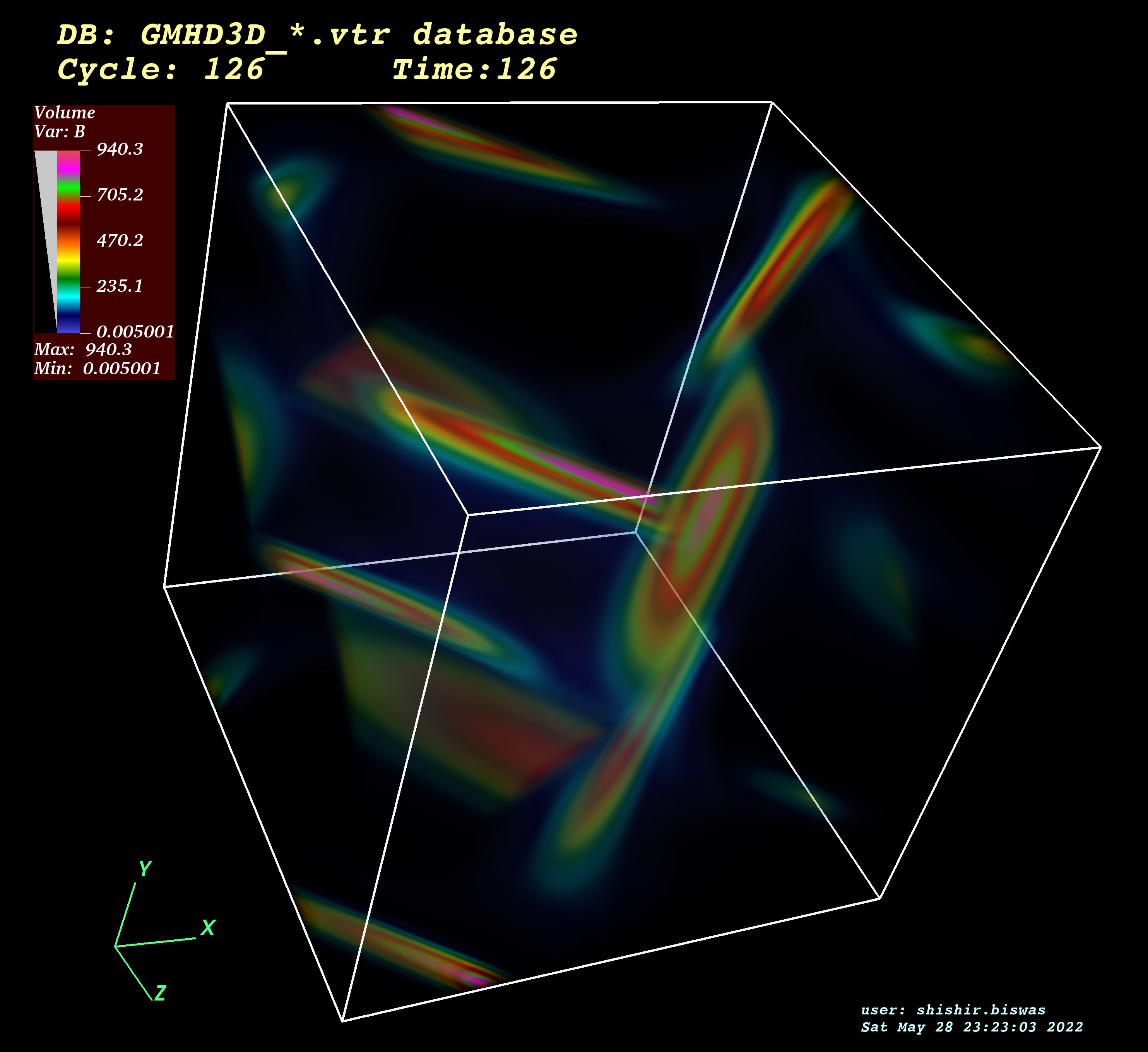}
		\caption{}
		\label{beta_0p0_volume}
	\end{subfigure}
	\begin{subfigure}{0.49\textwidth}
		\centering
		\includegraphics[scale=0.060]{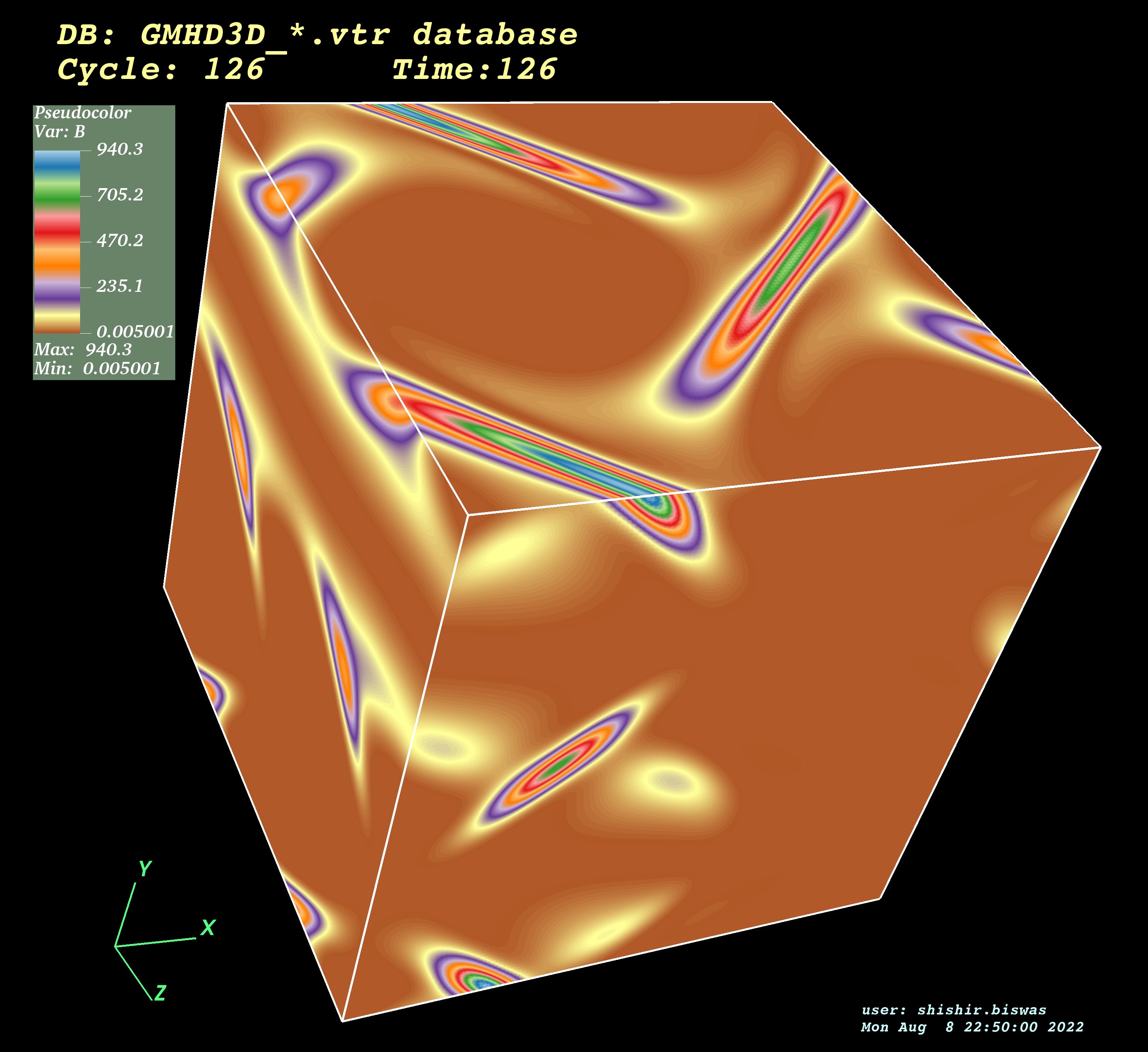}
		\caption{}
		\label{beta_0p0_pseudo}
	\end{subfigure}
	\caption{\textcolor{black}{Kinematic fast} Dynamo effect for different magnetic Reynolds number ($R_m$) using Yoshida-Morrison (YM) flow with \textcolor{black}{$\beta = 0.0$}. (a) The magnetic energy $\left(\sum\limits_{V} \frac{B^2(x,y,z,t)}{2} \right)$ growth is seen to be unaffected with increase of $R_m$. The (b) three-dimensional (3D) magnetic energy iso-surface \textcolor{black}{[Supplementary movie added]}, (c) three-dimensional (3D) volume rendering view of magnetic energy iso-surfaces and (d) three-dimensional (3D) pseudo-color view of magnetic energy iso-surfaces (for magnetic Reynolds number $R_m = 50$) is seen identical as a untwisted ``magnetic ribbon''. Simulation details: grid resolution $256^3$, stepping time dt = $10^{-4}$, magnitude of fluid velocity $u_0 = 1.0$, Alfven mach number $M_A = 10.0$.}
\end{figure*}

We study the \textcolor{black}{kinematic fast} dynamo action for various $\beta$ value starting from $0.0$ to $1.0$. For \textcolor{black}{$\beta = 0.0$} value, YM flow leads to EPI-2 dimensional flow \citep{EPI2D:2017}.  EPI-2 dimensional flow makes a topological bridge between \textcolor{black}{quasi-}2D and 3D class of flows \citep{EPI2D:2017}. We perform numerical runs for a wide range of magnetic Reynolds numbers ($R_m$). \textcolor{black}{From Fig. \ref{beta_0p0_energy} it is observed that, for $\beta = 0.0$ there is no prominent dynamo action, even at highest magnetic magnetic Reynolds numbers ($R_m = 5000$).} \textcolor{black}{At late times, magnetic energy curve is flat with respect to time, indicating that the growth rate is zero.} The key mechanism behind the \textcolor{black}{fast} dynamo action is believed to be stretching, twisting and folding (STF) of magnetic field lines \textcolor{black}{\citep{Zeldovich:1972}}, that generates \textcolor{black}{exponential growth} of magnetic energy and hence it is necessary that, the flow should be able to ``twist'' \textcolor{black}{(i.e., should contain enough chaoticity)} and hence should contain nonzero fluid helicity to generate exponential growth of magnetic energy. From our earlier discussion it is already evident that, YM flow with $\beta = 0.0$ has zero fluid helicity and hence is not able to amplify the magnetic field in exponential manner i.e, not able to generate dynamo action.

 \textcolor{black}{It is important to note that, there are examples of non-helical flows, that can produce fast dynamo action, that is due to Cross-helicity effect or the so-called Yoshizawa effect \citep{galloway_nature:1992, Sur:2009}. Also as discussed in the Introduction, several authors have reported dynamo action using 2-dimensional time dependent non-helical flows. For example, GP flow \citep{galloway_nature:1992, Hughes_PLA:1996}, has been used to study kinematic fast dynamo action. The flow profile goes as $\vec{u} (x,y,t)$, so the initial magnetic field is modeled as $B(x,y,t) \times e^{ik_zz}$, for a fixed $k_z$, the resulting problem for the magnetic field is two-dimensional. Using these initial condition fast dynamo action has been reported for non-helical flows.}

 Let us come back to YM flow with $\beta$ as the control parameter for initial helicity injected. We study of magnetic energy iso-surfaces for the $\beta = 0.0$ case and observe that most of the energy is concentrated in elongated almost two-dimensional un-twisted structures [See Fig. \ref{beta_0p0_iso}]. To obtain better visualization, we also provide volume rendered view [See Fig. \ref{beta_0p0_volume}] of magnetic energy iso-surfaces and pseudo-color view [See Fig. \ref{beta_0p0_pseudo}] of magnetic energy iso-surfaces. \textcolor{black}{These several visualizations (iso-surface view, volume rendering view, and pseudo-color view) aid in understanding the nature of iso-surfaces.} Such magnetic structures have been reported earlier for kinematic dynamo simulations by \citet{Alexakis_PRE:2011} and are referred to as ``magnetic ribbons''. It is clear that the structures do not have any twisting, which is due to the absence of fluid helicity in the flow field. 

\begin{figure*}
	\begin{subfigure}{0.49\textwidth}
		\centering
		\includegraphics[scale=0.55]{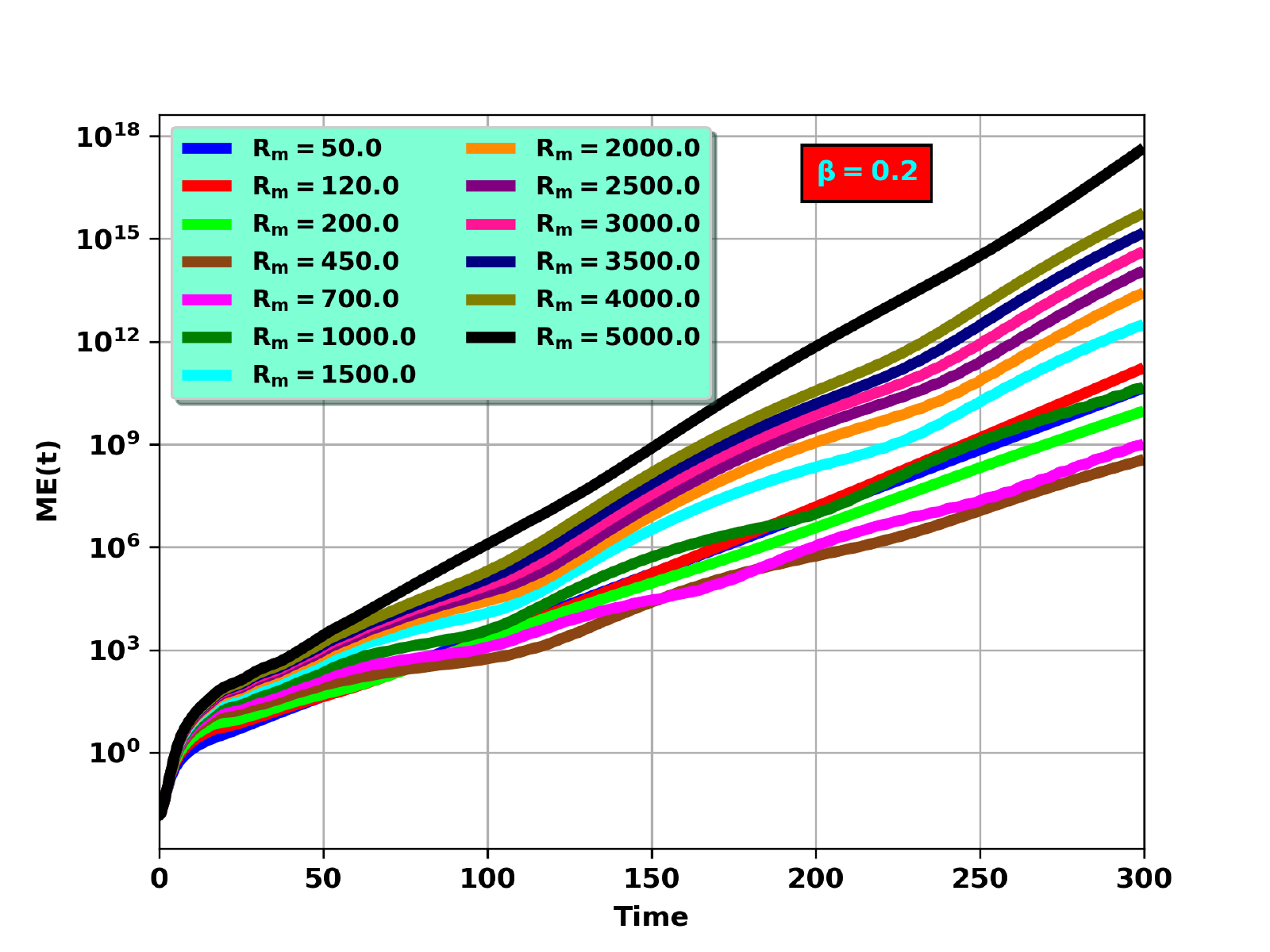}
		\caption{}
		\label{beta_0p2_energy}
	\end{subfigure}
	\begin{subfigure}{0.49\textwidth}
		\centering
		\includegraphics[scale=0.055]{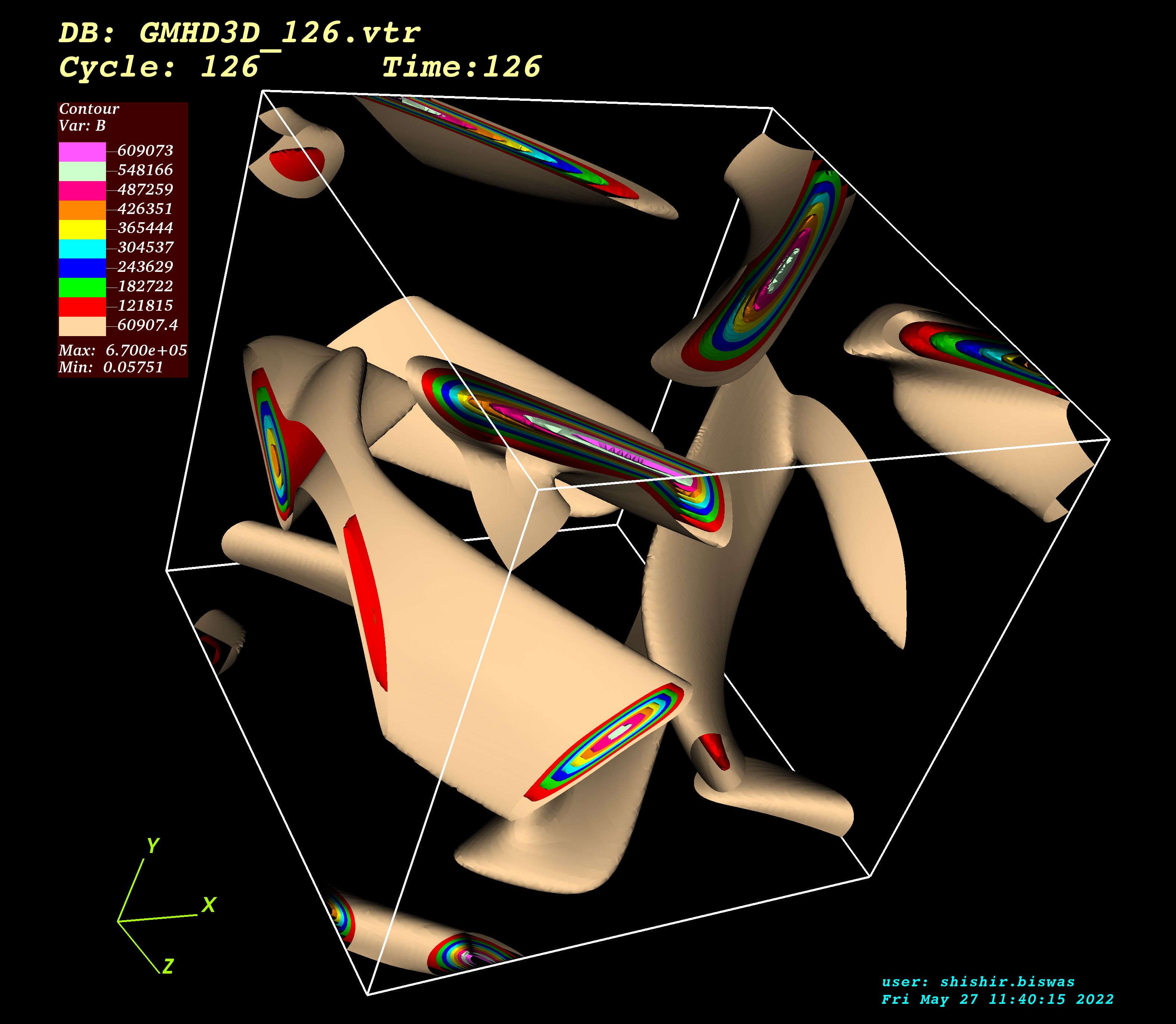}
		\caption{}
		\label{beta_0p2_iso}
	\end{subfigure}
	\begin{subfigure}{0.49\textwidth}
		\centering
		\includegraphics[scale=0.060]{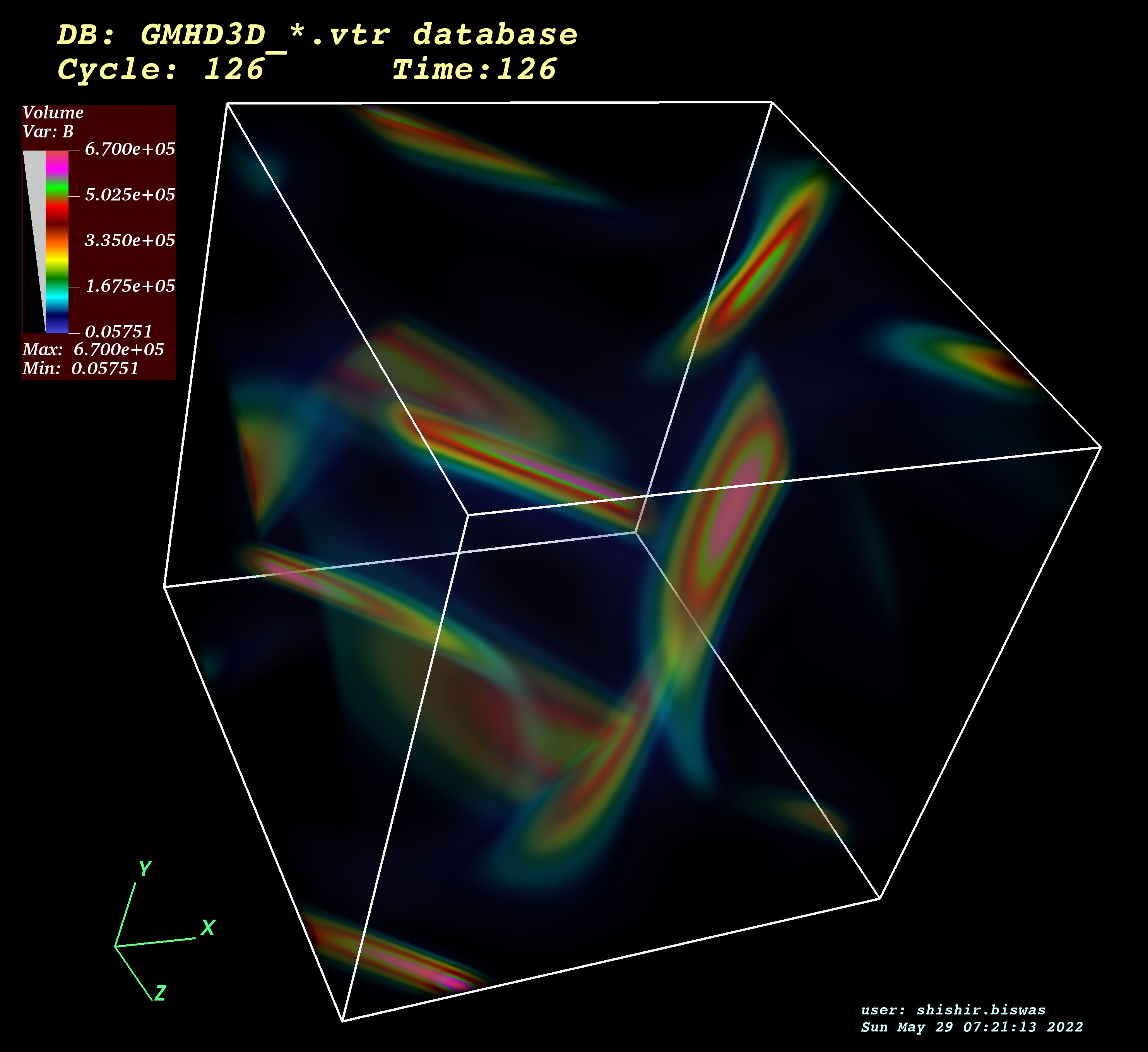}
		\caption{}
		\label{beta_0p2_volume}
	\end{subfigure}
	\begin{subfigure}{0.49\textwidth}
		\centering
		\includegraphics[scale=0.060]{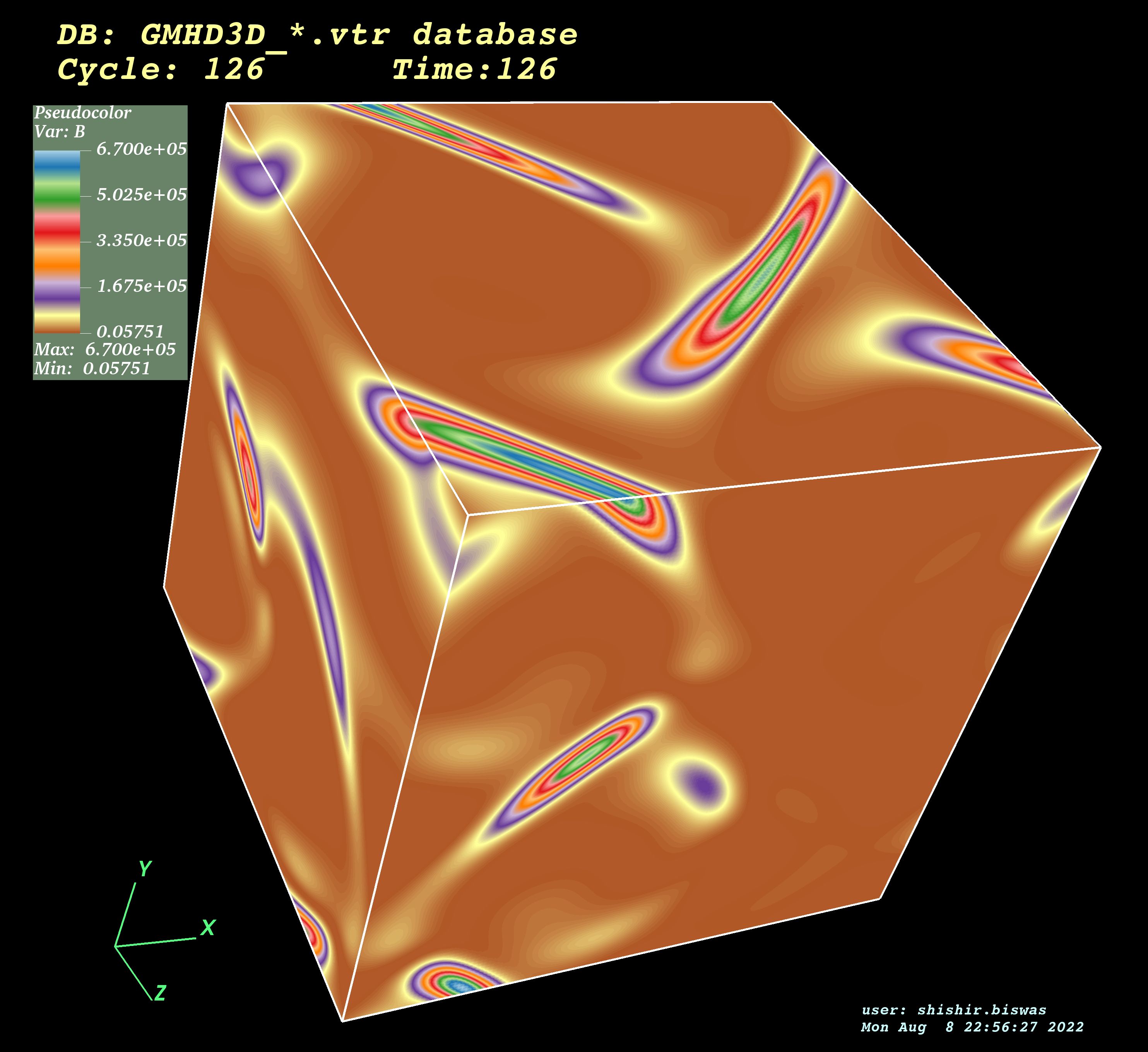}
		\caption{}
		\label{beta_0p2_pseudo}
	\end{subfigure}
	\caption{\textcolor{black}{Kinematic fast} Dynamo effect for different magnetic Reynolds number ($R_m$) using Yoshida-Morrison (YM) flow with \textcolor{black}{$\beta = 0.2$}. (a) The magnetic energy $\left(\sum\limits_{V} \frac{B^2(x,y,z,t)}{2} \right)$ growth is observed for different $R_m$ values. The (b) three-dimensional (3D) magnetic energy iso-surface \textcolor{black}{[Supplementary movie added]}, (c) three-dimensional (3D) volume rendering view of magnetic energy iso-surfaces and (d) three-dimensional (3D) pseudo-color view of magnetic energy iso-surfaces (for magnetic Reynolds number $R_m = 50$) is seen to be twisted ``ribbon'' like nature. Note that the twisting indicates the presence of fluid helicity in the flow. Simulation details: grid resolution $256^3$, stepping time dt = $10^{-4}$, magnitude of fluid velocity $u_0 = 1.0$, Alfven mach number $M_A = 10.0$.}
\end{figure*}

It was discussed earlier that, the injection of normalized fluid helicity in the system is possible by systematically varying some meaningful physical parameter, i.e. $\beta$ value for YM flow. In the following, we perform numerical runs and present simulation results  with $\beta = 0.2$ for various magnetic Reynolds numbers ($R_m$). The stretching, twisting and folding (STF) of magnetic field lines \textcolor{black}{\citep{Zeldovich:1972}} doubles the magnetic field per cycle because of flux freezing. Presence of higher diffusivity in the system makes the flux freezing condition to be violated, which in turns results in, suppression of exponential growth of magnetic energy. Hence, the fast dynamo action gets hindered for a higher diffusive system. From Fig. \ref{beta_0p2_energy} it is observed that, the YM flow (\textcolor{black}{$\beta = 0.2$}) efficiently amplifies the magnetic field in exponential manner for a wide range of $R_m$ value and exhibits  dynamo action. A careful observation of magnetic energy iso-surface shows that,  the energy is concentrated in long and flat structures or  ``ribbon'' like structures [See Fig. \ref{beta_0p2_iso}, \ref{beta_0p2_volume}, \ref{beta_0p2_pseudo}]. Although small scale structures are also present here, they appear to be more ``organized'' to form \textcolor{black}{special geometric} patterns, which is reported earlier \citep{Alexakis_PRE:2011}. It is notable that, there is visible twisting in the ``ribbons'' that indicates the presence of fluid helicity in the flow. With respect to earlier case, it is identified that by varying $\beta$ value from $0$ to $0.2$ we are able to produce \textcolor{black}{fast} dynamo action efficiently. For $\beta = 0.2$, there is fluid helicity present in the system and the presence of fluid helicity twists the ribbon which produces dynamo action.

\begin{figure*}
	\begin{subfigure}{0.49\textwidth}
		\centering
		\includegraphics[scale=0.55]{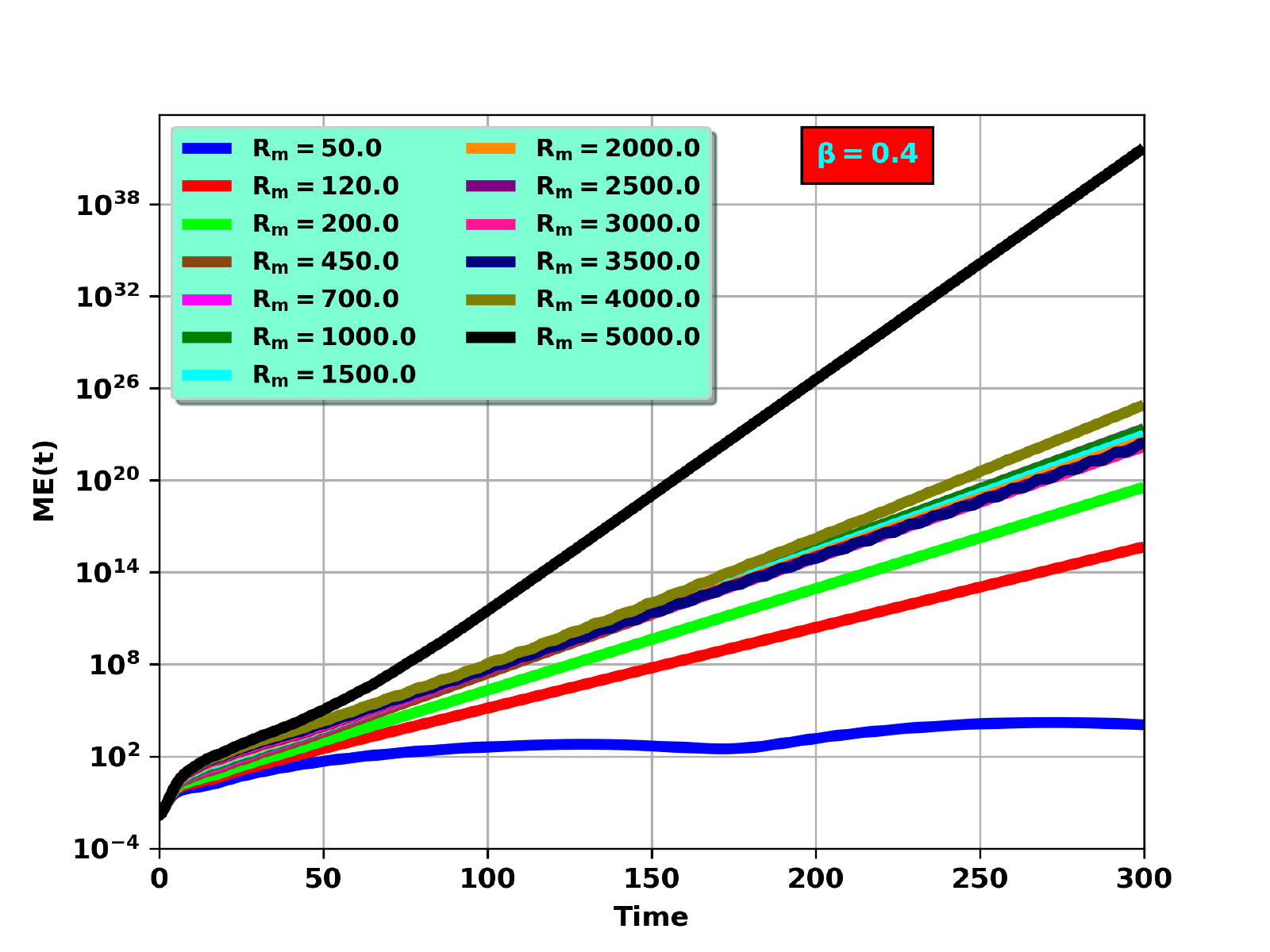}
		\caption{}
		\label{beta_0p4_energy}
	\end{subfigure}
	\begin{subfigure}{0.49\textwidth}
		\centering
		\includegraphics[scale=0.055]{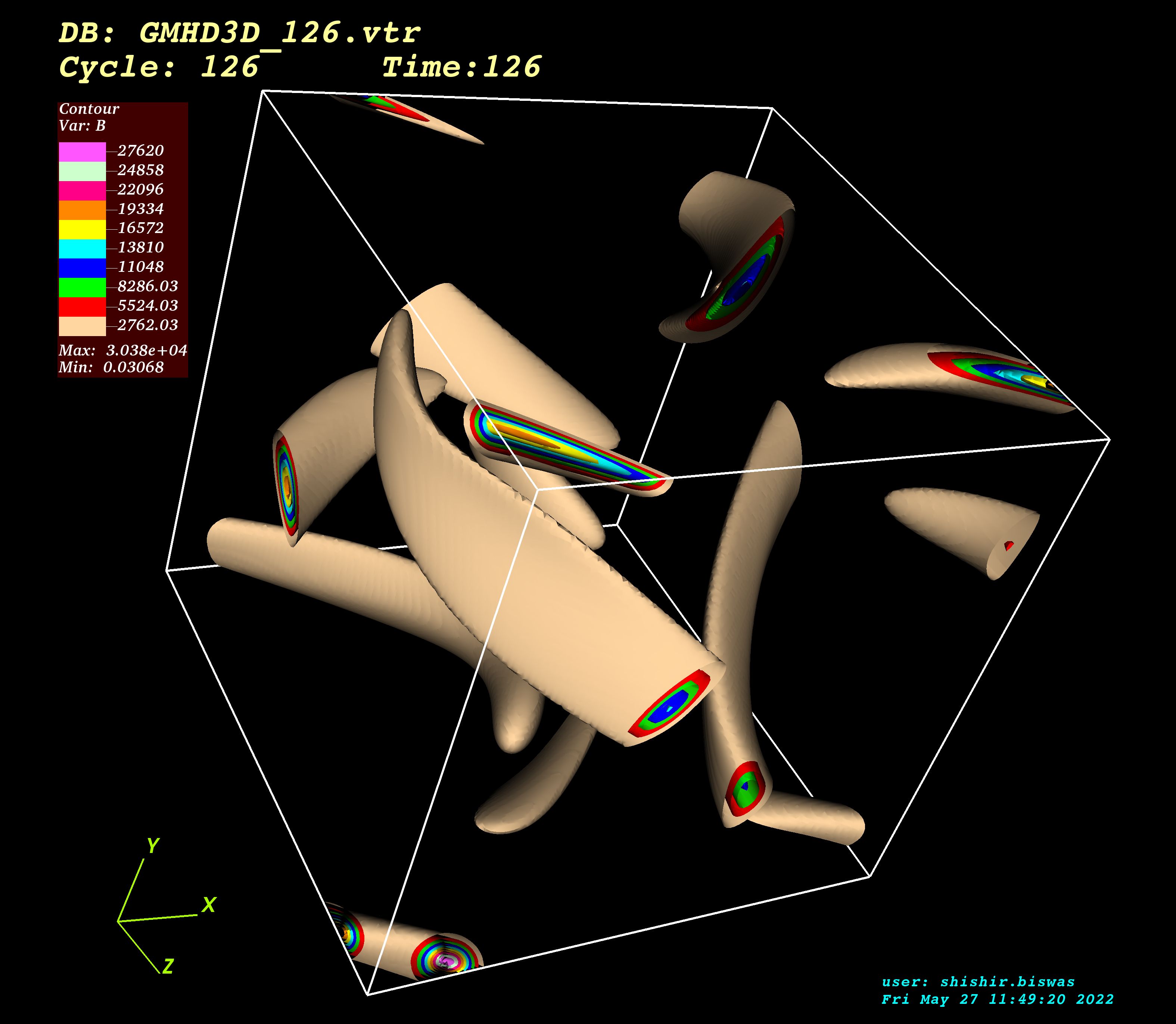}
		\caption{}
		\label{beta_0p4_iso}
	\end{subfigure}
	\begin{subfigure}{0.49\textwidth}
		\centering
		\includegraphics[scale=0.060]{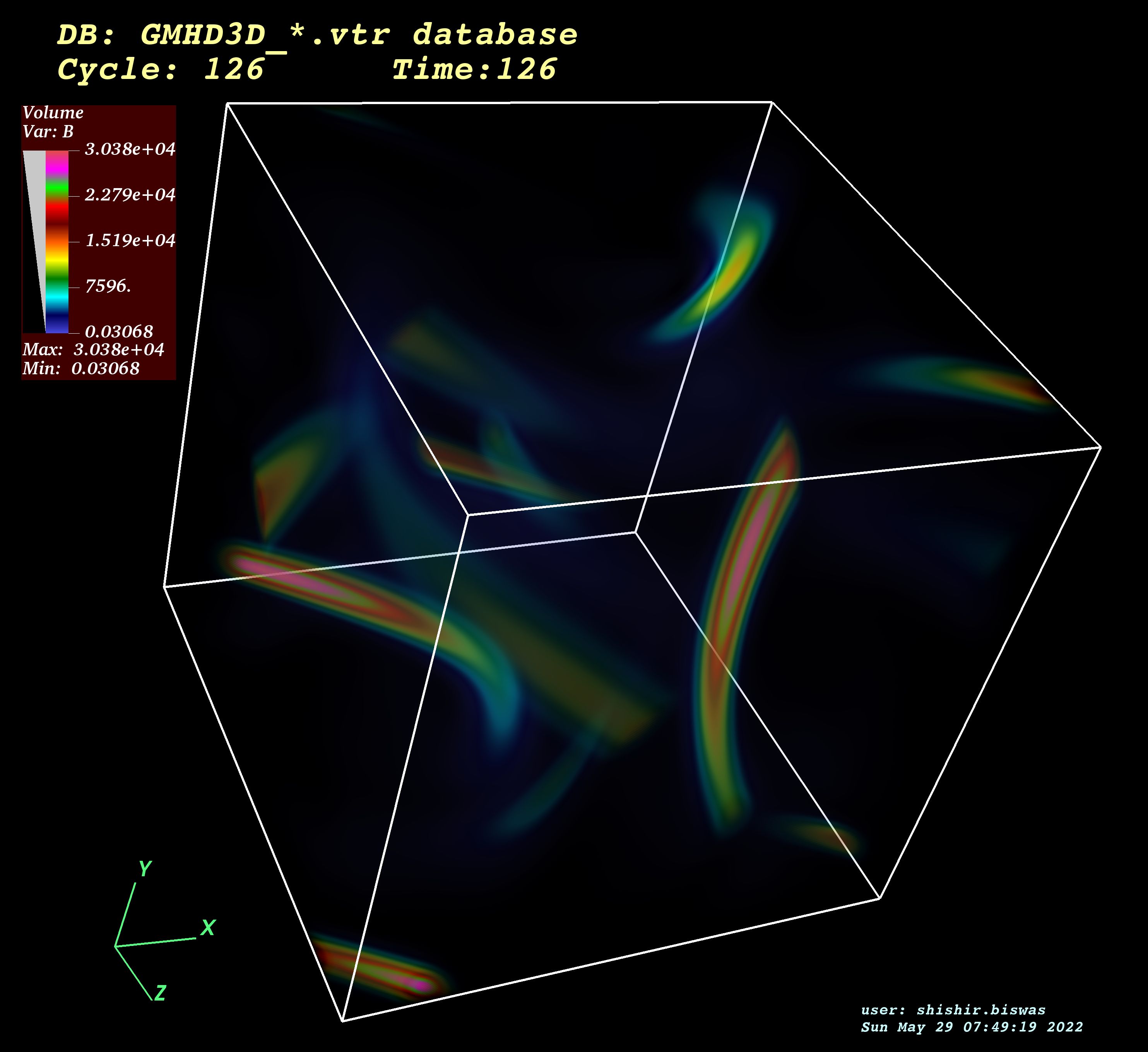}
		\caption{}
		\label{beta_0p4_volume}
	\end{subfigure}
	\begin{subfigure}{0.49\textwidth}
		\centering
		\includegraphics[scale=0.060]{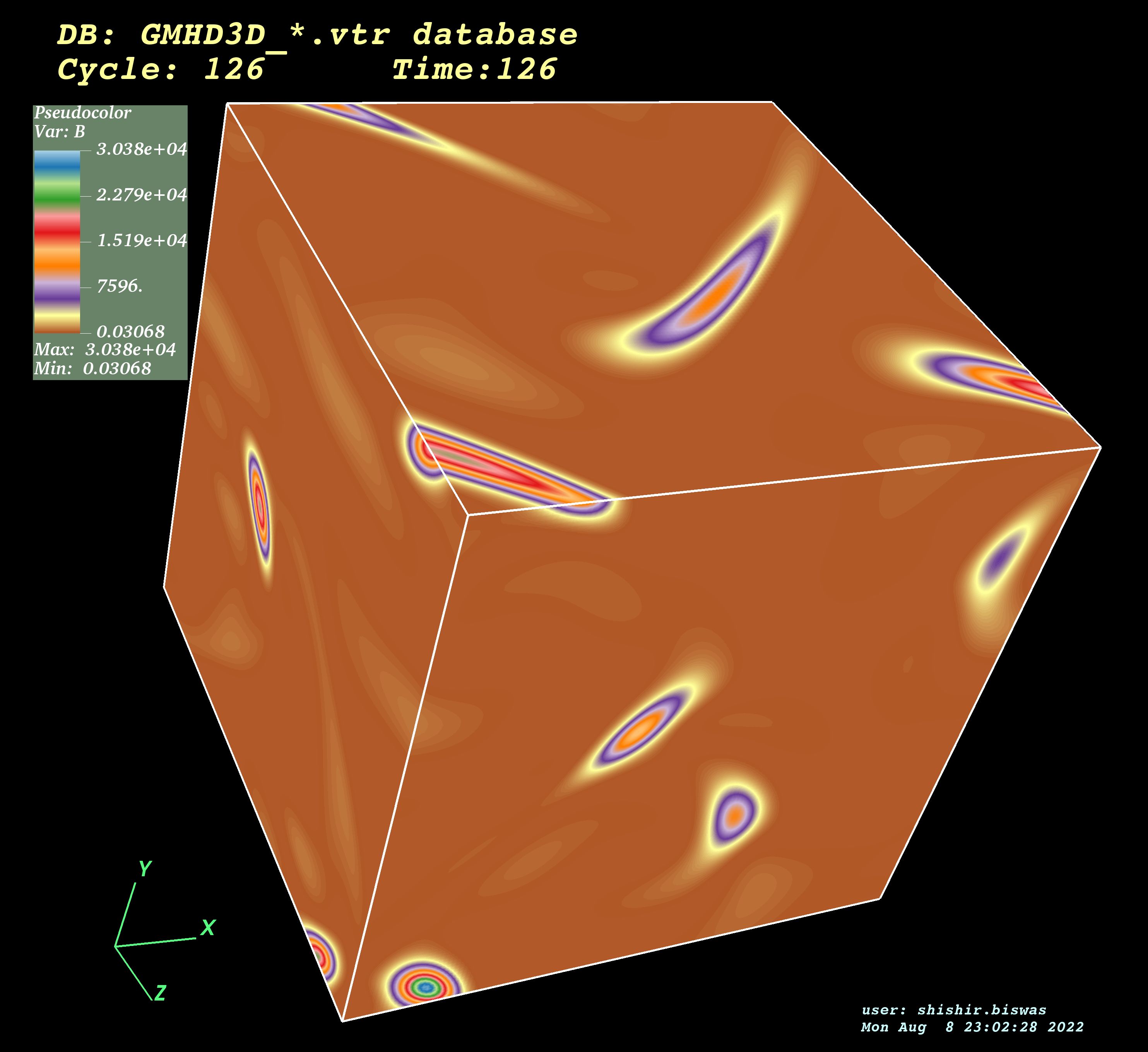}
		\caption{}
		\label{beta_0p4_pseudo}
	\end{subfigure}
	\caption{\textcolor{black}{Kinematic fast} Dynamo effect for different magnetic Reynolds number ($R_m$) using Yoshida-Morrison (YM) flow with \textcolor{black}{$\beta = 0.4$}. (a) The magnetic energy $\left(\sum\limits_{V} \frac{B^2(x,y,z,t)}{2} \right)$ growth is observed for different $R_m$ values. The (b) three-dimensional (3D) magnetic energy iso-surface, (c) three-dimensional (3D) volume rendering view of magnetic energy iso-surfaces and (d) three-dimensional (3D) pseudo-color view of magnetic energy iso-surfaces (for magnetic Reynolds number $R_m = 50$) is seen to be twisted ``ribbon'' like nature like earlier. Note that the twisting indicates the presence of helicity in the flow. Simulation details: grid resolution $256^3$, stepping time dt = $10^{-4}$, magnitude of fluid velocity $u_0 = 1.0$, Alfven mach number $M_A = 10.0$.}
\end{figure*}

To study the effect of fluid helicity in the context of \textcolor{black}{kinematic fast} dynamo action in detail, we explore various $\beta$ values such as 0.4, 0.6, 0.8 and 1.0. For $\beta = 0.4$, we observe that the generation of \textcolor{black}{exponential} magnetic energy \textcolor{black}{growth} is possible for a substantial range of magnetic Reynolds number ($R_m$) [See Fig. \ref{beta_0p4_energy}]. Due to the presence of finite fluid helicity in the system, \textcolor{black}{the flow} is able to produce dynamo action. From the magnetic energy iso-surface visualization, energy is seen to be concentrated in elongated structures [See Fig. \ref{beta_0p4_iso}, \ref{beta_0p4_volume}, \ref{beta_0p4_pseudo}]. These structures are part of sheetlike structures reported earlier \citep{Alexakis_PRE:2011}, much like the ``ribbons''  [See Fig. \ref{beta_0p4_iso}, \ref{beta_0p4_volume}, \ref{beta_0p4_pseudo}] observed for earlier cases as well.

\begin{figure*}
	\begin{subfigure}{0.49\textwidth}
		\centering
		\includegraphics[scale=0.55]{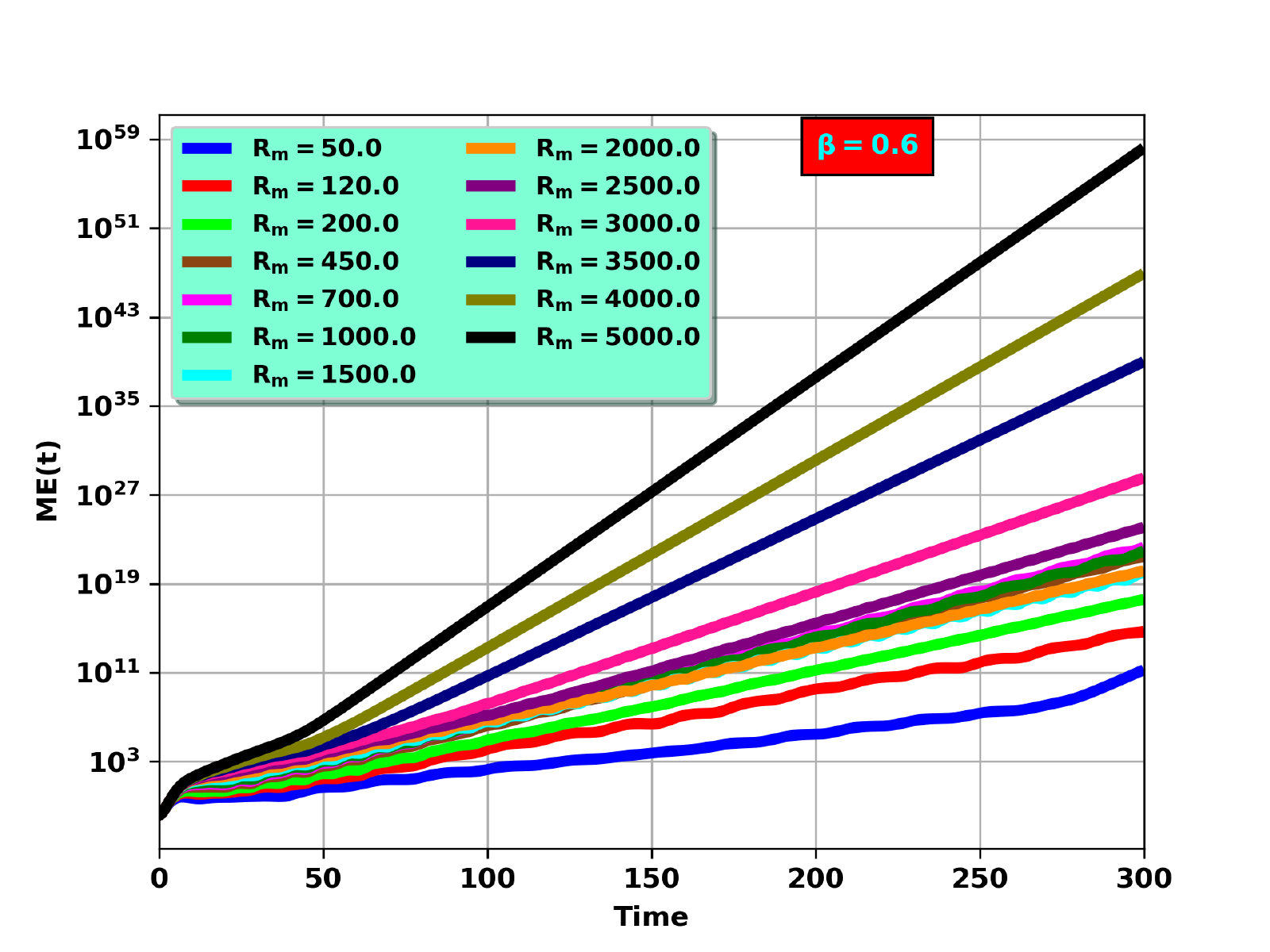}
		\caption{}
		\label{beta_0p6_energy}
	\end{subfigure}
	\begin{subfigure}{0.49\textwidth}
		\centering
		\includegraphics[scale=0.055]{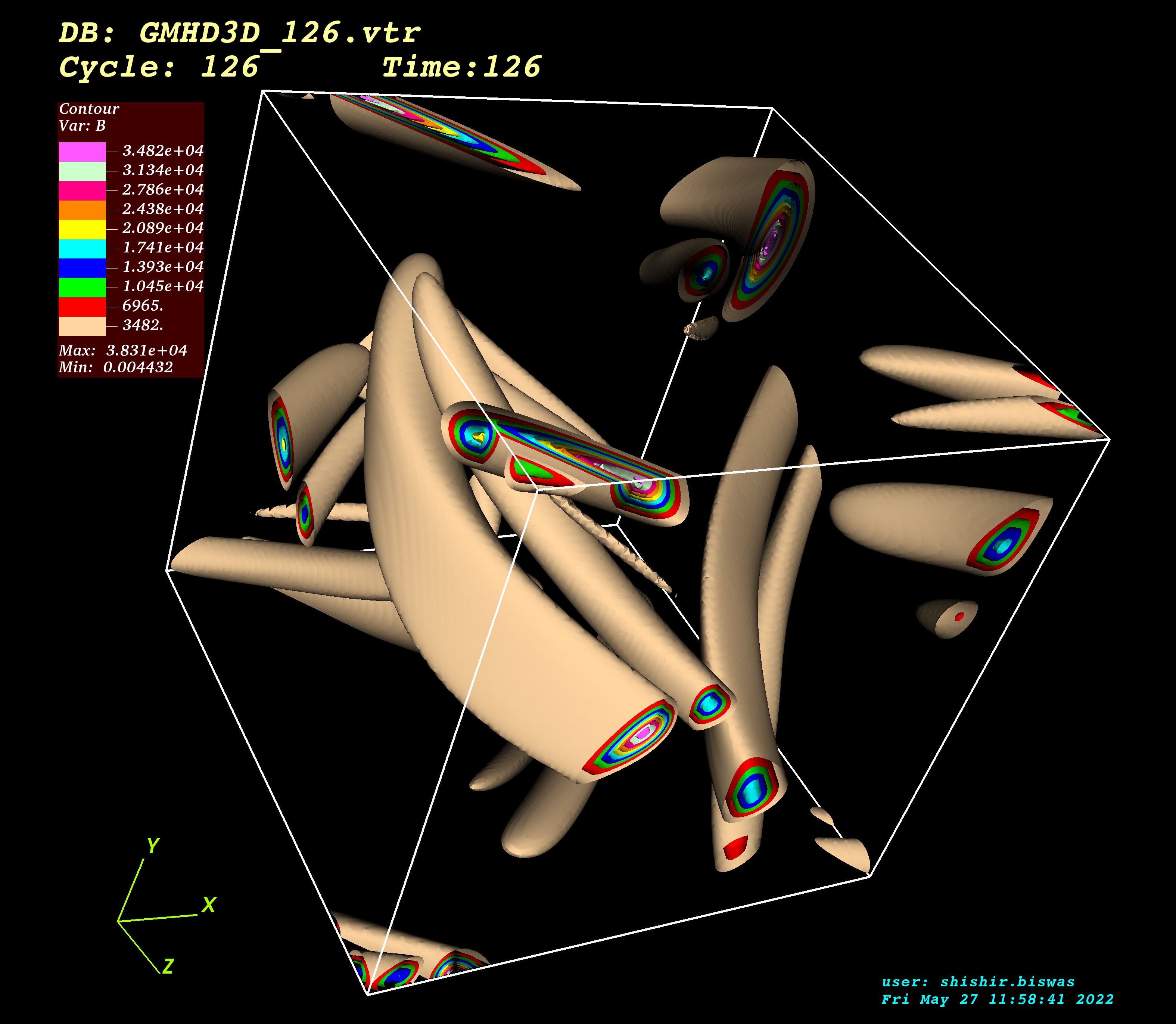}
		\caption{}
		\label{beta_0p6_iso}
	\end{subfigure}
	\begin{subfigure}{0.49\textwidth}
		\centering
		\includegraphics[scale=0.060]{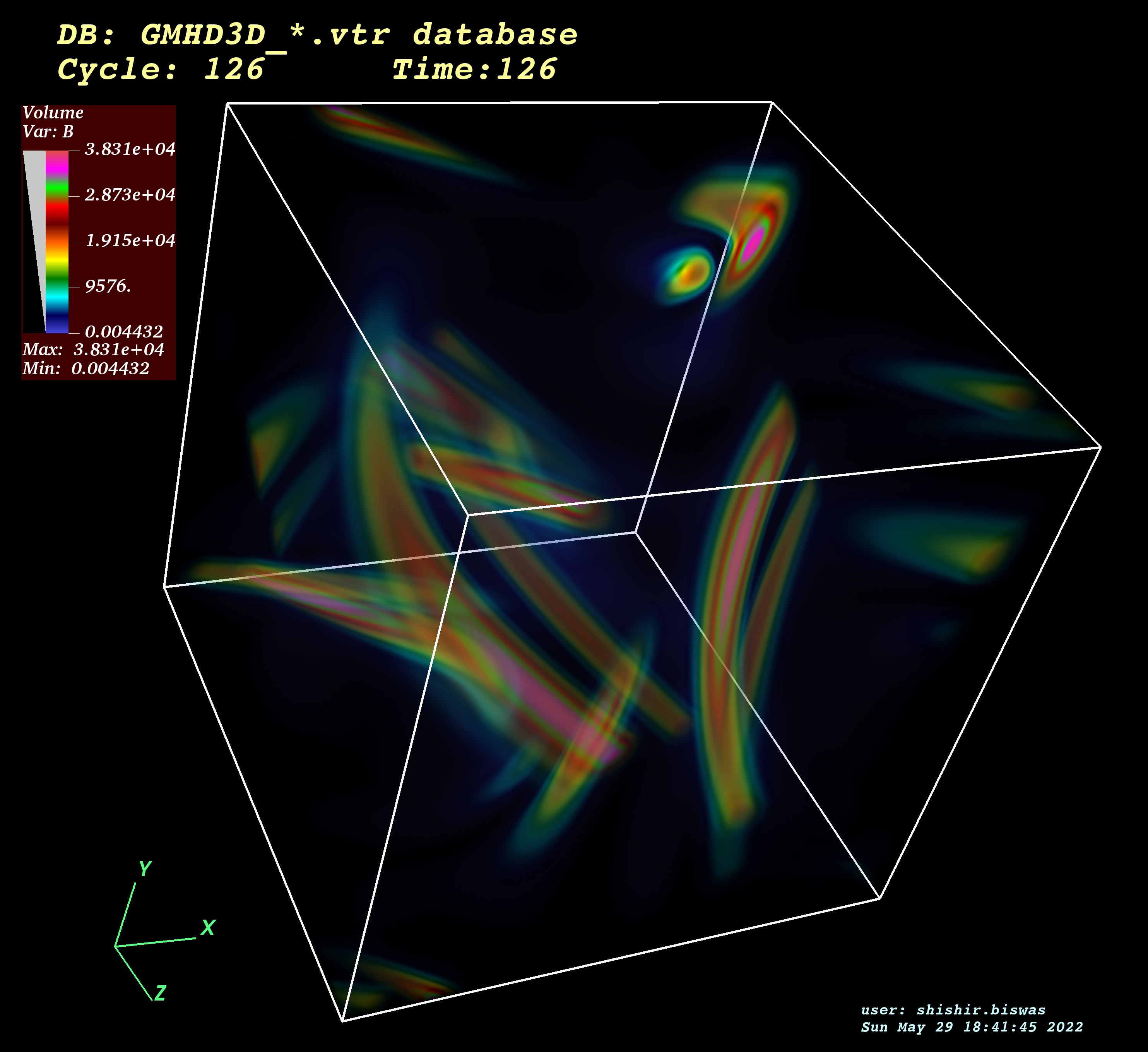}
		\caption{}
		\label{beta_0p6_volume}
	\end{subfigure}
	\begin{subfigure}{0.49\textwidth}
		\centering
		\includegraphics[scale=0.060]{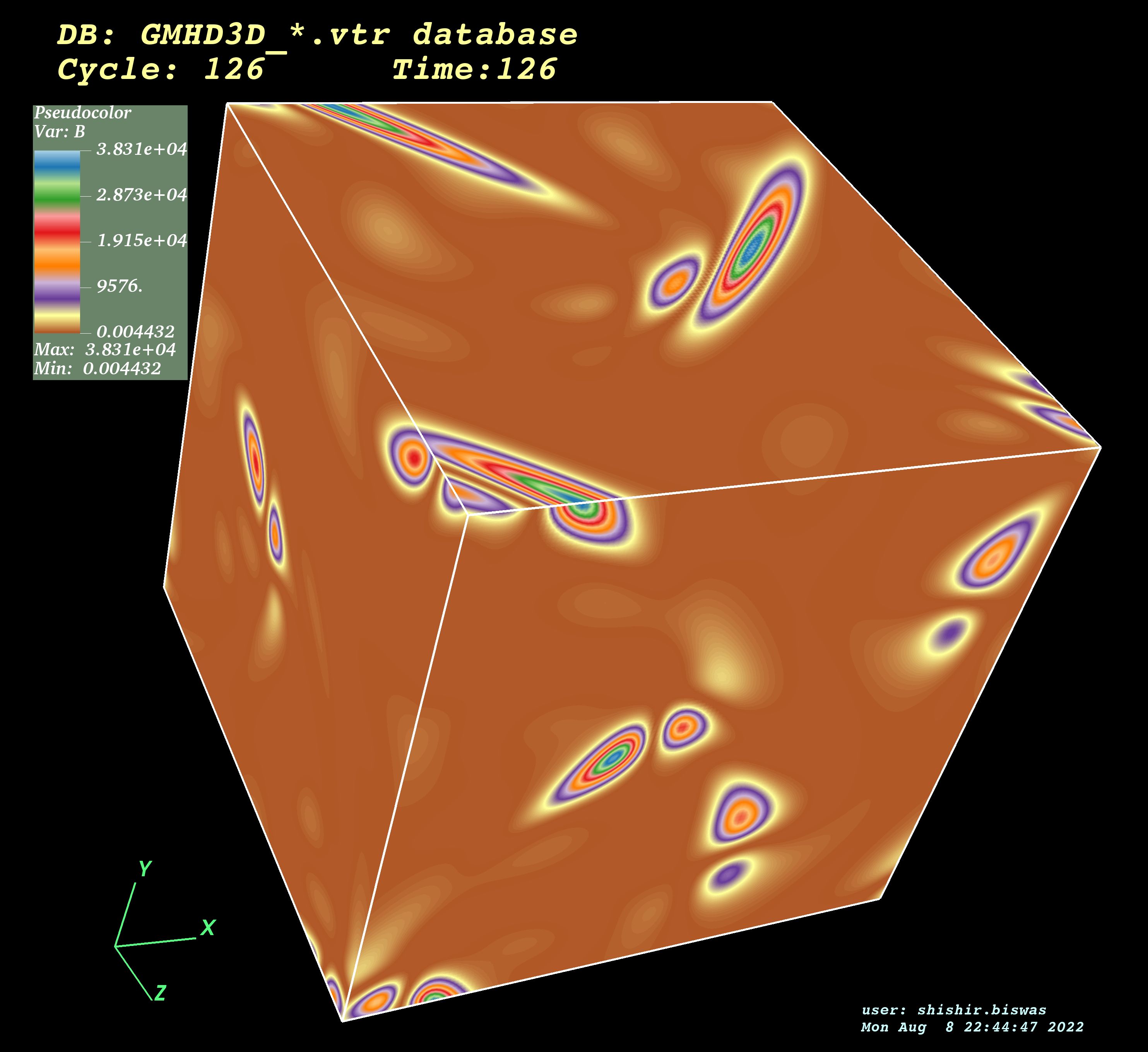}
		\caption{}
		\label{beta_0p6_pseudo}
	\end{subfigure}
	\caption{\textcolor{black}{Kinematic fast} Dynamo effect for different magnetic Reynolds number ($R_m$) using Yoshida-Morrison (YM) flow with \textcolor{black}{$\beta = 0.6$}. (a) The magnetic energy $\left(\sum\limits_{V} \frac{B^2(x,y,z,t)}{2} \right)$ growth is observed for different $R_m$ values. The (b) three-dimensional (3D) magnetic energy iso-surface \textcolor{black}{[Supplementary movie added]}, (c) three-dimensional (3D) volume rendering view of magnetic energy iso-surfaces and (d) three-dimensional (3D) pseudo-color view of magnetic energy iso-surfaces (for magnetic Reynolds number $R_m = 50$) is seen to be twisted ``ribbon'' like nature. Due to the presence of oscillation in the magnetic energy ``double ribbon'' structures are observed. Simulation details: grid resolution $256^3$, stepping time dt = $10^{-4}$, magnitude of fluid velocity $u_0 = 1.0$, Alfven mach number $M_A = 10.0$.}
\end{figure*}

For further higher $\beta$ value say $\beta = 0.6$, it is observed that magnetic energy grows exponentially with time for a wide range of $R_m$ values. We provide the results for magnetic Reynolds number ($R_m$) $50$ to $5000$ in small steps and observe that, for all the sets of numerical runs, there is prominent dynamo action [See Fig. \ref{beta_0p6_energy}]. This is due to the presence of fluid helicity \textcolor{black}{and twisting ability} in the system. As in the earlier cases, the presence of fluid helicity promotes the twisting ability of the flow and that results in dynamo action. The magnetic energy is seen to concentrate in elongated sheetlike or ``ribbon'' like structures [See Fig. \ref{beta_0p6_iso}, \ref{beta_0p6_volume}, \ref{beta_0p6_pseudo}] reported earlier as well \citep{Alexakis_PRE:2011}. The magnetic energy iso-surfaces form couple structures for some particular time. These structures are also seen similar to the ``double ribbon'' structures similar to the ``double cigar'' structures, reported earlier for ABC flow \citep{Dorch:2000}. In the simulation within the oscillatory regime, as the energy increases at the beginning of a `cycle', the secondary ribbon increases in size until the two ribbon become equal in size and strength midway through the cycle [See Fig. \ref{beta_0p6_iso}, \ref{beta_0p6_volume}, \ref{beta_0p6_pseudo}]. The primary ribbon now becomes smaller and actually vanishes at the end of the cycle, at which time the growth of the energy slows down. Towards the end of a cycle, the primary ribbon begins to decay, because the supply of flux to the primary ribbon is less than that to the secondary ribbon. At the end of one cycle the secondary ribbon is the only remaining structure.

\begin{figure*}
	\begin{subfigure}{0.34\textwidth}
		\centering
		\includegraphics[scale=0.43]{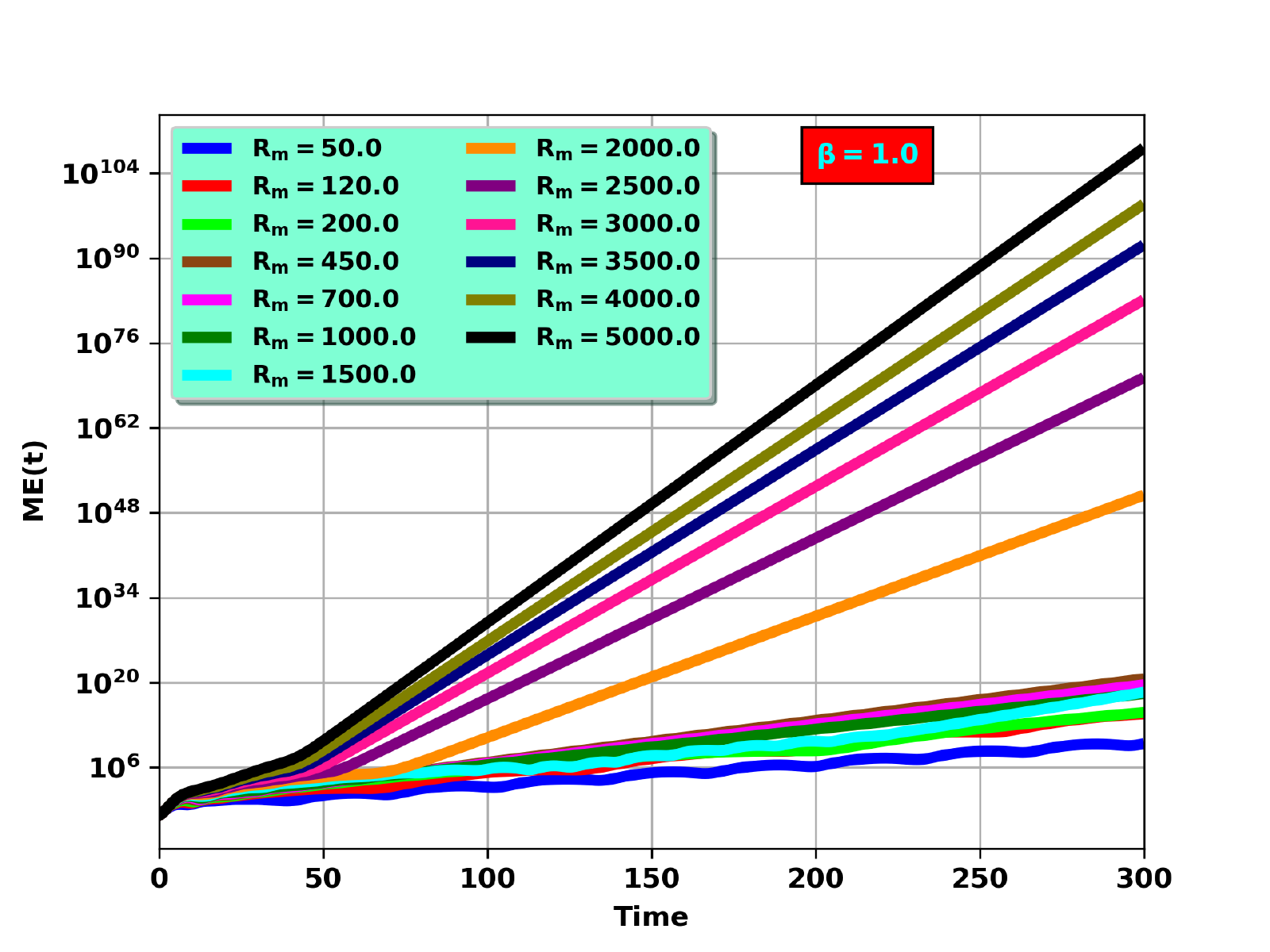}
		\caption{}
		\label{beta_1p0_energy}
	\end{subfigure}
	\begin{subfigure}{0.32\textwidth}
		\centering
		\includegraphics[scale=0.045]{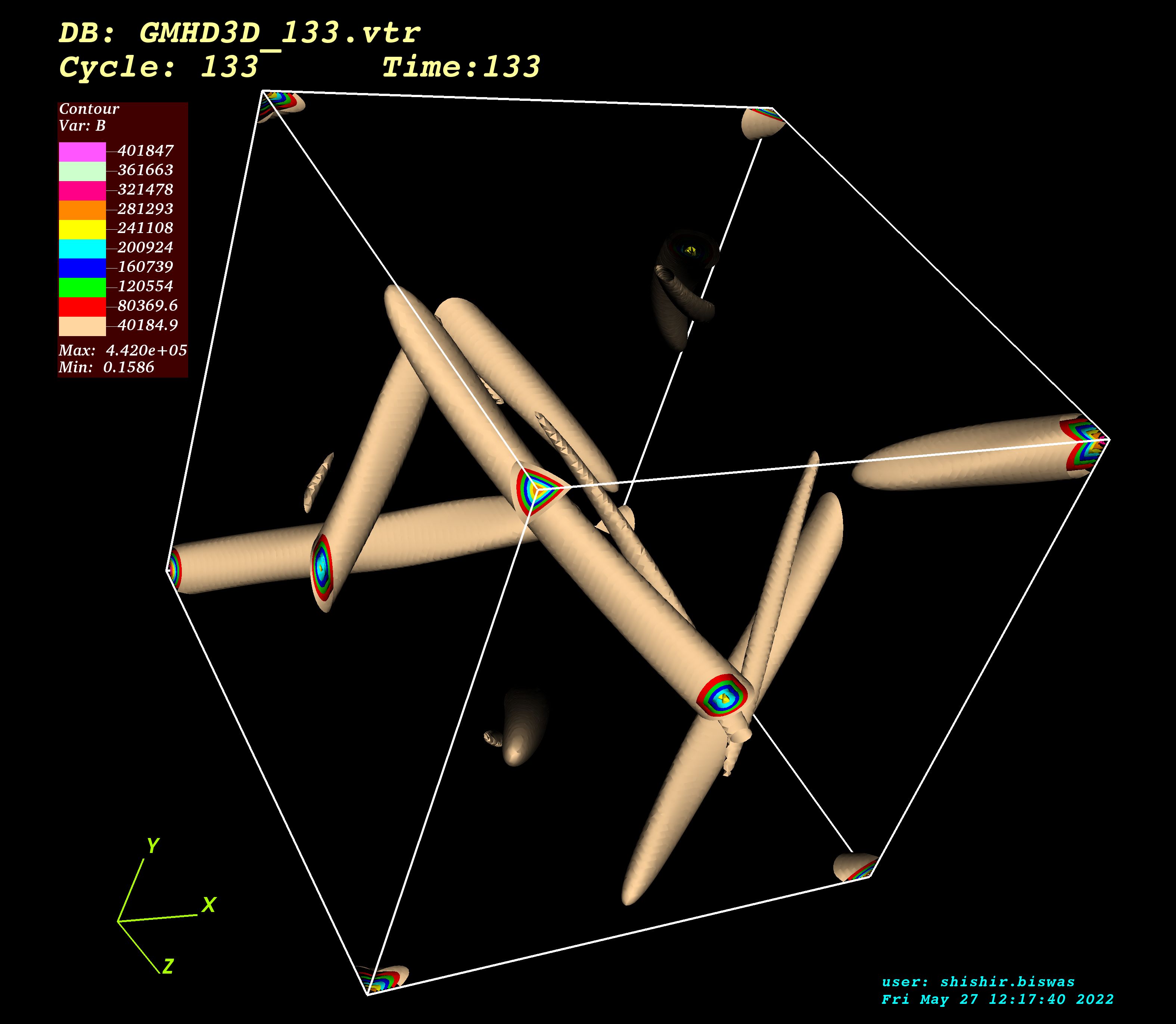}
		\caption{}
		\label{beta_1p0_iso}
	\end{subfigure}
	\begin{subfigure}{0.32\textwidth}
		\centering
		\includegraphics[scale=0.045]{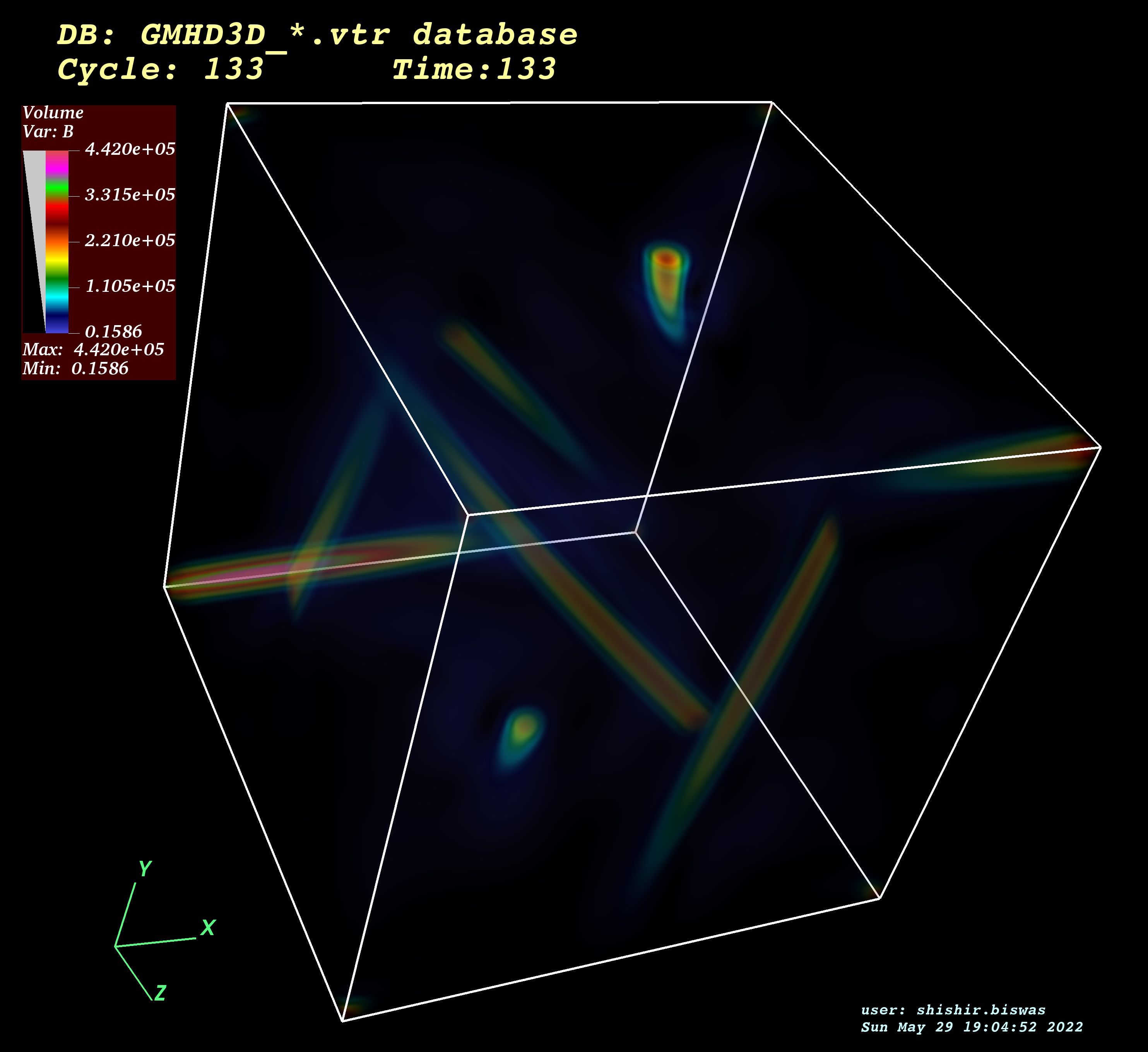}
		\caption{}
		\label{beta_1p0_volume}
	\end{subfigure}
	\begin{subfigure}{0.32\textwidth}
		\centering
		\includegraphics[scale=0.045]{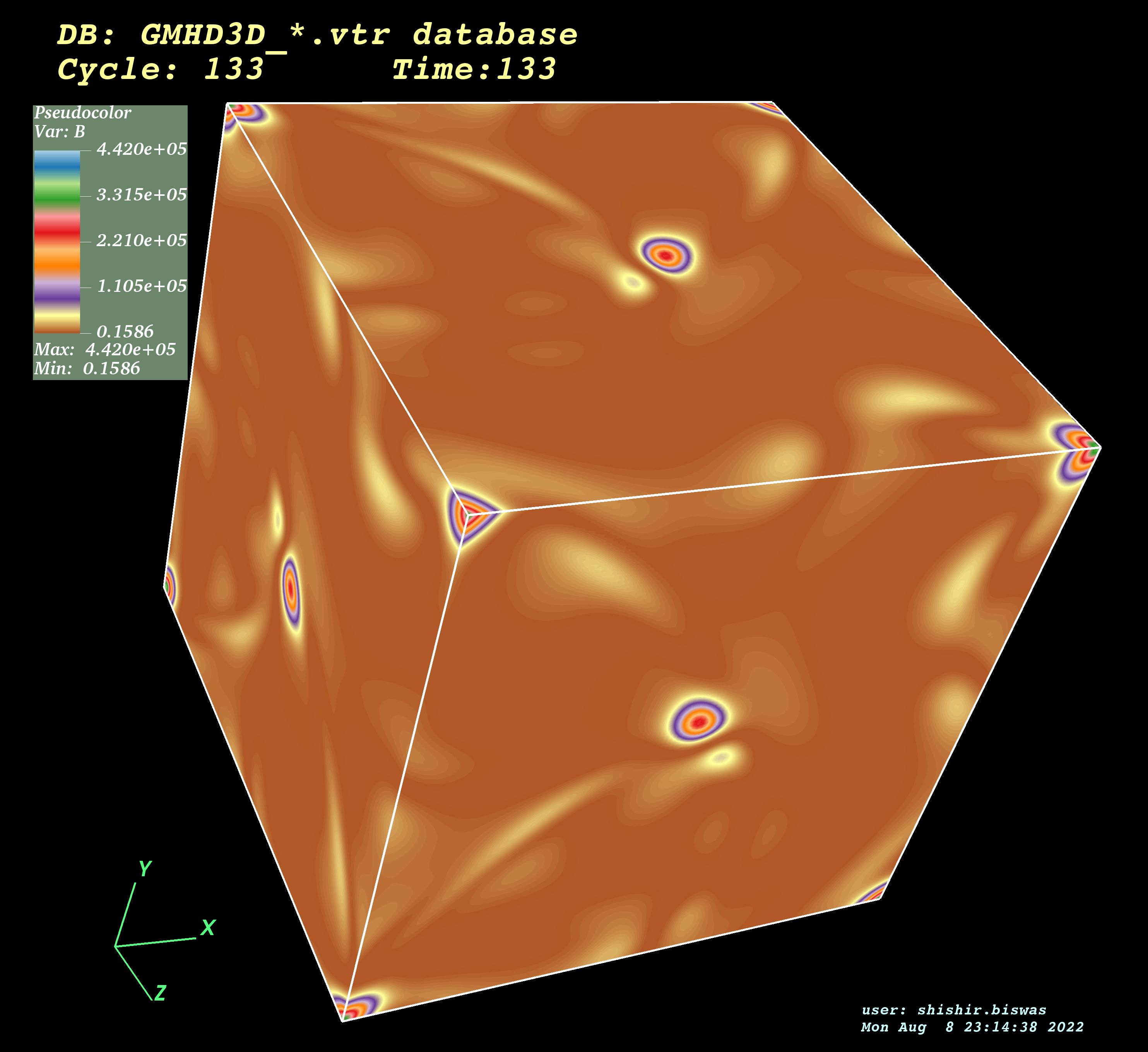}
		\caption{}
		\label{beta_1p0_pseudo}
	\end{subfigure}
	\begin{subfigure}{0.32\textwidth}
		\centering
		\includegraphics[scale=0.045]{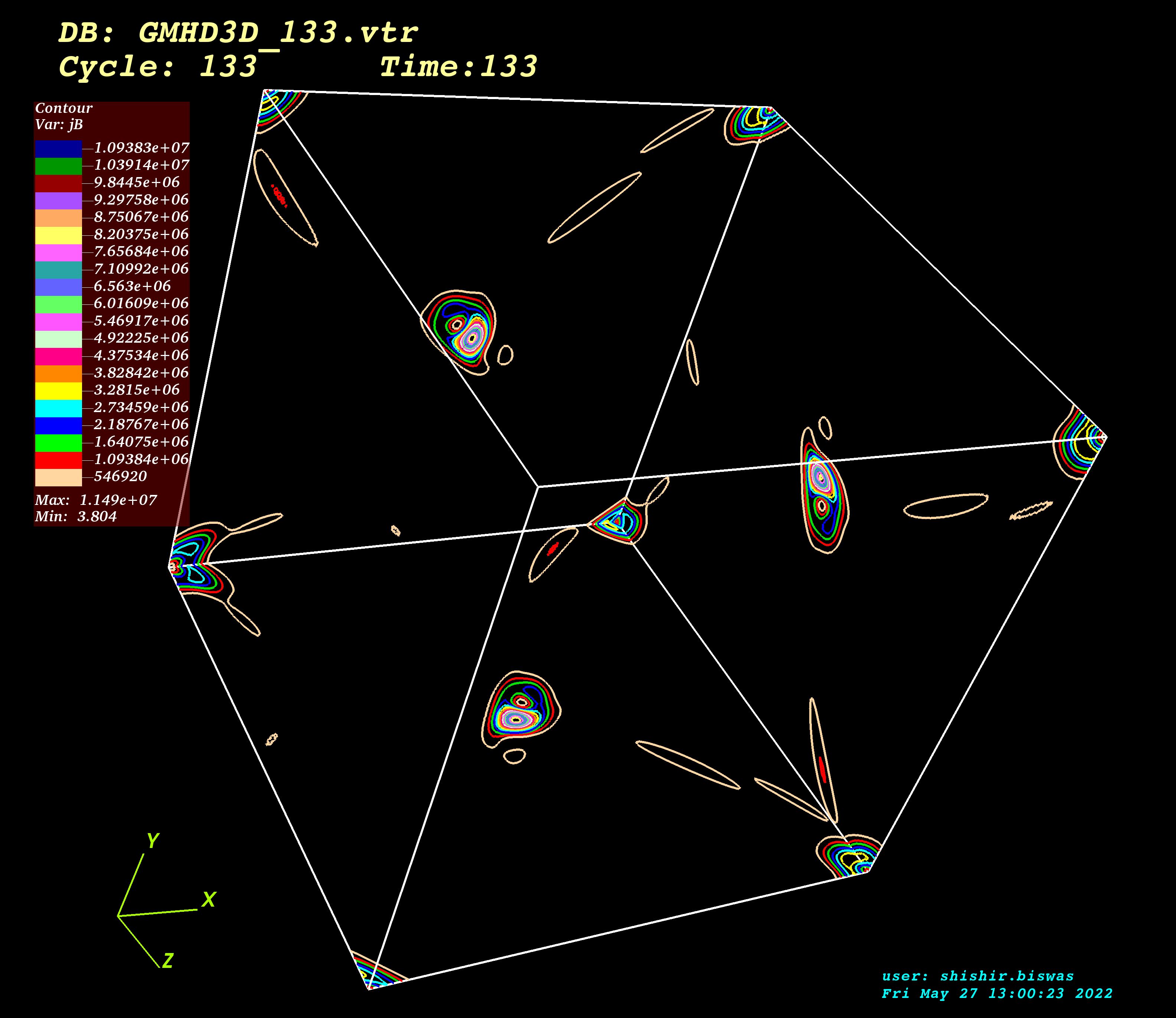}
		\caption{}
		\label{current density}
	\end{subfigure}
	\caption{\textcolor{black}{Kinematic fast} Dynamo effect for different magnetic Reynolds number ($R_m$) using Yoshida-Morrison (YM) flow with \textcolor{black}{$\beta = 1.0$}. (a) The magnetic energy $\left(\sum\limits_{V} \frac{B^2(x,y,z,t)}{2} \right)$ growth is observed for different $R_m$ values. The (b) three-dimensional (3D) magnetic energy iso-surface \textcolor{black}{[Supplementary movie added]}, (c) three-dimensional (3D) volume rendering view of magnetic energy iso-surfaces and (d) three-dimensional (3D) pseudo-color view of magnetic energy iso-surfaces (for magnetic Reynolds number $R_m = 50$) is seen to be ``cigar like'' in nature. (e) Strong localization of the current density is also observed. Simulation details: grid resolution $256^3$, stepping time dt = $10^{-4}$, magnitude of fluid velocity $u_0 = 1.0$, Alfven mach number $M_A = 10.0$.}
\end{figure*}

Finally, we explore the only left $\beta$ value, i.e. $\beta = 1.0$. As discussed earlier, YM flow with \textcolor{black}{$\beta = 1.0$} regenerates the classical Arnold-Beltrami-Childress [ABC] flow. As in the earlier cases (viz. $\beta = 0.0, 0.2, 0.4, 0.8$), we provide runs for a wide range of magnetic Reynolds number ($R_m$) starting from $R_m = 50$. As is expected, for $\beta=1$, for all the sets of numerical runs, we observe significant magnetic energy growth [See Fig. \ref{beta_1p0_energy}]. The visualization of this maximum helical ($\beta = 1.0$) flow leads to visible separatrix formation [See Fig. \ref{initial flow beta 1p0}], a signature of chaotic flow which drives  the exponential energy growth  for $\beta = 1.0$, as the flow becomes maximum helical similar like ABC flow. This exponential growth of magnetic energy using ABC flow via \textcolor{black}{kinematic fast} dynamo action is well explored so far by various authors \citep{Frish_Dynamo:1986,Dorch:2000,Archontis:2003,Bouya:2013}. Probably the ABC flow is the most explored flow, that has been used for dynamo action. We also produce identical magnetic energy growth like ABC flow starting from an initial negligible value, using YM flow with \textcolor{black}{$\beta = 1$} [See Fig. \ref{beta_1p0_energy}]. The study of magnetic energy iso-surface also leads to an interesting observation. It is observed that, the exponentially growing nature of magnetic energy generates \textcolor{black}{rigorous geometric} structures or likely to say ``cigar like'' structures [See Fig. \ref{beta_1p0_iso}, \ref{beta_1p0_volume}, \ref{beta_1p0_pseudo}], and it supported by literatures \citep{Frish_Dynamo:1986,Dorch:2000,Archontis:2003,Bouya:2013}. From our numerical experiments, we confirm that ``cigar like'' structure is a stable structure and its appearance is a signature for the onset of exponential growth in magnetic energy. Strong localization of the current density structures co-existing with magnetic ``cigars'', due to magnetic reconnection, is also observed [See Fig. \ref{current density}]. 

By systematically varying the helicity parameter $\beta$ in the YM flow \citep{EPI2D:2017} or in the other words by injecting normalized fluid helicity in the system, we have explored a systematic route that connects flat ``ribbon'' like structures to ``twisted ribbon'' and finally towards classical ``cigar'' like structures. The emergence of these \textcolor{black}{significant} structures (``ribbon'', ``twisted ribbon'', ``cigar'') due to the injected normalized fluid helicity are demonstrated via direct numerical simulation. Interestingly, though all these structures have been reported earlier, but the significant connection between these structures using a single fluid flow model, not explored so far. Here in this present study we connect all these different \textcolor{black}{well known} structures via normalized fluid helicity injection using YM flows.

\begin{figure*}
	\centering
	\begin{subfigure}{0.24\textwidth}
		\centering
		\includegraphics[scale=0.039]{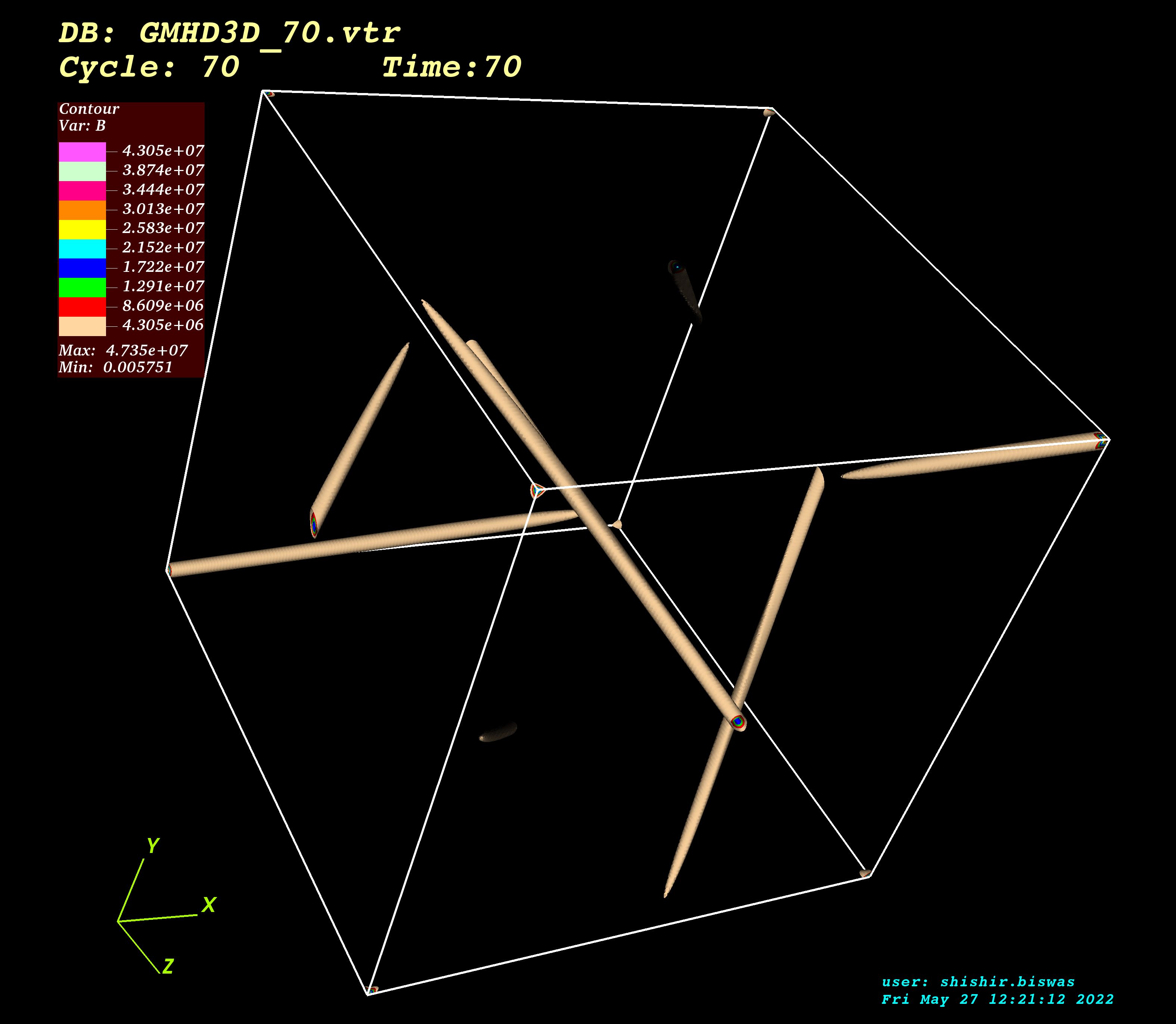}
		\caption{}
		\label{Rm_50}
	\end{subfigure}
	\begin{subfigure}{0.24\textwidth}
		\centering
		\includegraphics[scale=0.039]{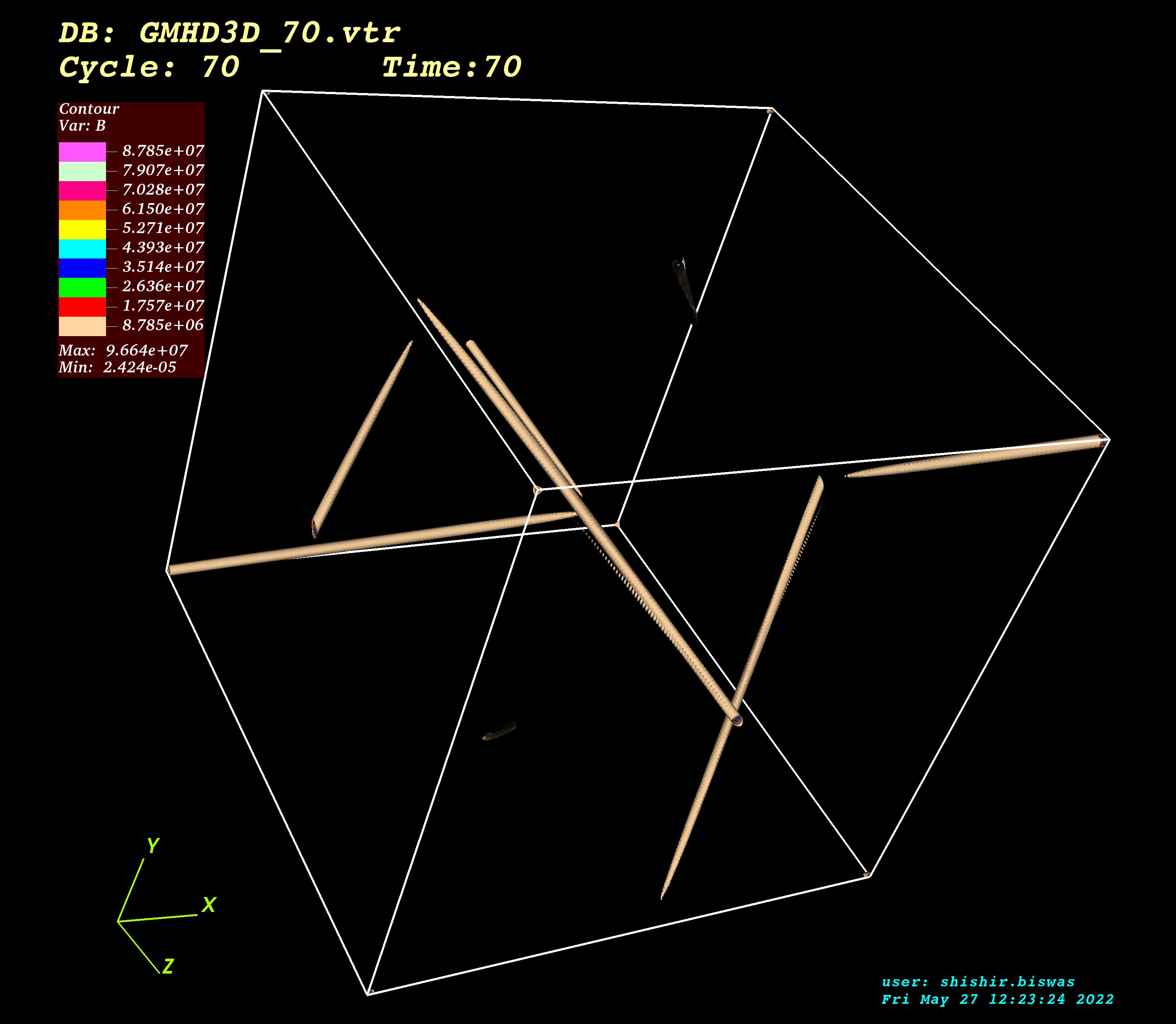}
		\caption{}
		\label{Rm_450}
	\end{subfigure}
	\begin{subfigure}{0.24\textwidth}
		\centering
		\includegraphics[scale=0.039]{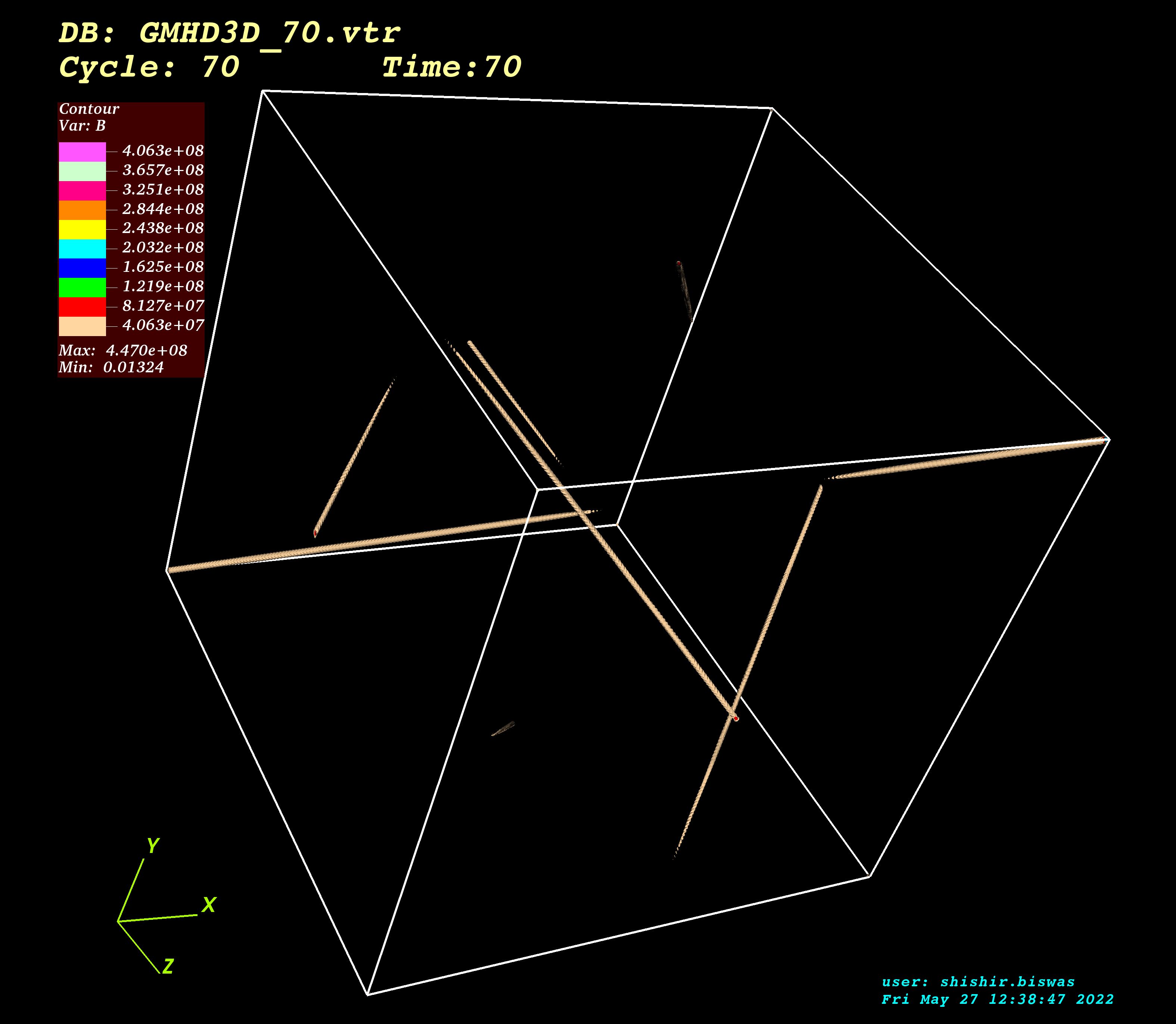}
		\caption{}
		\label{Rm_3000}
	\end{subfigure}
	\begin{subfigure}{0.24\textwidth}
		\centering
		\includegraphics[scale=0.40]{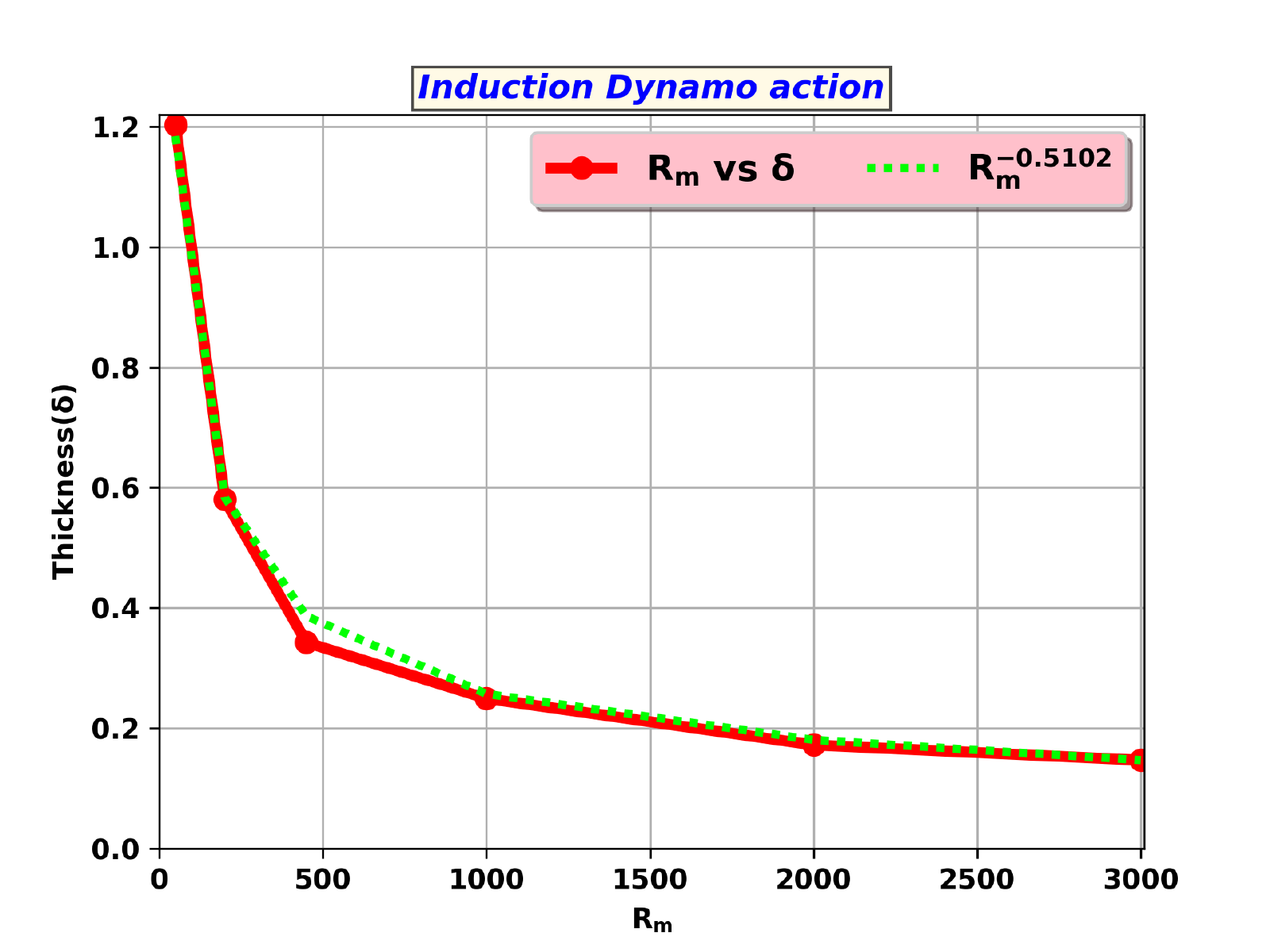}
		\caption{}
		\label{cigar thickness}
	\end{subfigure}
	\caption{Three-dimensional (3D) magnetic energy iso surface  for (a) $R_m = 450$, (b) $R_m = 1000$, (c) $R_m = 3000$ using Yoshida-Morrison (YM) flow with \textcolor{black}{$\beta = 1.0$}. (d) With the increase of $R_m$ value the thickness of magnetic cigars ($\delta$) are seen to be decreased with a strong scaling $\frac{1}{\sqrt{R_m}}$, which is observed earlier for ABC flow. }
	\label{cigars at various Rm}
\end{figure*}

Coming back to \textcolor{black}{$\beta = 1.0$} case, as discussed, this class of flow is identical to the well known Arnold-Beltrami-Childress [ABC] flow. It is shown that, the fastest magnetic energy growth or the exponential energy growth for this kind of flow is signified by generation of a special kind of structure called ``cigar'' like structure. We study the magnetic Reynolds number ($R_m$) effect on the ``cigar'' structure. Keeping all the parameters identical, we perform numerical runs for various magnetic Reynolds number ($R_m$) and visualize the magnetic energy iso-surfaces [See Fig. \ref{Rm_50}, \ref{Rm_450}, \ref{Rm_3000}]. From figures \ref{Rm_50}, \ref{Rm_450} and \ref{Rm_3000} it is identified that, the thickness of a ``cigar'' ($\delta$) decreases with the increase of magnetic Reynolds number ($R_m$). We also calculate the the ``cigar'' thickness ($\delta$) for various intermediate $R_m$ and plot it [See Fig. \ref{cigar thickness}]. From Fig. \ref{cigar thickness} it is observed that, the ``cigar'' thickness ($\delta$) decreases with $R_m$ with a strong scaling of $\frac{1}{\sqrt{R_m}}$. This $\frac{1}{\sqrt{R_m}}$ scaling is also identified by various authors \citep{Frish_Dynamo:1986, Archontis:2003} for ABC flow and suggests Sweet-Parker scaling. Here in this present study, we observe the identical scaling for YM flow (\textcolor{black}{$\beta = 1$}). As the change of magnetic Reynolds number ($R_m$) only changes the cigar thickness ($\delta$), hence to help visualize, we plot iso-surfaces for magnetic Reynolds number ($R_m = 50$) for all the case discussed above.

\begin{figure*}
	\centering
	\begin{subfigure}{0.49\textwidth}
		\centering
		\includegraphics[scale=0.55]{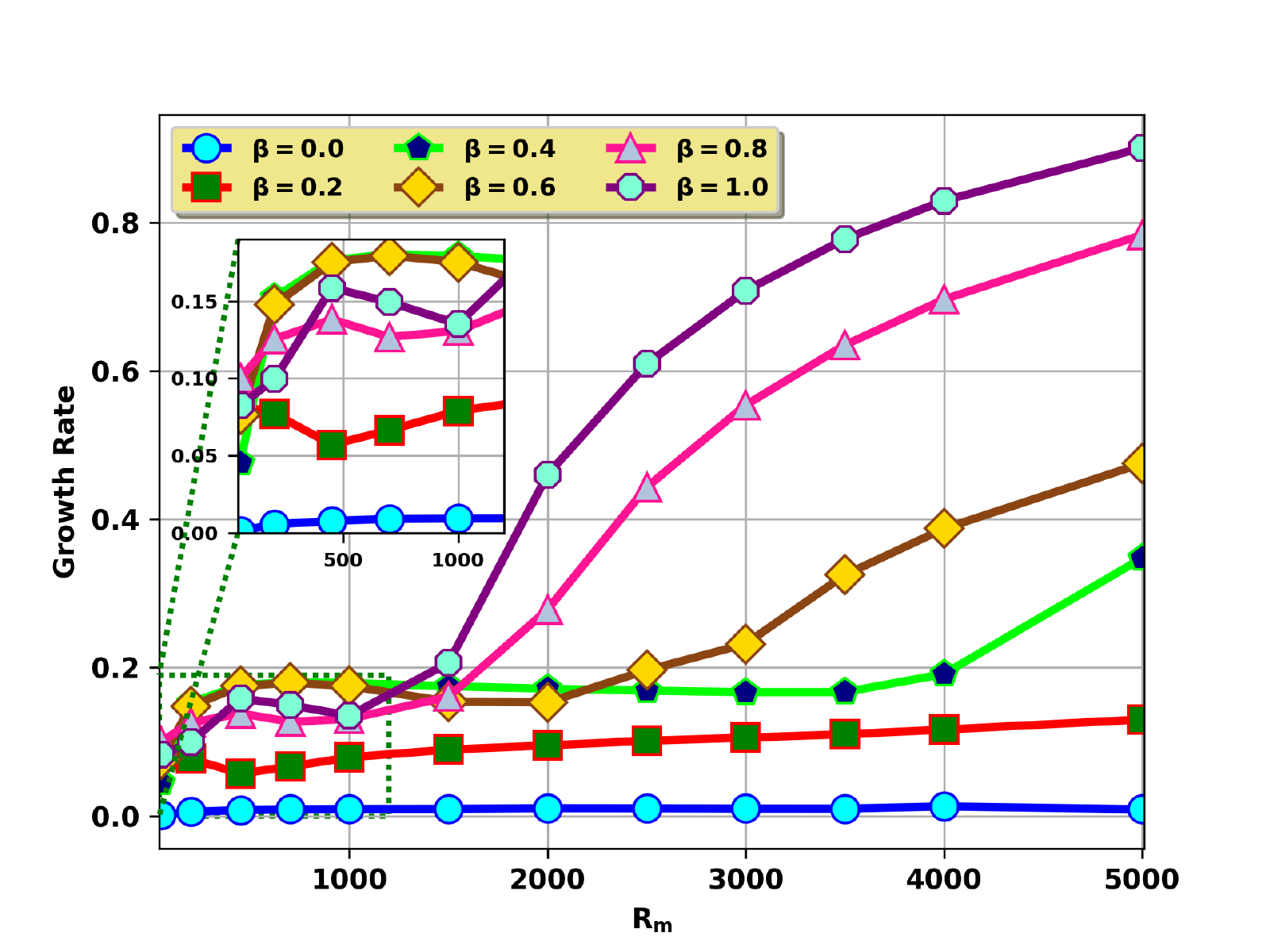}
	\end{subfigure}
	\caption{Magnetic energy growth rate for various class of Yoshida-Morrison (YM) flow at different magnetic Reynolds numbers ($R_m$). Importantly, for lower $R_m$ it is identified that ABC flow or Yoshida-Morrison (YM) flow with \textcolor{black}{$\beta = 1.0$} is not the best dynamo model.}
	\label{Rm vs Growth rate}
\end{figure*}

We also calculate the magnetic energy growth rate for all $\beta$ cases. It is seen from Fig. \ref{Rm vs Growth rate} that, for $\beta = 0$ as there is no fluid helicity present in the system, \textcolor{black}{additionally, the flow does not have enough stretching ability} , so there is no dynamo excitation. The growth rate is also seen to be flat (blue line in Fig. \ref{Rm vs Growth rate}) for higher and higher magnetic Reynolds number ($R_m$). The growth rate for $\beta = 0.2$ is seen to saturate over $R_m$, (red line in Fig. \ref{Rm vs Growth rate}) though the prominent dynamo action is observed for a large range of magnetic Reynolds number ($R_m$) value. For higher $\beta$ value, i.e, $\beta = 0.4, 0.6, 0.8$ it is noticed that, the magnetic energy growth rate increases as $R_m$ increases, after some critical value. For all the intermediate $\beta$ values, we observe significant magnetic energy growth [See Fig. \ref{Rm vs Growth rate}]. We estimate the energy growth rate for the well explored $\beta = 1.0$ case, which is homologous to the conventional ABC flow. For Fig. \ref{Rm vs Growth rate} it is observed that the the growth rate of magnetic energy for $\beta = 1.0$ case increases significantly with $R_m$. For lower magnetic Reynolds number ($R_m$), there is a dip in the growth rate curve (purple line in Fig. \ref{Rm vs Growth rate}), which is reported earlier as well \citep{Frish_Dynamo:1986}. It is also interesting to note that, for lower magnetic Reynolds number ($R_m$), the ABC dynamo model is not the best dynamo model [See Fig. \ref{Rm vs Growth rate}] in the sense of growth rate of magnetic energy. The magnetic energy growth rates for $\beta = 0.4$ \& $\beta = 0.6$ are seen to be larger than regular ABC flow or $\beta = 1.0$, flow for lower $R_m$ values. This observation suggests that for the YM class of flows, maximum fluid helicity does not imply the fastest dynamo action at lower $R_m$ values. 

\begin{figure*}
	\centering
	\begin{subfigure}{0.49\textwidth}
		\centering
		\includegraphics[scale=0.55]{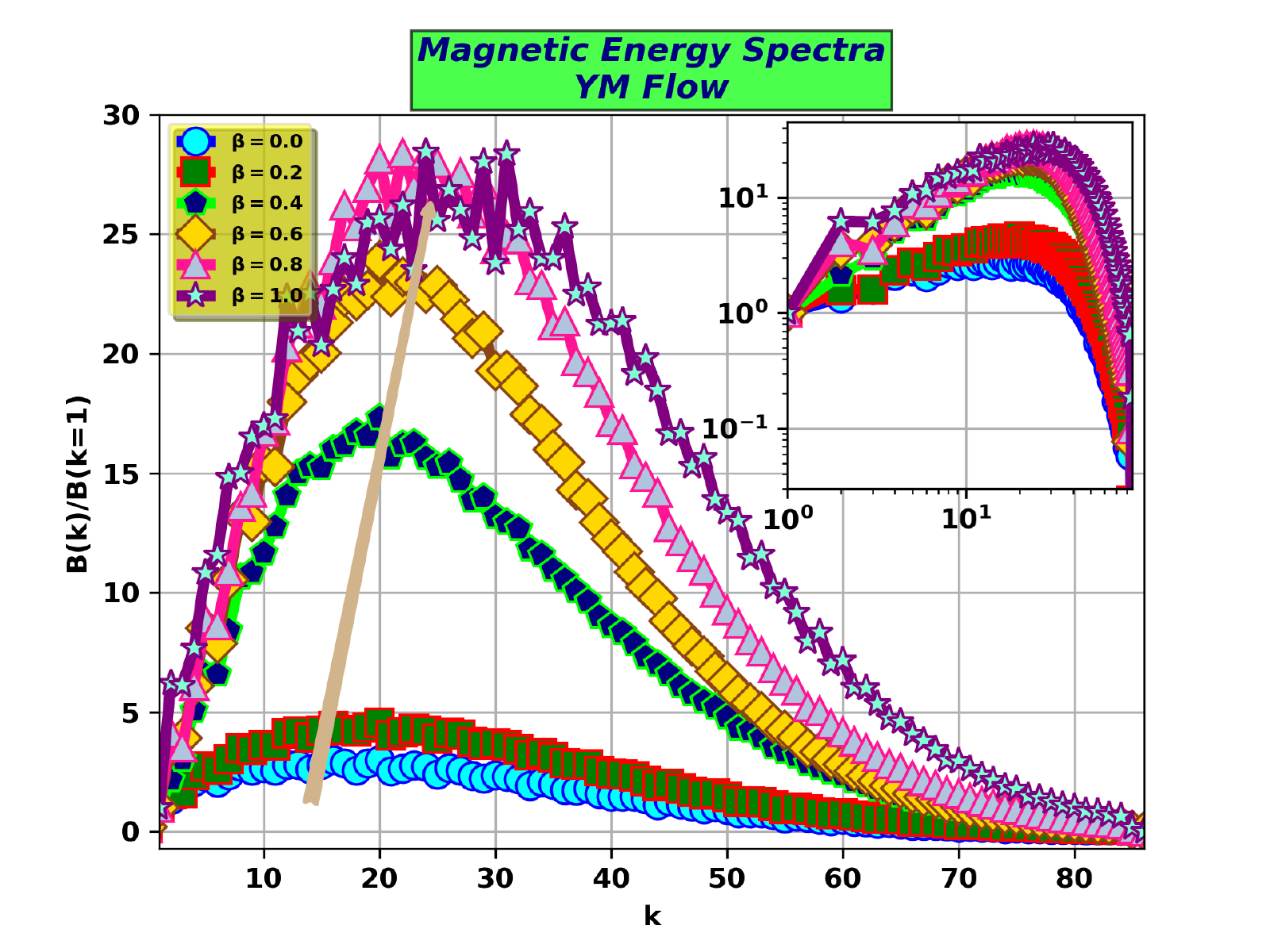}
	\end{subfigure}
	\caption{\textcolor{black}{Time averaged (time average taken from t = 41.0 to 100.0) magnetic energy spectral density $B(k)$ [such that $\int B(k,t)dk$ is the total energy at time t] for different classes of Yoshida-Morrison (YM) flow at fixed magnetic Reynolds numbers ($R_m = 450.0$) are shown in linear and log-log (inset view) scales. The spatial scale spectrum clearly indicates that the dynamo is a smale-scale fast dynamo.  Spectra are truncated at the mode number value ($k_{max} = 85$), and the peaks are in the range of $k = 15$ to $k = 40$, indicating that it is well resolved from  spectral point of view.}}
	\label{spectra}
\end{figure*}

\textcolor{black}{For each of the cases mentioned above, we have determined the time-averaged magnetic energy spectrum [$\int_{0}^{\infty} B(k) dk$]. The spectra has a peak at a higher mode number, which is the hallmark of small scale dynamo(SSD), as can be seen in  Fig. \ref{spectra}.  Additionally, all of the spectra are truncated at the mode value ($k_{max} = 85$), and the peaks are in the range of $k = 15$ to $40$, indicating that it is well resolved from a spectral perspective.
}

\begin{figure*}
	\centering
	\begin{subfigure}{0.32\textwidth}
		\centering
		\includegraphics[scale=0.38]{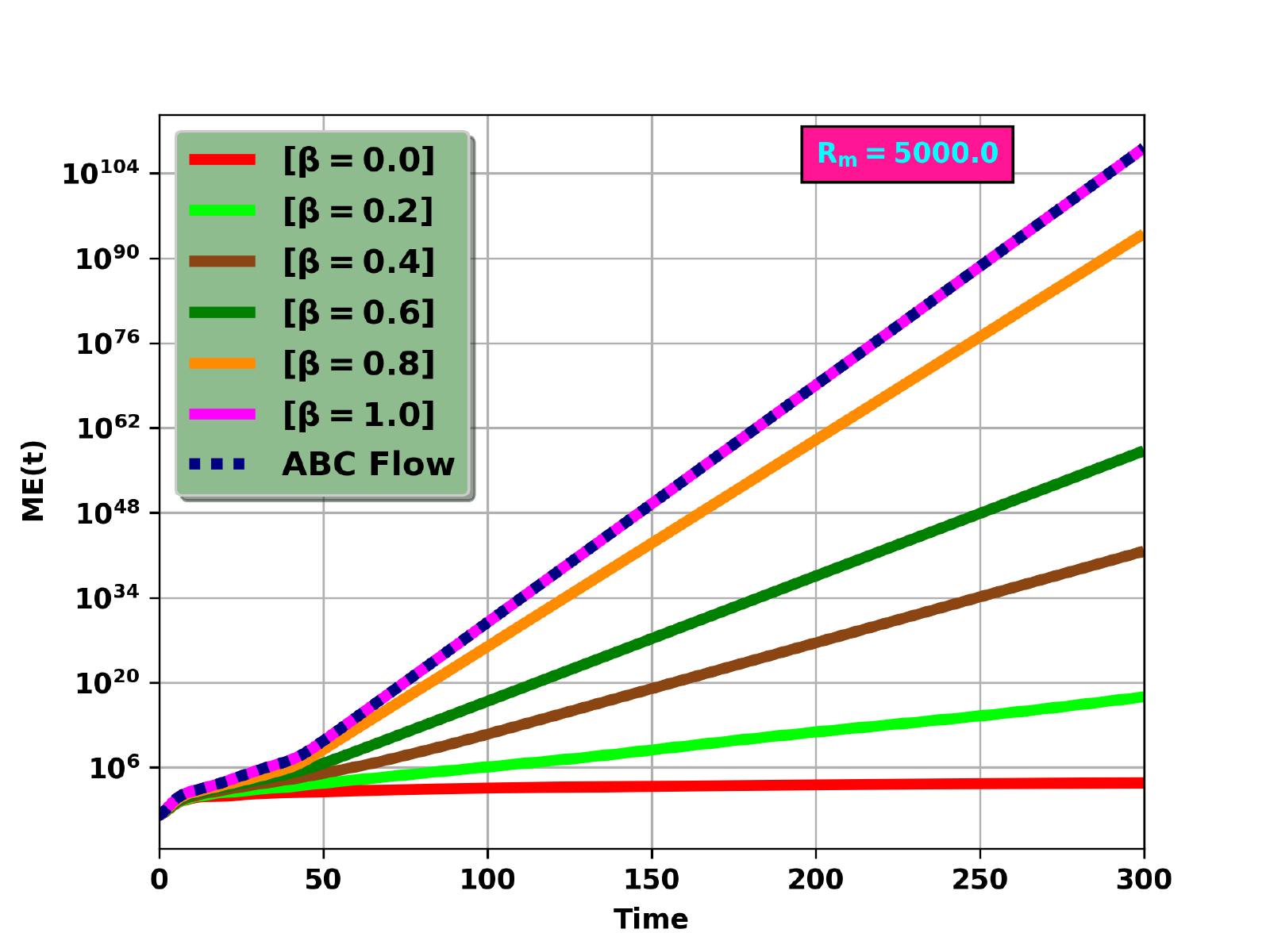}
		\caption{}
		\label{Rm = 5000}
	\end{subfigure}
	\begin{subfigure}{0.32\textwidth}
		\centering
		\includegraphics[scale=0.38]{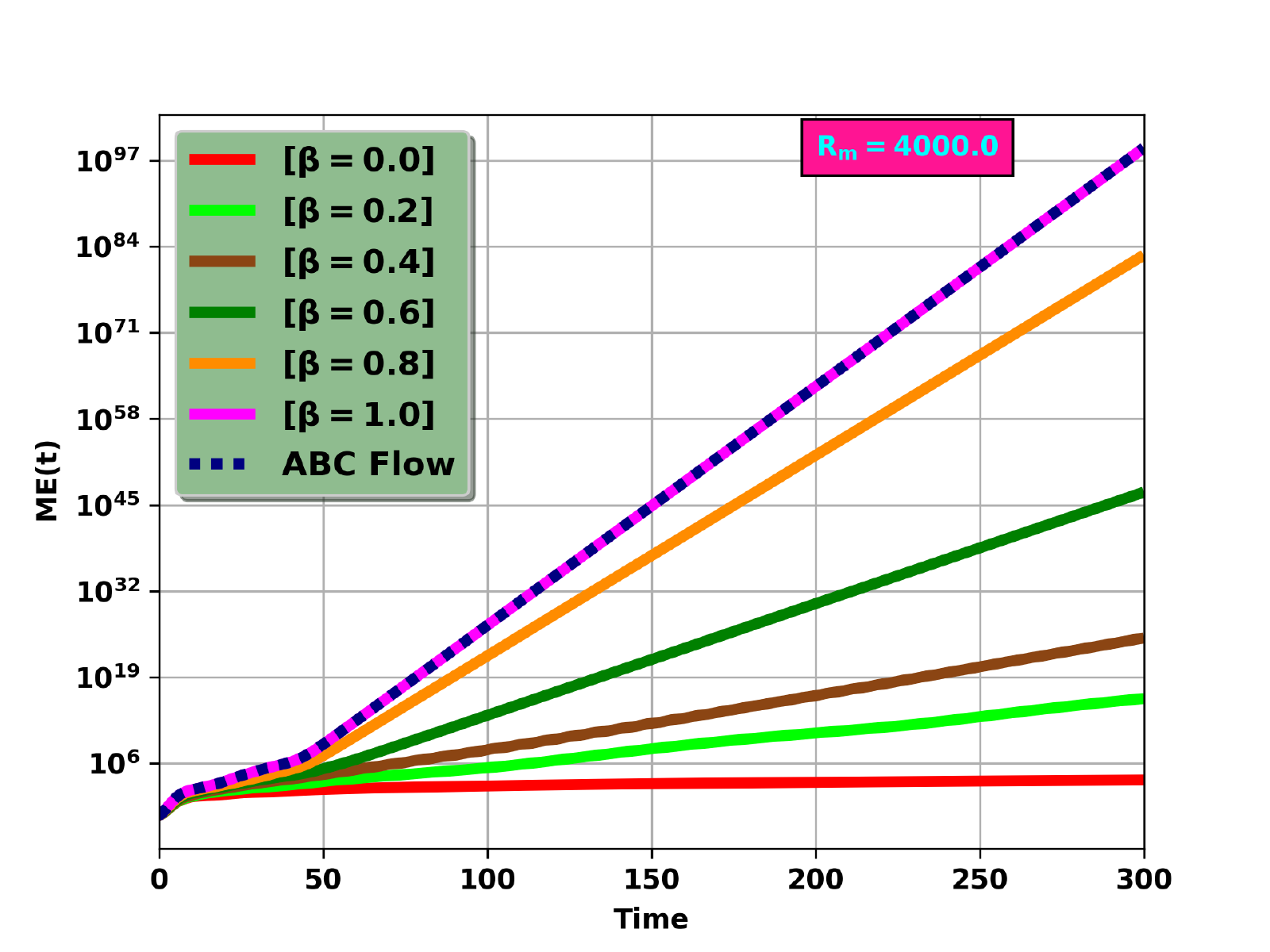}
		\caption{}
		\label{Rm = 4000}
	\end{subfigure}
	\begin{subfigure}{0.32\textwidth}
		\centering
		\includegraphics[scale=0.38]{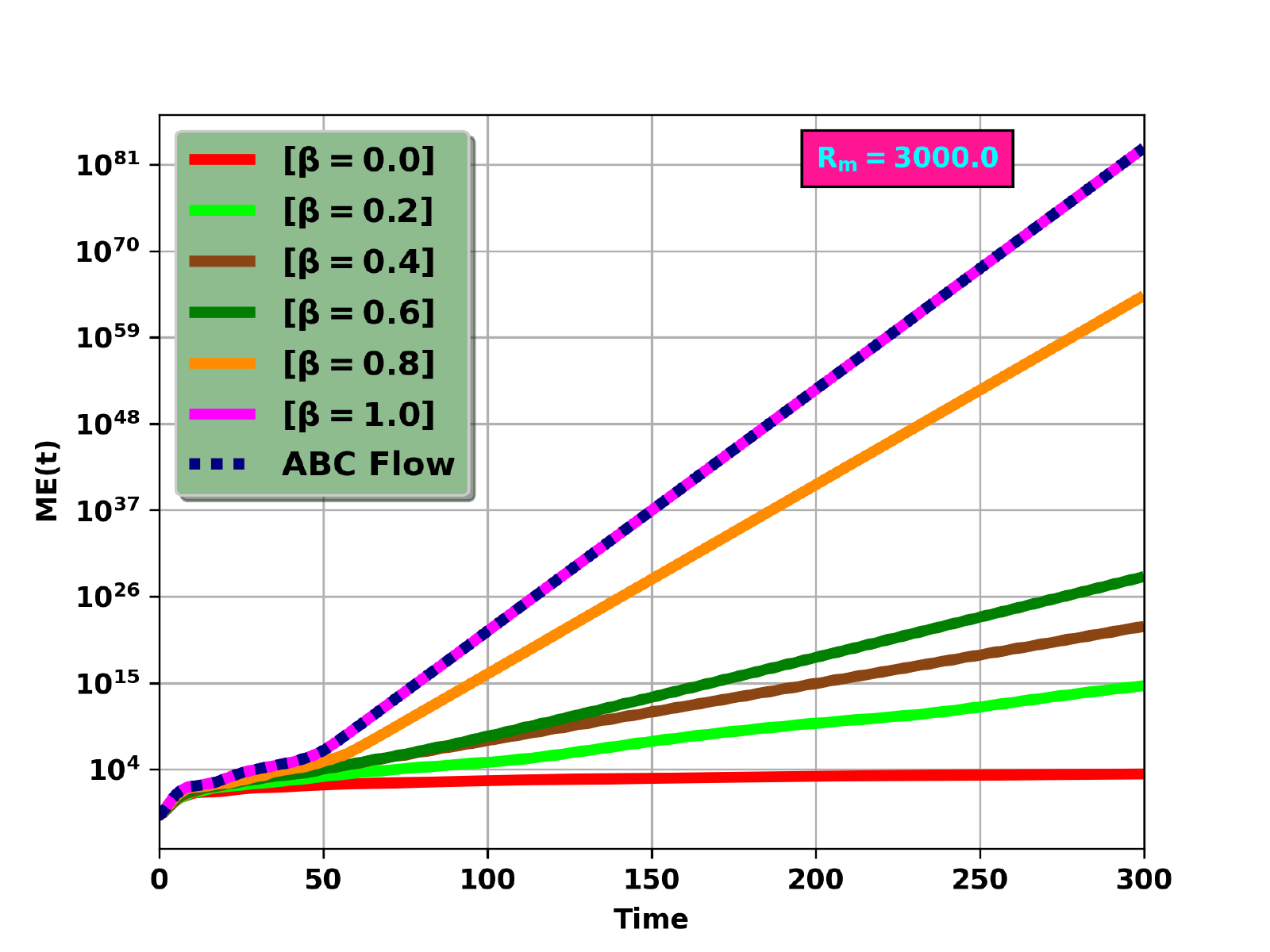}
		\caption{}
		\label{Rm = 3000}
	\end{subfigure}
	\begin{subfigure}{0.32\textwidth}
		\centering
		\includegraphics[scale=0.38]{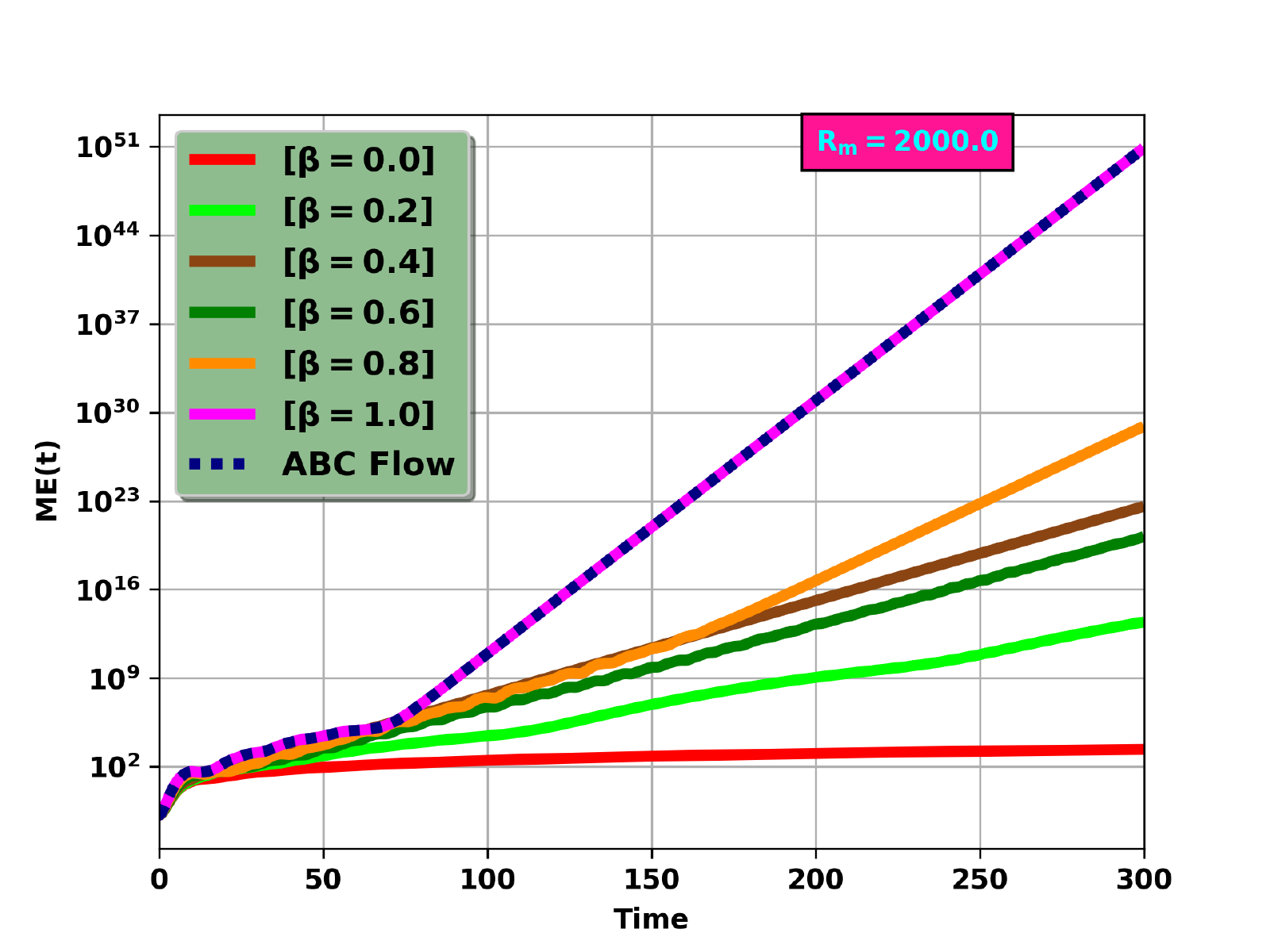}
		\caption{}
		\label{Rm = 2000}
	\end{subfigure}
	\begin{subfigure}{0.32\textwidth}
		\centering
		\includegraphics[scale=0.38]{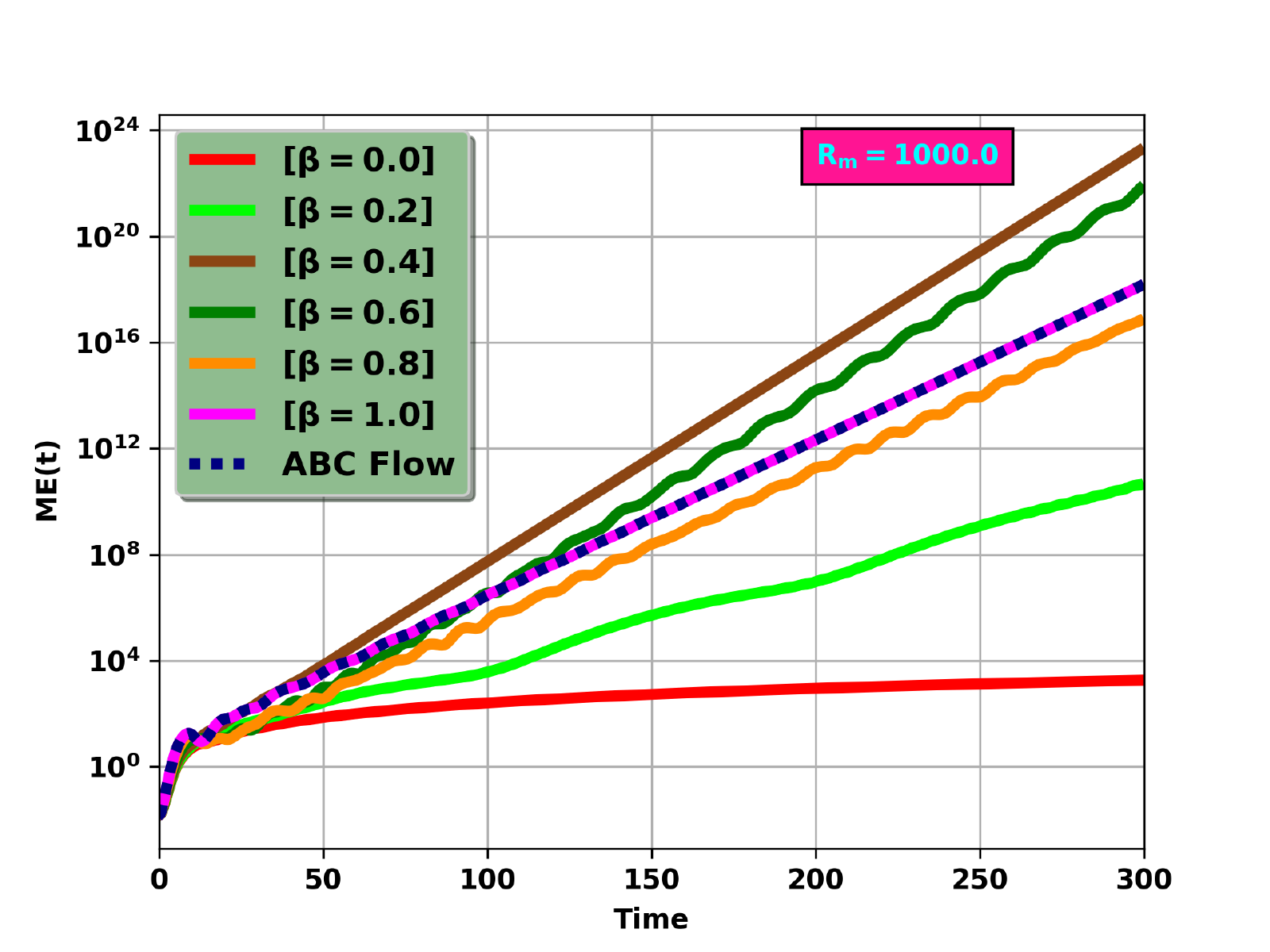}
		\caption{}
		\label{Rm = 1000}
	\end{subfigure}
	\begin{subfigure}{0.32\textwidth}
		\centering
		\includegraphics[scale=0.38]{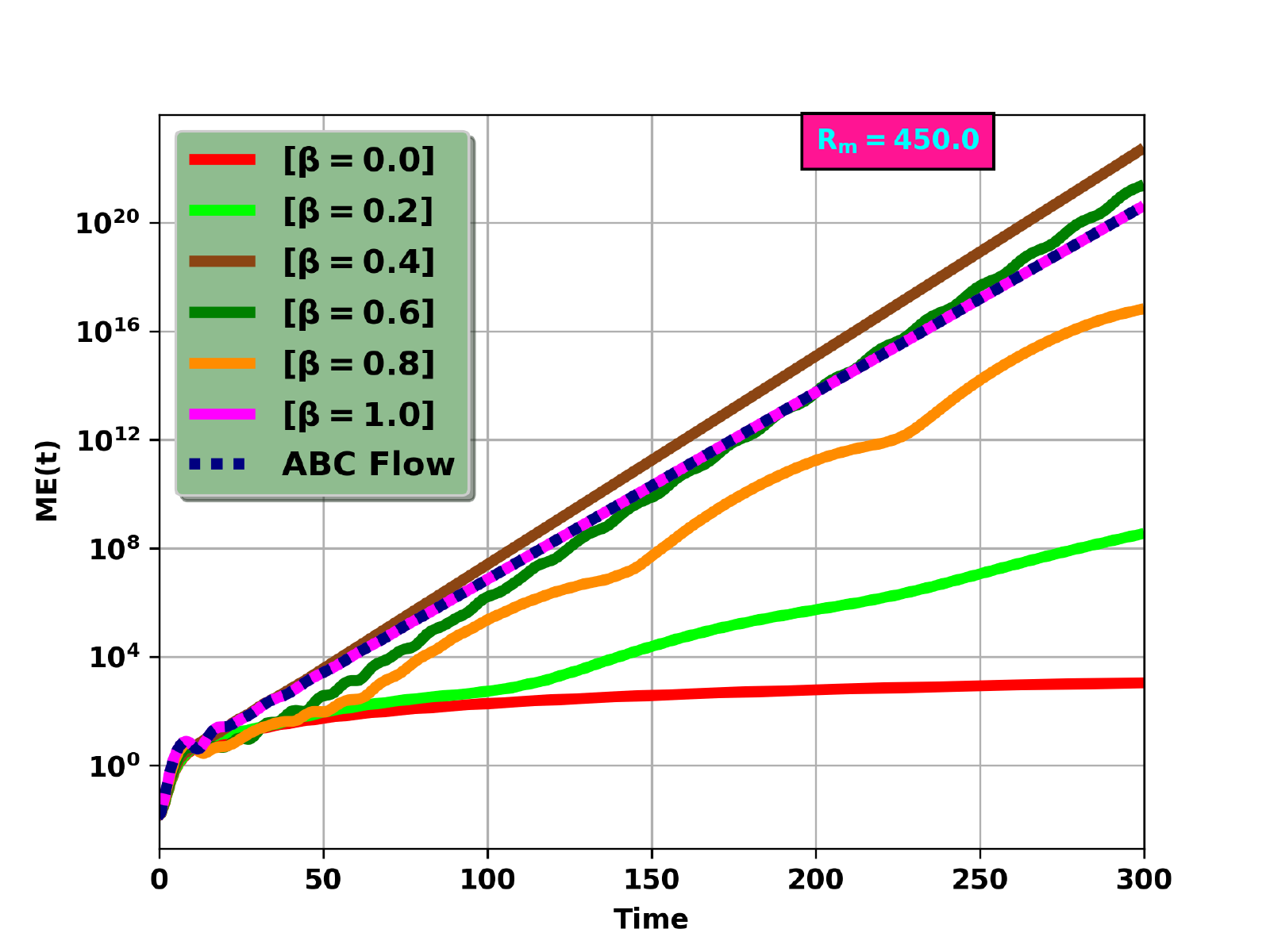}
		\caption{}
		\label{Rm = 450}
	\end{subfigure}
	\caption{Exponential growth of magnetic energy for (a) $R_m = 5000$, (b) $R_m = 4000$, (c) $R_m = 3000$, (d) $R_m = 2000$, (e) $R_m = 1000$, (f) $R_m = 450$ with various class of Yoshida-Morrison (YM) flow. A clear transition from non-dynamo regime to  dynamo regime is observed, at the cost of fluid helicity for higher magnetic Reynolds numbers ($R_m$). These figures significantly indicate the possible route that connects non-dynamo to dynamo phase.}
	\label{ME at different beta}
\end{figure*}

We have already discussed the the effect of magnetic Reynolds number ($R_m$) on different class of YM flow at different fluid helicity. We now present our results keeping a particular $R_m$ fixed, we study the effect of $\beta$ or the effect of fluid helicity on the flow. For $R_m = 5000$, it is observed that, by varying $\beta$ value from 0 to 1 magnetic energy generation rapidly enhances [See Fig. \ref{Rm = 5000}]. The reason behind this is explained via impact of fluid helicity on dynamo instability. For $\beta = 0.0$ there is no fluid helicity in the system, where as for $\beta = 1.0$ the flow is maximally helical, that generates dynamo action. The exponential energy growth for $\beta = 1.0$ flow or the ABC flow is well known so far, but here we identify a possible route that connects a non-dynamo regime ($\beta = 0.0$) to a dynamo regime ($\beta = 1.0$) via fluid helicity injection. From Fig. \ref{Rm = 5000} it is verified that, the transition from non-dynamo to dynamo regime is not abrupt, but is a continuous transition that is achieved via fluid helicity injection in small steps.

To bring out the importance of our numerical findings, we examine magnetic energy growth rate evolution with $R_m$ values $4000$ and $3000$ for all the class of YM flows (viz. $\beta = 0.0, 0.2, 0.4, 0.6, 0.8, 1.0$). From Fig. \ref{Rm = 4000}, \ref{Rm = 3000} the expected transition from non-dynamo to dynamo regime is identified via fluid helicity injection, as before. For $R_m = 2000$, a similar transition (non-dynamo to dynamo) is observed. From Fig. \ref{Rm = 2000} it is observed that, there is a cross over of energy between $\beta = 0.4$ and $\beta = 0.6$ case. The energy growth is higher for a flow which has lower fluid helicity for $R_m = 2000$. To understand this ``cross over'' we further investigate for few lower magnetic Reynolds numbers ($R_m$). In Fig. \ref{Rm = 1000} \& \ref{Rm = 450}, we plot the magnetic energy growth for $R_m = 1000$ \& $R_m = 450$ and observe that, the cross over is more prominent. It is also observed that, $\beta = 0.4$ \& $\beta = 0.6$ shows faster dynamo action than the regular ABC flow. This is striking and from Fig. \ref{Rm = 1000} \& \ref{Rm = 450} one can conclude that, the YM flow with \textcolor{black}{$\beta = 1.0$} or the classical ABC flow is not the best dynamo model at lower magnetic Reynolds numbers ($R_m$). This particular observation suggests existence of a non-monotonous relationship between flow field helicity and dynamo. It would be interesting to explore this parameter space in greater detail, in the future.

\begin{figure*}
	\centering
	\begin{subfigure}{0.49\textwidth}
		\centering
		\includegraphics[scale=0.55]{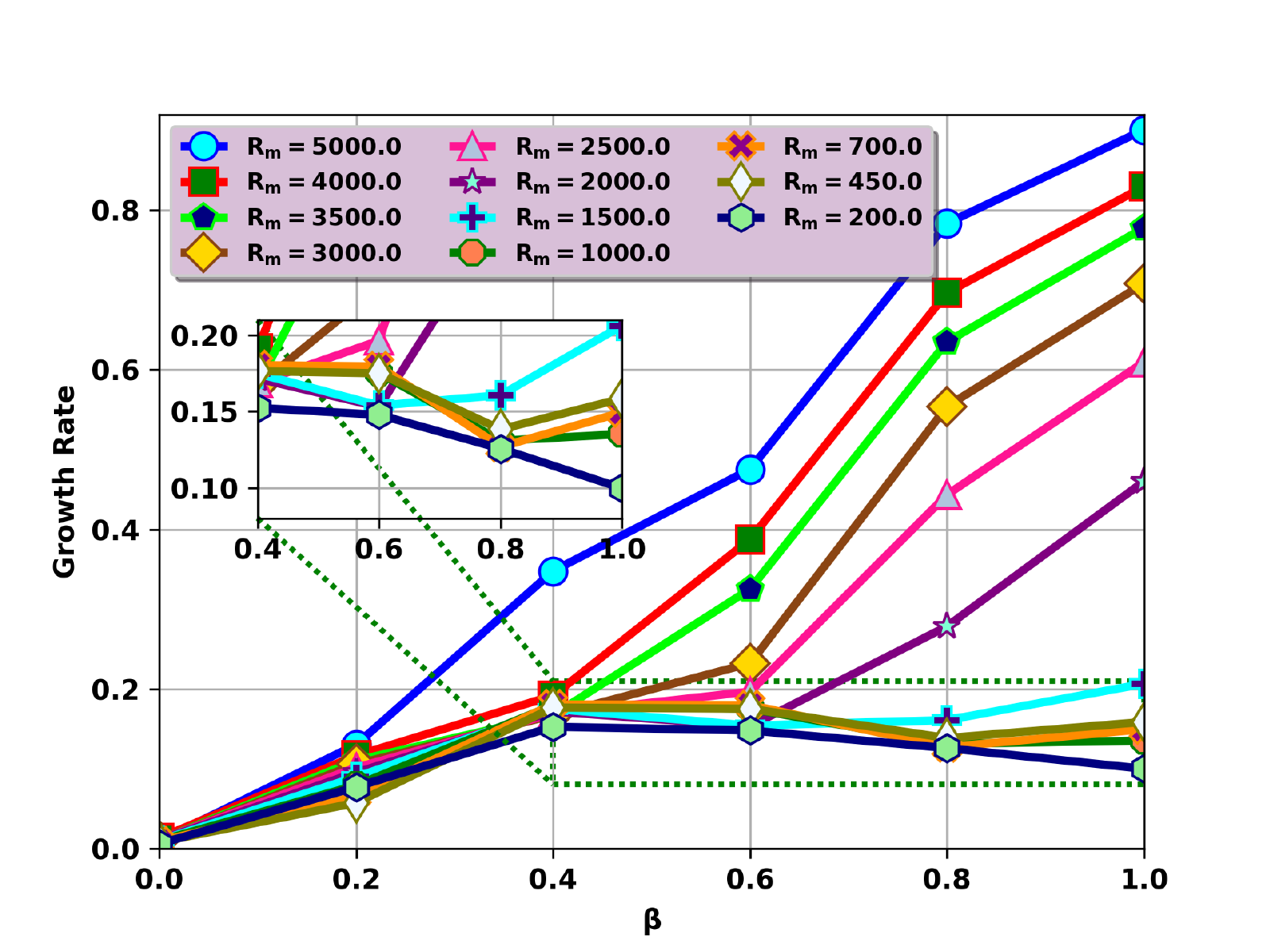}
	\end{subfigure}
	\caption{Magnetic energy growth rate for various class of Yoshida-Morrison (YM) flow at  fixed magnetic Reynolds numbers ($R_m$). A notable transition from non-dynamo regime to  dynamo regime is observed in growth rate, at the cost of fluid helicity injection, for higher magnetic Reynolds number ($R_m$).}
	\label{growth vs beta}
\end{figure*}

We also calculate the magnetic energy growth rate for the above discussed cases. As discussed above, a clear transition from non-dynamo regime to  dynamo regime is observed in growth rate as a function of fluid helicity, for higher magnetic Reynolds number ($R_m$) [See Fig. \ref{growth vs beta}]. From Fig. \ref{growth vs beta} it is also observed that the for lower magnetic Reynolds number ($R_m$) the ABC model is not the best dynamo model.

\textcolor{black}{It is important to indicate that, for this present study addressed so far, we have fixed $u_0 = 1.0$ and changed $\beta$, which basically changes the kinetic energy of the driving flow. We have also considered another possible approach and performed few test runs, in which we have kept the average kinetic energy ($<E>$) constant and normalized the magnetic Reynolds number using $<E>$. The same unambiguous shift from the non-dynamo to the dynamo regime is seen in our numerical experiment using this initial condition (not shown here). Thus, the major finding is unaffected by maintaining a constant value for the average kinetic energy.}

\section{Summary and Conclusion}
In this work, we have performed direct numerical simulations of 3-dimensional magnetohydrodynamic plasmas at sufficiently high grid resolutions.

We have analyzed \textcolor{black}{kinematic fast} dynamo model using YM flow recently proposed by \citep{EPI2D:2017}. An interesting and useful aspect of this flow is that, it is possible to inject finite fluid helicity [$\int_{V}\vec{u}.(\vec{\nabla} \times \vec{u}) dV$] in the system, by systematically varying certain physically meaningful parameter.\\

Our major findings are: \\

$\bullet$ Using a simple \textcolor{black}{kinematic fast} dynamo model, we demonstrate, a fast dynamo action primarily due to normalized fluid helicity injection. By injecting normalized fluid helicity in the system systematically, we have explored a systematic route that connects ``non-dynamo'' to ``dynamo'' regime, via direct numerical simulation.  

$\bullet$ The study of magnetic field iso-surface is seen to exhibit untwisted ribbon like structures, when there is no helicity injection, where as, helicity injection is shown to introduce twisting in the iso-surfaces that generates dynamo. 

$\bullet$ We demonstrate as to, how an untwisted ribbon like non-dynamo iso-surface gets converted into cigar like fast dynamo iso-surface  with the injection of fluid helicty.

$\bullet$ \textcolor{black}{We have calculated the time-averaged magnetic energy spectrum and it is observed that, the spectra contain a discernible maxima at a higher mode number, which is the distinguishing feature of small scale dynamo (SSD).}

$\bullet$ \textcolor{black}{The conventional understanding is that a lack of reflectional symmetry (e.g., non-zero fluid helicity) is necessary for large scale dynamo action (LSD), whereas, for small scale dynamo (SSD), it is not. In our work, for the flow considered here, we demonstrate how helical structure of the flow does strongly controls the SSD growth and spectral structure. SSD. This is one of the interesting findings of the present work.}

$\bullet$ We also identify that the most studied ABC dynamo model is not the best suited dynamo model for lower magnetic Reynolds number problems.\\

To conclude, with high enough resolution and at various magnetic Reynolds number, we investigate a systematic route that connects non-dynamo to dynamo regime via fluid helicity injection. Our observation is seen to conform with the magnetic iso-surface dynamics. We have shown that, how an untwisted ribbon like non-dynamo iso-surface slowly converts to twisted ribbon like dynamo iso-surface and finally leads to cigar like fast dynamo iso-surface  with injection of fluid helicty. \textcolor{black}{Our work also indicates that, the fluid helicity does affect the
	dynamics of small-scale dynamos.} We believe that, this work brings out unambiguously, the role of fluid helicity \textcolor{black}{on small-scale dynamos and on the transition} from ``non-dynamo'' to ``dynamo'' regime, via direct numerical simulation.

In our present work, it is demonstrated that the role of fluid helicity is the key factor to exhibit dynamo action.  However, in the past, existence of non-helical flows (i.e, flows with zero fluid helicity) to be able to generate dynamo, in \textcolor{black}{Kinematic Fast} and Saturation (with magnetic feedback) dynamos models \citep{galloway_nature:1992, Sur:2009}, wherein cross-helicity (also called Yoshizawa effect) is shown to play a crucial role. However the investigation of the role of cross-helicity (also called Yoshizawa effect) on YM flow and others, in the context of various dynamo action is beyond the scope of the present work and will be addressed in future communication. The effect of fluid helicity injection in the context of recurrence \citep{RM_recurrence:2019} and quasi-recurrence \citep{biswas:2022} phenomenon is also an interesting study to investigate out, that will also be reported soon. \textcolor{black}{The role of hydro-dynamically unstable shear flow \citep{Shishir_POF:2022} at the largest scale is interesting to investigate in this context. We hope to attempt this problem in the near future.}

\section{ACKNOWLEDGMENTS}
The simulations and visualizations presented here are performed on GPU nodes and visualization nodes of Antya cluster at the Institute for Plasma Research (IPR), INDIA. One of the author S.B is thankful to Dr. Rupak Mukherjee at Central University of Sikkim (CUS), Gangtok, Sikkim, India for providing an initial version of \textit{GMHD3D} code. S.B thanks N. Vydyanathan, Bengaluru and B. K. Sharma at NVIDIA, Bengaluru, India, for extending their help with  basic GPU methods. S.B is grateful to Mr. Soumen De Karmakar at IPR, India for many helpful discussions regarding GPUs, and HPC support team of IPR for extending their help related to ANTYA cluster. The authors would also like to thank Dr. Jugal Chowdhury from IPR for careful reviewing the manuscript and for his valuable suggestions.

\section{DATA AVAILABILITY}
The data underlying this article will be shared on reasonable request to the corresponding author.

\section{Conflict of Interest}
The authors have no conflicts to disclose.

\section{SUPPORTING INFORMATION}
Supplementary movies:
\begin{itemize}
	\item \textbf{YM\_beta\_0p0\_B\_Field\_IsoB.mp4}
	\item \textbf{\textcolor{black}{YM\_beta\_0p2\_B\_Field\_IsoB.mp4}}
	\item \textbf{YM\_beta\_0p6\_B\_Field\_IsoB.mp4}
	\item \textbf{YM\_beta\_1p0\_B\_Field\_IsoB.mp4}
\end{itemize}
 are added.


\bibliographystyle{mnras}
\bibliography{ShishirB_RGanesh_Dynamo_MNRAS} 

\begin{thebibliography}{}
\makeatletter
\relax
\def\mn@urlcharsother{\let\do\@makeother \do\$\do\&\do\#\do\^\do\_\do\%\do\~}
\def\mn@doi{\begingroup\mn@urlcharsother \@ifnextchar [ {\mn@doi@}
  {\mn@doi@[]}}
\def\mn@doi@[#1]#2{\def\@tempa{#1}\ifx\@tempa\@empty \href
  {http://dx.doi.org/#2} {doi:#2}\else \href {http://dx.doi.org/#2} {#1}\fi
  \endgroup}
\def\mn@eprint#1#2{\mn@eprint@#1:#2::\@nil}
\def\mn@eprint@arXiv#1{\href {http://arxiv.org/abs/#1} {{\tt arXiv:#1}}}
\def\mn@eprint@dblp#1{\href {http://dblp.uni-trier.de/rec/bibtex/#1.xml}
  {dblp:#1}}
\def\mn@eprint@#1:#2:#3:#4\@nil{\def\@tempa {#1}\def\@tempb {#2}\def\@tempc
  {#3}\ifx \@tempc \@empty \let \@tempc \@tempb \let \@tempb \@tempa \fi \ifx
  \@tempb \@empty \def\@tempb {arXiv}\fi \@ifundefined
  {mn@eprint@\@tempb}{\@tempb:\@tempc}{\expandafter \expandafter \csname
  mn@eprint@\@tempb\endcsname \expandafter{\@tempc}}}

\bibitem[\protect\citeauthoryear{Alexakis}{Alexakis}{2011}]{Alexakis_PRE:2011}
Alexakis A.,  2011, \mn@doi [Phys. Rev. E] {10.1103/PhysRevE.83.036301}, 83,
  036301

\bibitem[\protect\citeauthoryear{{Archontis, V.}, {Dorch, S. B. F.}  \&
  {Nordlund, \AA{}.}}{{Archontis, V.} et~al.}{2003}]{Archontis:2003}
{Archontis, V.} {Dorch, S. B. F.}  {Nordlund, \AA{}.} 2003, \mn@doi [A\&A]
  {10.1051/0004-6361:20021568}, 397, 393

\bibitem[\protect\citeauthoryear{Arnold \& Korkina}{Arnold \&
  Korkina}{1983}]{arnold:1983}
Arnold V.,  Korkina E.,  1983, Matem. Mekh., 3, 43

\bibitem[\protect\citeauthoryear{Beresnyak}{Beresnyak}{2019}]{beresnyak_mhd:2019}
Beresnyak A.,  2019, Living Reviews in Computational Astrophysics, 5, 1

\bibitem[\protect\citeauthoryear{Biskamp}{Biskamp}{2003}]{biskamp:2003}
Biskamp D.,  2003, Magnetohydrodynamic turbulence.
Cambridge University Press

\bibitem[\protect\citeauthoryear{Biswas \& Ganesh}{Biswas \&
  Ganesh}{2022}]{Shishir_POF:2022}
Biswas S.,  Ganesh R.,  2022, \mn@doi [Physics of Fluids] {10.1063/5.0092212},
  34, 065101

\bibitem[\protect\citeauthoryear{Biswas, Ganesh  \& et al.}{Biswas
  et~al.}{2022a}]{GTC}
Biswas S.,  Ganesh R.,   et al. 2022a, {GPU Technology Conference 2022},
  \url{https://www.nvidia.com/en-us/on-demand/session/gtcspring22-s41199/}

\bibitem[\protect\citeauthoryear{Biswas, Ganesh, Mukherjee  \& Sen}{Biswas
  et~al.}{2022b}]{biswas:2022}
Biswas S.,  Ganesh R.,  Mukherjee R.,   Sen A.,  2022b, ``Three Dimensional
  MagnetoHydroDynamics of EPI Two Dimensional Flows: Quasi-recurrence and
  nonlinear Alfven wave oscillations'', Manuscript under preparation

\bibitem[\protect\citeauthoryear{Bouya \& Dormy}{Bouya \&
  Dormy}{2013}]{Bouya:2013}
Bouya I.,  Dormy E.,  2013, \mn@doi [Physics of Fluids] {10.1063/1.4795546},
  25, 037103

\bibitem[\protect\citeauthoryear{Brachet, Meneguzzi, Politano  \&
  Sulem}{Brachet et~al.}{1988}]{Brachet:1988}
Brachet M.~E.,  Meneguzzi M.,  Politano H.,   Sulem P.~L.,  1988, \mn@doi
  [Journal of Fluid Mechanics] {10.1017/S0022112088003015}, 194, 333–349

\bibitem[\protect\citeauthoryear{Cattaneo, Kim, Proctor  \& Tao}{Cattaneo
  et~al.}{1995}]{Cattaneo_PRL:1995}
Cattaneo F.,  Kim E.-j.,  Proctor M.,   Tao L.,  1995, \mn@doi [Phys. Rev.
  Lett.] {10.1103/PhysRevLett.75.1522}, 75, 1522

\bibitem[\protect\citeauthoryear{Cattaneo, Hughes  \& Kim}{Cattaneo
  et~al.}{1996}]{Cattaneo_PRL:1996}
Cattaneo F.,  Hughes D.~W.,   Kim E.-j.,  1996, \mn@doi [Phys. Rev. Lett.]
  {10.1103/PhysRevLett.76.2057}, 76, 2057

\bibitem[\protect\citeauthoryear{Childress \& Gilbert}{Childress \&
  Gilbert}{1995}]{childress_STF:1995}
Childress S.,  Gilbert A.~D.,  1995, Stretch, twist, fold: the fast dynamo.
 Vol. 37, Springer Science \& Business Media

\bibitem[\protect\citeauthoryear{Choudhuri}{Choudhuri}{1990}]{choudhuri:1990}
Choudhuri A.~R.,  1990, The Astrophysical Journal, 355, 733

\bibitem[\protect\citeauthoryear{Courvoisier, Hughes  \& Tobias}{Courvoisier
  et~al.}{2006}]{alpha_effect:2006}
Courvoisier A.,  Hughes D.~W.,   Tobias S.~M.,  2006, \mn@doi [Phys. Rev.
  Lett.] {10.1103/PhysRevLett.96.034503}, 96, 034503

\bibitem[\protect\citeauthoryear{Dombre, Frisch, Greene, Hénon, Mehr  \&
  Soward}{Dombre et~al.}{1986}]{dombre:1986}
Dombre T.,  Frisch U.,  Greene J.~M.,  Hénon M.,  Mehr A.,   Soward A.~M.,
  1986, \mn@doi [Journal of Fluid Mechanics] {10.1017/S0022112086002859}, 167,
  353–391

\bibitem[\protect\citeauthoryear{Dorch}{Dorch}{2000}]{Dorch:2000}
Dorch S. B.~F.,  2000, \mn@doi [Physica Scripta]
  {10.1238/physica.regular.061a00717}, 61, 717

\bibitem[\protect\citeauthoryear{Galanti, Sulem  \& Pouquet}{Galanti
  et~al.}{1992}]{galanti:1992}
Galanti B.,  Sulem P.~L.,   Pouquet A.,  1992, \mn@doi [Geophysical \&
  Astrophysical Fluid Dynamics] {10.1080/03091929208229056}, 66, 183

\bibitem[\protect\citeauthoryear{Galloway \& Frisch}{Galloway \&
  Frisch}{1984}]{Frish_Dynamo:1984}
Galloway D.,  Frisch U.,  1984, \mn@doi [Geophysical \& Astrophysical Fluid
  Dynamics] {10.1080/03091928408248180}, 29, 13

\bibitem[\protect\citeauthoryear{Galloway \& Frisch}{Galloway \&
  Frisch}{1986}]{Frish_Dynamo:1986}
Galloway D.,  Frisch U.,  1986, \mn@doi [Geophysical \& Astrophysical Fluid
  Dynamics] {10.1080/03091928608208797}, 36, 53

\bibitem[\protect\citeauthoryear{Galloway \& Proctor}{Galloway \&
  Proctor}{1992}]{galloway_nature:1992}
Galloway D.~J.,  Proctor M.~R.,  1992, Nature, 356, 691

\bibitem[\protect\citeauthoryear{Gholami, Hill, Malhotra  \& Biros}{Gholami
  et~al.}{2016}]{Accfftw:2016}
Gholami A.,  Hill J.,  Malhotra D.,   Biros G.,  2016, \mn@doi
  [http://arxiv.org/abs/1506.07933] {arXiv:1506.07933}

\bibitem[\protect\citeauthoryear{Hughes, Cattaneo  \& jin Kim}{Hughes
  et~al.}{1996}]{Hughes_PLA:1996}
Hughes D.~W.,  Cattaneo F.,   jin Kim E.,  1996, \mn@doi [Physics Letters A]
  {https://doi.org/10.1016/S0375-9601(96)00722-0}, 223, 167

\bibitem[\protect\citeauthoryear{Kitware}{Kitware}{2022}]{paraview}
Kitware 2022, {Paraview}, \url{https://www.paraview.org/}

\bibitem[\protect\citeauthoryear{Kraichnan}{Kraichnan}{1965}]{Kraichnan:1965}
Kraichnan R.~H.,  1965, \mn@doi [The Physics of Fluids] {10.1063/1.1761412}, 8,
  1385

\bibitem[\protect\citeauthoryear{LLNL}{LLNL}{2020}]{visit}
LLNL 2020, {VisIt},
  \url{https://wci.llnl.gov/simulation/computer-codes/visit/releases/release-notes-3.1.2}

\bibitem[\protect\citeauthoryear{Mukherjee}{Mukherjee}{2019}]{rupak_thesis:2019}
Mukherjee R.,  June,2019, ``Turbulence, Flows and Magnetic Field Generation in
  Plasmas using a Magnetohydrodynamic Model'': HBNI Phd Thesis

\bibitem[\protect\citeauthoryear{Mukherjee, Ganesh  \& Sen}{Mukherjee
  et~al.}{2019}]{RM_recurrence:2019}
Mukherjee R.,  Ganesh R.,   Sen A.,  2019, \mn@doi [Physics of Plasmas]
  {10.1063/1.5083052}, 26, 022101

\bibitem[\protect\citeauthoryear{Parker}{Parker}{1979}]{parker:1979}
Parker E.,  1979, Cosmical Magnetic Fields, Clarendon

\bibitem[\protect\citeauthoryear{Patterson \& Orszag}{Patterson \&
  Orszag}{1971}]{dealiasing:1971}
Patterson G.~S.,  Orszag S.~A.,  1971, \mn@doi [The Physics of Fluids]
  {10.1063/1.1693365}, 14, 2538

\bibitem[\protect\citeauthoryear{Paulo-herrera}{Paulo-herrera}{2021}]{VTK}
Paulo-herrera 2021, {PyEVTK}, \url{https://github.com/paulo-herrera/PyEVTK}

\bibitem[\protect\citeauthoryear{R\"adler \& Brandenburg}{R\"adler \&
  Brandenburg}{2003}]{Brandenburg_PRE:2003}
R\"adler K.-H.,  Brandenburg A.,  2003, \mn@doi [Phys. Rev. E]
  {10.1103/PhysRevE.67.026401}, 67, 026401

\bibitem[\protect\citeauthoryear{Rincon}{Rincon}{2019}]{rincon_dynamo:2019}
Rincon F.,  2019, \mn@doi [Journal of Plasma Physics]
  {10.1017/S0022377819000539}, 85, 205850401

\bibitem[\protect\citeauthoryear{Schekochihin}{Schekochihin}{2022}]{Schekochihin:2022}
Schekochihin A.~A.,  2022, \mn@doi [Journal of Plasma Physics]
  {10.1017/S0022377822000721}, 88, 155880501

\bibitem[\protect\citeauthoryear{Squire \& Bhattacharjee}{Squire \&
  Bhattacharjee}{2014a}]{amitava:2014a}
Squire J.,  Bhattacharjee A.,  2014a, \mn@doi [Phys. Rev. Lett.]
  {10.1103/PhysRevLett.113.025006}, 113, 025006

\bibitem[\protect\citeauthoryear{Squire \& Bhattacharjee}{Squire \&
  Bhattacharjee}{2014b}]{amitava:2014b}
Squire J.,  Bhattacharjee A.,  2014b, \mn@doi [The Astrophysical Journal]
  {10.1088/0004-637x/797/1/67}, 797, 67

\bibitem[\protect\citeauthoryear{Squire \& Bhattacharjee}{Squire \&
  Bhattacharjee}{2015}]{amitava_APJ:2015}
Squire J.,  Bhattacharjee A.,  2015, \mn@doi [The Astrophysical Journal]
  {10.1088/0004-637x/813/1/52}, 813, 52

\bibitem[\protect\citeauthoryear{Sur \& Brandenburg}{Sur \&
  Brandenburg}{2009}]{Sur:2009}
Sur S.,  Brandenburg A.,  2009, \mn@doi [Monthly Notices of the Royal
  Astronomical Society] {10.1111/j.1365-2966.2009.15254.x}, 399, 273

\bibitem[\protect\citeauthoryear{Tobias \& Cattaneo}{Tobias \&
  Cattaneo}{2013}]{tobias_shear:2013}
Tobias S.~M.,  Cattaneo F.,  2013, Nature, 497, 463

\bibitem[\protect\citeauthoryear{Vainshtein \& Zel'dovich}{Vainshtein \&
  Zel'dovich}{1972}]{Zeldovich:1972}
Vainshtein S.~I.,  Zel'dovich Y.~B.,  1972, \mn@doi [Soviet Physics Uspekhi]
  {10.1070/PU1972v015n02ABEH004960}, 15, 159

\bibitem[\protect\citeauthoryear{Yoshida \& Morrison}{Yoshida \&
  Morrison}{2017}]{EPI2D:2017}
Yoshida Z.,  Morrison P.~J.,  2017, \mn@doi [Phys. Rev. Lett.]
  {10.1103/PhysRevLett.119.244501}, 119, 244501

\bibitem[\protect\citeauthoryear{Biswas \& Ganesh}{ami}{}]{amitava_PRL:2015}


\bibitem[\protect\citeauthoryear{ami}{zel}{}]{zeldovich:1957}


\makeatother
\end{thebibliography}



\onecolumn 
\appendix
\section{Fluid helicity calculation for Yoshida-Morrison (YM) flow}\label{Appen A}
Fluid helicity is defined as, \begin{equation*}
	H_f = \int_{V}\vec{u}.(\vec{\nabla} \times \vec{u}) dV.
\end{equation*} 

Here, \begin{equation*}
	\vec{u} = \hat{i} \left( \alpha u_0 [ B \sin(k_0y) - C \cos(k_0z) ] \right) + \hat{j} \left(  \beta u_0 [ C \sin(k_0z) - A \cos(k_0x) ] \right) + \hat{k} \left(  u_0 [ \alpha A \sin(k_0x) - \beta B \cos(k_0y) ] \right).
\end{equation*}

Taking curl and putting the real constant value, $\alpha = A = B = C = 1.0$, $k_0 = 1.0$ , we get,
\begin{equation*}
	\vec{\nabla} \times \vec{u} = \hat{i} \left( \beta u_0 [\sin y - \cos z] \right) + \hat{j} \left( u_0 [\cos x - \sin z] \right) + \hat{k} \left(u_0 [\beta \sin x -  \cos y] \right)
\end{equation*}

Taking $\vec{u} \cdot$ on both side,

\begin{equation*}
	\begin{aligned}
		\vec{u} \cdot \vec{\nabla} \times \vec{u} =  &\left( u_0 [\sin y - \cos z ] \right)  \left( \beta u_0 [\sin y - \cos z] \right) + \left(  \beta u_0 [ \sin z -  \cos x ] \right) \left(  u_0 [\cos x - \sin z] \right) \\ 
		& + \left(  u_0 [   \sin x - \beta  \cos y ] \right) \left(u_0 [\beta \sin x -  \cos y] \right).
	\end{aligned}
\end{equation*}

or, \begin{equation*}
	\begin{aligned}
		\vec{u} \cdot \vec{\nabla} \times \vec{u} =   \beta u_0^{2} \left[  (\sin y - \cos z )^2 \right] -  \beta u_0^{2} \left[  (\cos x - \sin z )^2 \right] + u_0^{2} \left[  ( \sin x - \beta \cos y ) (\beta \sin x -  \cos y ) \right] 
	\end{aligned}
\end{equation*}

or, \begin{equation*}
	\begin{aligned}
		\vec{u} \cdot \vec{\nabla} \times \vec{u} = 
		&  \beta u_0^{2} \left[  (\sin^2 y + \cos^2 z  - 2\sin y  \cos z) \right] -  \beta u_0^{2} \left[  (\cos^2 x + \sin^2 z - 2 \cos x \sin z ) \right] \\
		&+ u_0^{2} \left[  ( \beta \sin^2 x -  \sin x \cos y - \beta^2 \sin x  \cos y +  \beta \cos^2 y ) \right] 
	\end{aligned}
\end{equation*}

Integrating over total volume,\\
\begin{equation*}
	\begin{aligned}
		\int_{V}\vec{u}.(\vec{\nabla} \times \vec{u}) dV = 
		&  \beta u_0^{2} \left[  (\int_{0}^{2\pi}\int_{0}^{2\pi}\int_{0}^{2\pi}\sin^2y dx dy dz + \int_{0}^{2\pi}\int_{0}^{2\pi}\int_{0}^{2\pi} \cos^2 z dx dy dz - \int_{0}^{2\pi}\int_{0}^{2\pi}\int_{0}^{2\pi} 2\sin y \cos z dx dy dz) \right]\\  
		& -  \beta u_0^{2} \left[  (\int_{0}^{2\pi}\int_{0}^{2\pi}\int_{0}^{2\pi}\cos^2 x dx dy dz + \int_{0}^{2\pi}\int_{0}^{2\pi}\int_{0}^{2\pi} \sin^2 z dx dy dz -\int_{0}^{2\pi}\int_{0}^{2\pi}\int_{0}^{2\pi}2 \cos x \sin z dx dy dz) \right] \\
		&+  \beta u_0^{2} \left[  ( \int_{0}^{2\pi}\int_{0}^{2\pi}\int_{0}^{2\pi}\sin^2 x dx dy dz \right] -   u_0^{2} \left[\int_{0}^{2\pi}\int_{0}^{2\pi}\int_{0}^{2\pi} \sin x \cos y dx dy dz\right]\\
		&- \beta^2 u_0^{2} \left[\int_{0}^{2\pi}\int_{0}^{2\pi}\int_{0}^{2\pi} \sin x  \cos y dx dy dz \right]+  \beta u_0^{2} \left[\int_{0}^{2\pi}\int_{0}^{2\pi}\int_{0}^{2\pi}\cos^2 y dx dy dz  \right]
	\end{aligned}
\end{equation*}
Finally after simplification one obtains, \begin{equation*}
	\int_{V}\vec{u}.(\vec{\nabla} \times \vec{u}) dV =   8\pi^3\beta u_0^2
\end{equation*}
Considering velocity to be normalized,

\begin{equation*}
	H_f =  \beta 
\end{equation*}
So normalized fluid helicity for YM flow depends upon constant $\beta$ value of the flow.

\section{Grid size scaling study for ABC flow}\label{Appen B}
\begin{figure}
	\begin{subfigure}{0.49\textwidth}
		\centering
		\includegraphics[scale=0.54]{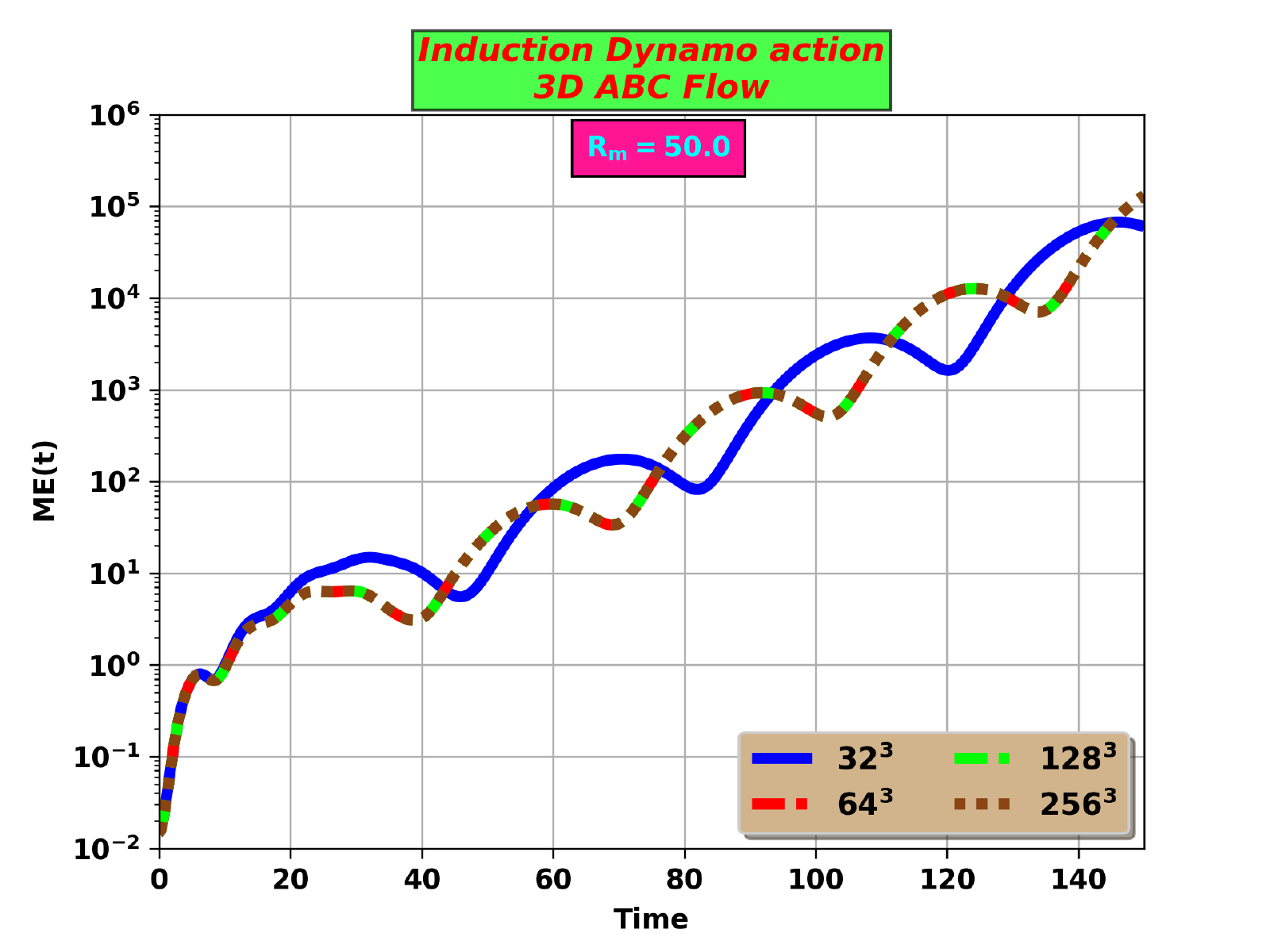}
		\caption{}
		\label{Scaling_Rm50}
	\end{subfigure}
	\begin{subfigure}{0.49\textwidth}
		\centering
		\includegraphics[scale=0.54]{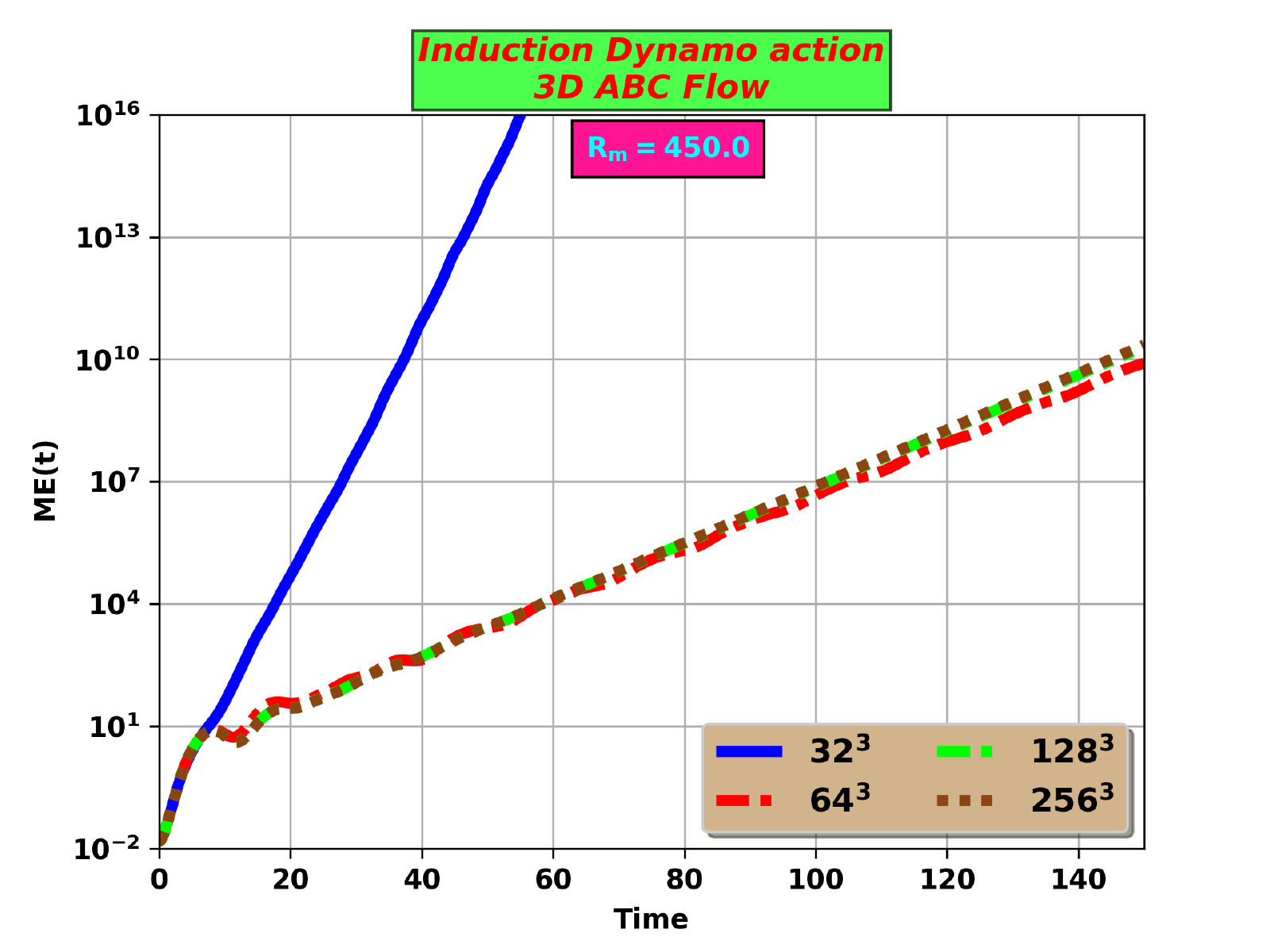}
		\caption{}
		\label{Scaling_Rm450}
	\end{subfigure}
	\caption{\textcolor{black}{Fast} Dynamo effect following Galloway et al. \citep{Frish_Dynamo:1984} using Arnold-Beltrami-Childress (ABC) flow for magnetic Reynolds numbers (a) $R_m = 50.0$ (b) $R_m = 450.0$ at different grid resolutions.}
\end{figure}

Grid size scaling study have been performed using Arnold-Beltrami-Childress (ABC) flow [ Eq. \ref{ABC Flow} ] at different magnetic Reynolds numbers. It is seen that $256^3$ grid resolution is more than enough for this problem [See Fig. \ref{Scaling_Rm50}, \ref{Scaling_Rm450}].
\nocite{*}


\label{lastpage}
\end{document}